\documentclass[11pt]{article}
\usepackage{jheppub}
\usepackage{amsmath,amssymb,amsfonts,amsthm,graphicx,physics}
\usepackage{setspace}
\usepackage{subcaption}
\usepackage{dsfont}
\usepackage{leftindex}
\usepackage{bbm}

\usepackage[left=1in,right=1in,top=1in,bottom=1.2in]{geometry}
\usepackage{xcolor}
\usepackage{simpler-wick}

\usepackage{tikz}
    \usepackage{amssymb,amsfonts,amsmath}
    \usepackage{tkz-euclide}
        \usetikzlibrary{arrows,calc,patterns}
\usepackage{pgfplots}

\definecolor{darkblue}{rgb}{0.1,0.1,.7}
\definecolor{purple}{rgb}{0.6,0,0.6}
\definecolor{orange}{rgb}{0.9,0.6,0}
\definecolor{llgray}{rgb}{0.9,0.9,1}
\definecolor{dgreen}{rgb}{0,0.5,0}
\definecolor{outsideyellow}{rgb}{0.96,0.86,0.41}
\definecolor{insideyellow}{rgb}{0.99,0.96,0.82}
\definecolor{observerred}{rgb}{0.93,0.27,0.13}
\definecolor{mattergreen}{rgb}{0,0.63,0.29}
\usepackage[]{latexsym}
\usepackage[utf8]{inputenc}
\usepackage{geometry}
\usepackage{amscd}
\usepackage[all,cmtip]{xy}
\usepackage{mathrsfs}

\usepackage[margin=10pt,font=small,labelfont=bf]{caption}
\usepackage{changepage}
\usepackage{setspace}
\usepackage{tikz}
\usepackage{subcaption}
\usepackage{ marvosym }
\usepackage{tensor}
\usepackage{CJKutf8}
\usepackage{empheq}
\usepackage{color,graphicx}

\usepackage[ragged]{footmisc}
\definecolor{webgreen}{rgb}{0, 0.5, 0}
\definecolor{webblue}{rgb}{0, 0, 0.5}
\definecolor{webred}{rgb}{0.5, 0, 0}
\definecolor{darkgreen}{rgb}{0,0.5,0}

\def\ben{\begin{equation}}
	\def\een{\end{equation}}

     \let\r=v

\def\be{\begin{equation}}
	\def\ee{\end{equation}}
\def\ba{\begin{array}}
	\def\ea{\end{array}}

\def\dalemb#1#2{{\vbox{\hrule height .#2pt
			\hbox{\vrule width.#2pt height#1pt \kern#1pt
				\vrule width.#2pt}
			\hrule height.#2pt}}}

\newcommand{\bea}{\begin{eqnarray}}
	\newcommand{\eea}{\end{eqnarray}}

\let\tilde=\widetilde

\numberwithin{equation}{section}

\newcommand{\island}{I}

\title{\begin{center}
    An apologia for islands
\end{center}} 
\author[1]{Stefano Antonini,}
\author[1]{Chang-Han Chen,}
\author[2]{Henry Maxfield,}
\author[1]{Geoff Penington}
\affiliation[1]{Leinweber Institute for Theoretical Physics and Department of Physics, University of California, Berkeley, California 94720, U.S.A.}
\affiliation[2]{Stanford Institute for Theoretical Physics, Stanford University, Stanford, CA 94305, USA}
\emailAdd{santonini@berkeley.edu}
\emailAdd{changhanc@berkeley.edu}
\emailAdd{henrym@stanford.edu}
\emailAdd{geoffp@berkeley.edu}

\abstract{Entanglement islands have played a key role in the recent derivation of the Page curve and other progress on the black hole information problem. Arising from the inclusion of connected wormhole saddles in a gravitational replica trick, islands signal that degrees of freedom in the black hole interior are not microscopically independent of the exterior Hawking radiation. Islands were originally discovered in the context of AdS/CFT coupled to an external, nongravitating reservoir, where the coupling gives graviton excitations an anomalous boundary scaling dimension (or ``mass''). It has been claimed in the literature that this mass is crucial for the existence of islands and even the Page curve itself.  In this paper, however, we explain how entanglement islands can also appear in setups with massless gravitons and no external reservoir, giving a number of examples including the entanglement wedges of boundary CFT regions, of radiation at null infinity in asymptotically flat spacetimes, and of radiation inside a semiclassical but gravitating spacetime. In each case, the Page curve is physically observable and can be determined with sufficiently careful experiments on many copies of the black hole. We give general arguments for the existence of gauge-invariant operators in gravity which are  compactly supported  to all orders in perturbation theory (whenever no isometries of the background spacetime exist) and refine a recently-proposed explicit construction of such operators. When applied to islands, these results -- together with entanglement wedge reconstruction -- guarantee that semiclassical operators in the island can be approximated by nonperturbative operators on the Hawking radiation. 
}

\date{\today}

\begin{document}

\maketitle

\section{Introduction}
\label{sec:intro}

In perturbative semiclassical gravity, Hawking radiation emitted by a black hole is in a thermal mixed state (up to well understood greybody factors) and is purified by partner modes inside the black hole \cite{Hawking:1975vcx,Almheiri:2020cfm}. As a result, the entanglement entropy of this radiation appears to grow monotonically as the black hole evaporates. After the ``Page time'', roughly halfway through the evaporation process, this entropy becomes inconsistent with the unitarity of quantum mechanics, if one assumes that the number of possible microstates of the black hole is controlled by the Bekenstein-Hawking entropy $S_{BH} = A_{\rm hor}/4G$ \cite{Bekenstein:1972tm,Bekenstein:1973ur,Hawking:1974rv,Hawking:1975vcx}. As the black hole grows smaller, and more and more radiation is emitted, there are simply too few black hole microstates to be able to purify the radiation.

There is a different way of calculating the entropy of Hawking radiation that gives a more satisfying answer. Describing the full state of the Hawking radiation requires computing a density matrix with exponentially many entries, which means that it can only be used in controlled entanglement entropy computations if the individual matrix elements are known to exponential precision. Such precision is impossible without a nonperturbative theory of quantum gravity. However, one can  instead directly compute the R\'enyi entropy $S_n(\rho) = \frac{1}{1-n} \log \Tr(\rho^n)$ using a gravitational path integral on $n$ copies or ``replicas'' of the system \cite{Penington:2019npb,Almheiri:2019psf, Penington:2019kki,Almheiri:2019qdq}. The entanglement entropy $S(\rho) = - \Tr(\rho \log \rho)$ can then be found by analytically continuing $n\to 1$.

After taking this limit, the replica trick calculation reduces to the ``island rule'', which says that
\begin{align}\label{eq:island}
S(R) =  \underset{I}{\rm min\,\, ext} \left[S_{\rm gen} (R \cup I)\right] = \underset{I}{\rm min\,\, ext} \left[\frac{\mathcal{A}(\partial I)}{4G} + S_{\rm eff} (R \cup I)\right].
\end{align}
Here $S(R)$ is the entropy of the radiation found using the gravitational path integral, while $S_{\rm eff}(R)$ would represent the hypothetical entropy of the same quantum fields computed in effective field theory on a fixed spacetime background (i.e. the entropy computed by Hawking). Crucially, however, the entropy appearing on the right-hand side of \eqref{eq:island} is instead $S_{\rm eff}(R \cup I)$, which includes in addition all quantum fields in a particular spacetime region, called the island $I$. Moreover, in addition to the entropy $S_{\rm eff}(R \cup I)$, we have to include a separate contribution proportional to the area $\mathcal{A}(\partial I)$ of the edge of the island. The combination of these two terms is called the generalised entropy $S_{\rm gen} (R \cup I)$ of the radiation and island. To find the location of the island, we first need to look for all candidate islands $I'$ where $S_{\rm gen} (R \cup I')$ is invariant under first-order perturbations of the edge $\partial I$. Subregions $I'$ satisfying this property are said to be ``quantum extremal''. The true island $I$ is found by minimising $S_{\rm gen} (R \cup I)$ within this set.

If we apply the island rule before the Page time, one finds that the island is empty and $S(R) = S_{\rm eff}(R)$ as Hawking expected. 
After the Page time, however, one finds an entanglement island that includes most of the black hole interior. The existence of this island reflects the fact that for integer $n > 1$ the gravitational replica trick is dominated by saddles containing spacetime wormholes \cite{Penington:2019kki,Almheiri:2019qdq}. Because the island includes almost all of the modes that purify the radiation, we have $S_{\rm eff}(R \cup I) = O(1)$ and hence
\be
S(R) = S_{\rm gen}(R \cup I) \approx \frac{\mathcal{A}(\partial I)}{4G} \approx \frac{\mathcal{A}_{\rm hor}}{4G}.
\ee
We see that, rather than increasing indefinitely, the entropy $S(R)$ of the radiation saturates at precisely the Bekenstein-Hawking entropy $S_{BH}$ and then begins to decrease as the black hole continues to evaporate. The radiation is said to follow the ``Page curve'', exactly as needed to save unitarity.

How can entropy of the radiation decrease while its entanglement with the interior of the black hole continues to increase? The answer appears to be that, in a complete microscopic description of quantum gravity, e.g. the CFT in AdS/CFT \cite{tHooft:1993dmi,Susskind:1994vu,Maldacena:1997re,Witten:1998qj,Aharony:1999ti}, modes in the island are not independent degrees of freedom from the radiation. Instead, operators that act on the island in perturbative semiclassical gravity can be approximated, to all orders in perturbation theory, by operators acting only on the radiation in the complete theory. For historical reasons, we say that the island can be ``reconstructed'' from the radiation. Importantly, the reconstructed operators are exponentially complex \cite{Harlow:2013tf, Brown:2019rox} and, on appropriate code subspaces \cite{Hayden:2018khn, Akers:2019wxj, Akers:2021fut}, appear to commute to all orders in perturbation theory with simple operators acting on the radiation so that the semiclassical picture remains valid for all practical purposes \cite{Engelhardt:2021mue, Akers:2022qdl}.

In a number of papers \cite{Geng:2020fxl,Geng:2020qvw,Geng:2021hlu,Raju:2020smc,Geng:2023zhq,Geng:2025rov}, it has been argued that  entanglement islands are intimately connected to the existence of `massive gravitons'. The principal claims of \cite{Geng:2020fxl,Geng:2020qvw,Geng:2021hlu,Raju:2020smc,Geng:2023zhq,Geng:2025rov} are the following. Firstly, they argue that, in all known examples where islands have been found, gravitons have an anomalous boundary scaling dimension (the analogue in asymptotically AdS spacetimes of a nonzero pole mass), and that we should therefore question whether islands can exist at all in massless gravity.\footnote{One version of this argument was given in \cite{Geng:2020qvw}, where the authors showed that in a particular limit where the graviton mass goes to zero, the island stops becoming dominant for black holes that have been left to evaporate for any fixed, finite time. However, as acknowledged in \cite{Geng:2020qvw}, this occurs simply because the black hole evaporation time (and hence also the Page time) diverge in the limit they are taking. Consequently, at finite times, we would not expect or want islands to exist.} Secondly, they suggest that operator reconstruction in an island may be impossible in the absence of massive gravitons, because gravitational constraints prevent gauge-invariant operators from being localised in a compact region such as an island. 
A final, related assertion is that the evaporation of black holes in asymptotically flat spacetimes (with massless gravity) is not described by a Page curve (instead the relevant entropy is constant) and so there is no reason that islands are needed in that setting \cite{Laddha_2021,Raju:2020smc}.

The goal of this paper is to respond to the claims of \cite{Laddha_2021,Geng:2020fxl,Geng:2020qvw,Geng:2021hlu,Raju:2020smc,Geng:2023zhq,Geng:2025rov} and to argue that islands exist, and thereby reveal the unitarity of black hole evaporation, in any reasonable theory of quantum gravity. We start in Section \ref{sec:review} by reviewing the derivation of entanglement islands from the gravitational replica trick and explaining how the appearance of islands is a generic mechanism that rescues unitarity when an information paradox could arise, independent of the specific details of the setup. We also clarify the regime of validity of entanglement wedge reconstruction, and in what sense `massive gravitons' arise when coupling gravitational systems to non-gravitational external reservoir.

In Section \ref{sec:masslessislands} we discuss the existence of islands in theories of massless gravity. First, in Section \ref{sec:boundryislands} we propose setups in which islands for the Hawking radiation of an evaporating black hole exist in asymptotically AdS spacetimes not coupled to external reservoirs, thereby avoiding the appearance of massive gravitons. In Section \ref{sec:flat} we discuss asymptotically flat spacetimes. In this context, one can study the entropy of proper subalgebras describing Hawking radiation defined at future null infinity $\mathscr{I}^+$ prior to some retarded time $u_0$. As $u_0$ increases, the algebra describes a larger portion of Hawking radiation. The entropy is expected to follow a Page curve as a function of $u_0$. We explain how to construct these proper subalgebras for matter quantum fields and gravitons, and point out that their entropy does indeed follow a Page curve. We constrast this algebra with a larger algebra, containing the ADM Hamiltonian, that was used in \cite{Laddha_2021} to obtain a flat entropy curve. We argue that the latter algebra includes information about nonradiative degrees of freedom and was never supposed to follow a Page curve. In Section \ref{sec:bulkradiation}, we describe how to define subalgebras and entropies for radiation that has not reached asymptotic infinity, focusing on the case of small black holes in AdS/CFT.  Finally, in Section \ref{sec:experiment} we discuss in some detail how the Page curve of Hawking radiation could in principle be measured in massless gravity theories by a bulk observer with a fault-tolerant quantum computer (independent of asymptotics). We also discuss how this experiment is robust against the loss of a large fraction of the radiation.

We then move on to the issue of the existence of compactly supported, gauge-invariant operators in massless theories of gravity. It is unlikely that such operators can exist non-perturbatively (even if one can make sense of compact support in a non-perturbative setting), see e.g. \cite{Donnelly:2015hta,Donnelly:2016rvo,Giddings:2005id}. There are also obstructions to their existence in perturbation theory around backgrounds with isometries such as Minkowski or AdS spacetimes. However, they can exist at all orders in perturbation theory in $G$ (which is what we require for reconstruction results) around a  background spacetime which breaks all symmetries. In Section \ref{sec:linearconstraints}, we show that any compactly supported source of energy-momentum can be `dressed' by compactly supported metric perturbations so that the gravitational constraints are satisfied in perturbation theory around any background which breaks all symmetries. We then explain how this can be used to construct a diffeomorphism-invariant dressed (quantum) operator from an arbitrary matter operator, order by order in perturbation theory. The intuitive idea is that, if the spacetime breaks all symmetries, operators can be made gauge invariant by dressing them to features of the spacetime such as a lump of matter (or, more generically, features of the quantum gravity state) instead of asymptotic data. As a result, at the perturbative level, the operator is supported on a compact bulk subregion---which includes the point where we want to act with the operator as well as the feature to which we are dressing the operator---and therefore commutes with the asymptotic charges. We also address certain important subtleties when the background spacetime has approximate symmetries, such as time-translation for an evaporating black hole. Our arguments imply the existence of gauge-invariant operators localised in the island for generic black hole spacetimes (which do break all symmetries), in the perturbative sense required for reconstruction theorems. The non-existence of such operators was the main argument against the existence of islands in massless gravity theories given in \cite{Geng:2021hlu}.

In Section \ref{sec:BBPSVops} we focus on asymptotically AdS spacetimes and discuss a generic, explicit microscopic procedure to construct all-order gauge-invariant operators dressed to features of the state. The goal of this construction, which was first introduced by the authors of \cite{Bahiru:2022oas,Bahiru:2023zlc} (BBPSV), is to start with a boundary-dressed operator $\phi$---which does not commute with asymptotic charges and acts at leading order like the corresponding local, non-gauge invariant operator---and obtain an operator $\hat{\phi}$ which a) commutes with all boundary charges at all orders in perturbation theory in a given code subspace, and b) still acts like $\phi$ at leading order. In Section \ref{sec:bbpsvreview} we review the original BBPSV construction and point out that the operators $\hat{\phi}$ therein constructed satisfy the requirement a) but fail to satisfy b). In Section \ref{sec:bbpsvrefinement} we propose a refinement of the BBPSV construction to address this issue. We also discuss how the resulting operators, when localised inside the island, commute with all boundary-dressed operators localised outside the island.

Finally, in Appendix \ref{app:KM} we discuss an additional example in which islands appear in a massless gravity setup involving Kourkoulou-Maldacena black hole microstates \cite{Kourkoulou:2017zaj}. This setup has the nice property that reconstruction of operators localised inside the island is particularly simple. 

In the course of writing this paper, we had extensive discussions with the authors of \cite{Geng:2021hlu} about the ideas presented here and in their work. In general, we were able to reach broad (although not complete) agreement on the technical points involved. However, there are also aspects where we all agree on the relevant physics at a technical level, but disagree on (or at least differ in choice of emphasis regarding) its interpretation. Consequently, the authors of \cite{Geng:2021hlu} intend to present their own perspective on these questions in upcoming work.

\section{Background and review}
\label{sec:review}

In this section we will review known results about entanglement islands, discuss their generality, and clarify some possible misunderstandings. We will start in Section \ref{sec:review_island} by reviewing the island rule and its derivation from the gravitational path integral, emphasising its generality in gravitational theories, and commenting on sufficient conditions for the existence of islands in generic settings which were first studied in \cite{Bousso:2021sji}. In Section \ref{sec:ewrdressingreview} we will discuss the regime of validity of entanglement wedge reconstruction. Finally, in Section \ref{sec:massivegravitons} we will clarify in what sense `massive gravitons' can arise when specific boundary conditions, such as those used in toy models for black hole evaporation \cite{Penington:2019npb,Almheiri:2019psf}, are imposed in asymptotically AdS spacetimes.

Most of the material in this section can be safely skipped by expert readers. However it may be helpful, even for those readers, to briefly review the following definitions, which will be used throughout this paper, since small variants of all these definitions have been used elsewhere in the literature.

\begin{description}
    \item[Entanglement island:] Let $W(A)$ be the entanglement wedge of some microscopic degrees of freedom $A$, which may describe a boundary subregion in a holographic CFT or form part of an external, non-gravitational bath. We say that $W(A)$ contains an island $\island$ if $\island \subseteq W(A)$ is a compact subregion of the gravitating spacetime that is disconnected from the remainder of $W(A)$.\footnote{\label{ft:islanddef} Note that this definition differs from that given in \cite{Geng:2021hlu}, where $W(A)$ is not allowed to include any part of the gravitating spacetime other than $\island$. We believe that this alternative does not make a natural or physically well-motivated distinction. Our notion of `island' correlates with the appearance of wormhole topologies in the replica calculations, and also with the additional challenges of defining diffeomorphism-invariant observables on $\island$ when dressing directly to an asymptotic region is impossible. In fact, the definition used in \cite{Geng:2021hlu} is too narrow to include the original islands considered in \cite{Penington:2019npb}, where the radiation was extracted from the original black hole spacetime and was then thrown into a different gravitating spacetime with smaller $G$.} See Section \ref{sec:review_island} for further details.
   \item[Entanglement wedge reconstruction:] If a semiclassical operator $\phi$ is localised within $W(A)$ (including its gravitational dressing) and is defined to some order $O(G^n)$ in perturbation theory, the claim of entanglement wedge reconstruction is that there always exists a microscopic operator $\phi_A$, acting only on $A$, whose action on any state in the code subspace used to define $W(A)$ is the same as $\phi$ up to higher-order $o(G^n)$ corrections.  See Section \ref{sec:ewrdressingreview} for details.
   \item[Massive gravitons:] The clearest notion of massive gravity is a modification to general relativity which explicitly violates diffeomorphism invariance, giving a massive dispersion relation when linearised around flat space. But as we discuss in Section \ref{sec:massivegravitons}, such a theory is not what concerns us: we are interested in a more subtle version, so must take care to clearly define precisely what `massive graviton' means for us here. Adopting a commonly-used terminology (including e.g.~\cite{Geng:2021hlu}) we say that an asymptotically-AdS spacetime has a massless graviton if and only if the boundary gravitational stress-energy tensor $T^{ab}_{\rm grav}$ is conserved ($\nabla_a T^{ab}_{\rm grav} = 0$) and consequently has scaling dimension $\Delta = d$. We emphasise that this is explicitly a statement about the asymptotic near-boundary properties of the theory, and need not imply anything about physics deep in the bulk.
\end{description}

\subsection{Entanglement islands and the gravitational replica trick}\label{sec:review_island}

In AdS/CFT, the entanglement entropy of a boundary CFT region $B$ in a semiclassical bulk state can be computed using the QES prescription \cite{Ryu2006a,Ryu2006b,Hubeny:2007xt,Engelhardt:2014gca}
\begin{equation}\label{eq:QES}
    S(B)= S_{\rm gen}(W(B)) = \underset{b}{\rm min ~ext}\left[\frac{\mathcal{A}(\partial b)}{4 G} + S_{\rm eff}(b)\right],
\end{equation}
where, as in the introduction, $S_{\rm eff}(b)$ is the entropy of all bulk quantum fields in the region $b$.\footnote{The computation of the bulk entropy $S_{\rm eff}$ for graviton excitations is subtle, and it is not completely clear that $S_{\rm eff}$ can be defined for an arbitrary spacetime region $b$. See \cite{Colin-Ellerin:2025dgq} for the current status of computing graviton contributions to the QES prescription.} The entropy $S_{\rm eff}$ and the gravitational constant $G$ are both UV divergent, but these divergences are expected to cancel so that the generalised entropy $S_{\rm gen}$ is UV finite. We first look for all bulk subregions $b$ with conformal boundary $B$ whose  generalised entropy $S_{\rm gen}$ is constant at linear order in any small perturbation of the edge $\partial b$. We then minimise $S_{\rm gen}$ over this set of extremal subregions $b$. The resulting subregion $W(B)$ is called the entanglement wedge.

\begin{figure}
    \centering
    \includegraphics[width=0.4\linewidth]{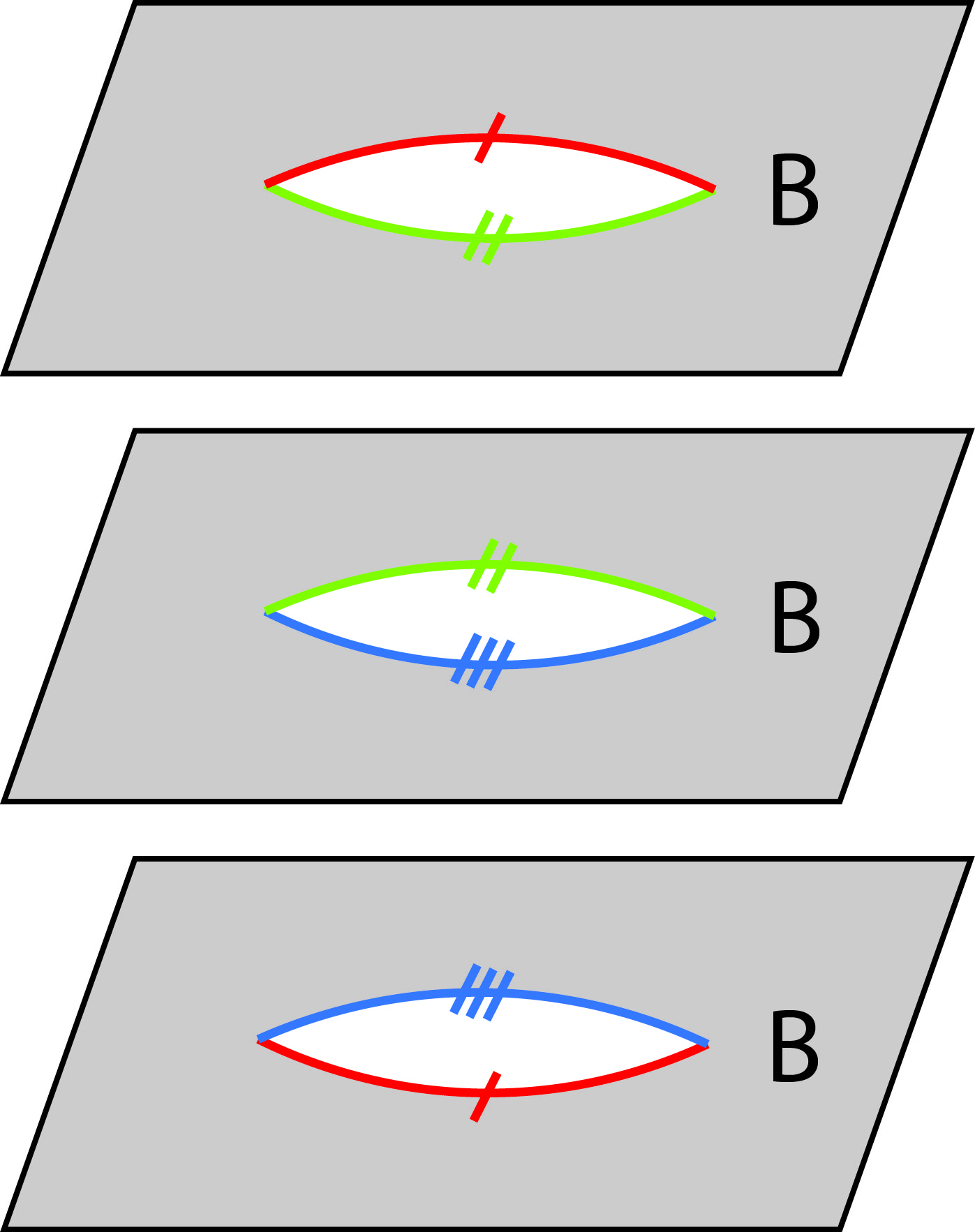}
    \caption{When computing the entanglement entropy of a subregion $B$ with the replica trick, one considers $n$ copies of the system cyclically glued along subregion $B$ as indicated by the different colors of each side of the cut along $B$. Here we represent an $n=3$ replica geometry. The path integral on the resulting Riemanniann manifold computes $\Tr(\rho_B^n)$, which is related to the entanglement entropy by equation \eqref{eq:replica}.}
    \label{fig:replica1}
\end{figure}

The prescription \eqref{eq:QES} can be derived using the gravitational replica trick \cite{Lewkowycz:2013nqa,Faulkner:2013ana}. We can write the entanglement entropy as 
\begin{align} \label{eq:replica}
S(B)=\lim_{n\to 1}\frac{1}{1-n}\log\Tr(\rho_B^n).
\end{align} 
For integer $n$, the right-hand side of \eqref{eq:replica} can be computed by the CFT path integral on $n$ copies (or ``replicas'') with the region $B$ cyclically glued between the different replicas, as shown in Figure \ref{fig:replica1}.
In the semiclassical limit, we can compute this path integral using a gravitational saddle whose asymptotic boundary conditions match the geometry of the boundary path integral. To analytically continue to noninteger $n$, we assume that the bulk saddle preserves the $\mathbb{Z}_n$ replica symmetry of the asymptotic boundary. We can then quotient by this symmetry to produce a single replica spacetime geometry with a conical singularity with opening angle $2\pi/n$ at the fixed point of the $\mathbb{Z}_n$ symmetry. Taking the limit $n \to 1$ is equivalent to taking the limit where the opening angle approaches $2\pi$ and the singularity vanishes. In this limit, the replica fixed point approaches the edge of the entanglement wedge $W(B)$ and one obtains the prescription \eqref{eq:QES}.

An obvious question is whether the entanglement wedge $W(B)$ needs to be connected or, alternatively, whether it is allowed to contain a component $I\subset W(B)$, called an \emph{entanglement island}, that is not connected to the asymptotic boundary $B$. It turns out that the entanglement wedge is allowed to contain an island if and only if the gravitational replica trick is allowed to contain spacetime wormholes \cite{Penington:2019kki,Almheiri:2019qdq}. Generally, it is assumed that the gravitational path integral should include arbitrary spacetime topologies and hence islands should be allowed.

\begin{figure}
    \centering
    \includegraphics[width=\linewidth]{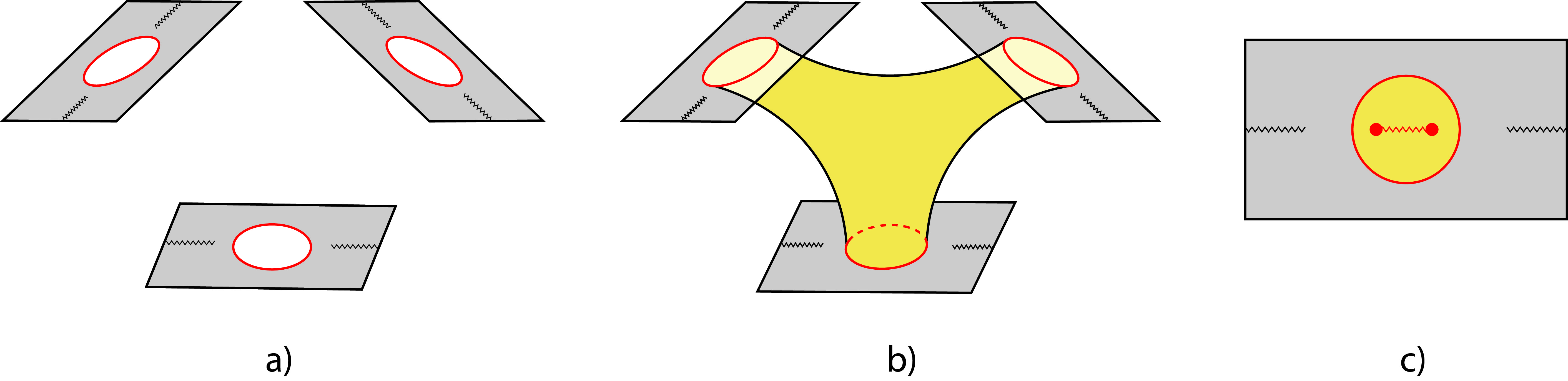}
    \caption{Replica trick to compute the entanglement entropy of a subset $R$ of non-gravitational degrees of freedom coupled to a holographic system. a) The different replicas (here we take $n=3$) are cyclically glued along region $R$, depicted by a black wiggled line. The holographic system is depicted by a red circle. b) We can use the gravitational path integral to fill in these boundary conditions. Keeping only $\mathbb{Z}_n$ replica-symmetric bulk contributions, we obtain two bulk geometries: one fully disconnected and the fully connected replica wormhole depicted here in yellow. The latter is the dominant saddle when $R$ is the Hawking radiation after the Page time. c) Quotienting by the $\mathbb{Z}_n$ replica symmetry and taking the $n\to 1$ limit, the replica fixed point yields the QES. In the case of the replica wormhole geometry depicted in panel b), there is an island, depicted by a red wiggled line in the gravitational region (depicted in yellow).}
    \label{fig:replica2}
\end{figure}

We can generalise \eqref{eq:QES} beyond CFT boundary subregions to compute the entropy of any external quantum degrees of freedom $R$ that are entangled with a gravitational system \cite{Penington:2019npb,Almheiri:2019psf}. The replica trick derivation works in essentially the same way. The only difference is that asymptotic boundary conditions no longer glue together different replicas within a boundary subregion $B$; instead we simply swap the external degrees of freedom $R$ between replicas and allow the gravitational spacetime to adapt to minimise the action (e.g. by forming a spacetime wormhole), see Figure \ref{fig:replica2}. 
After analytic continuation in $n$, the resulting prescription is
\begin{equation}\label{eq:islandrule}
    S(R)= S_{\rm gen}(R\cup W(R)) = \underset{I}{\rm min ~ext}\left[\frac{\mathcal{A}(\partial I)}{4 G} + S_{\rm eff}(I \cup R)\right]
\end{equation}
  Note that, in contrast to \eqref{eq:QES}, the region $I = W(R)$ is now necessarily a compact bulk subregion with no asymptotic boundary -- the entire entanglement wedge (assuming it is nonempty) is therefore an island. For this reason, \eqref{eq:islandrule} is often known as the island rule.  We also emphasise that it is important to distinguish between $S_{\rm eff}$, which is the (somewhat hypothetical) entropy of the quantum fields in a given region if gravity was switched off so that we were studying QFT in curved spacetime, and the true entropy $S(R)$ of the subsystem $R$ entangled with a state in quantum gravity. We will further discuss this distinction in Section \ref{sec:ewrdressingreview}. In particular, the true entropy $S(R)$ will not be equal to the corresponding QFT entropy $S_{\rm eff}(R)$ whenever the island $I$ is nonempty.

Because the asymptotic boundary conditions no longer connect different replicas, there is no reason that the gravitational system being considered in \eqref{eq:islandrule} needs to have asymptotically AdS boundary conditions. Indeed, as we shall discuss in Section \ref{sec:masslessislands}, \eqref{eq:islandrule} also applies to radiation at null infinity in asymptotically flat spacetimes and to EFT modes within the gravitating spacetime (so long as those modes can, in principle, be extracted to an arbitrarily weakly gravitating region).

Importantly, \eqref{eq:islandrule} always leads to a nontrivial island whenever the entanglement between $R$ and the interior of a black hole exceeds the Bekenstein-Hawking entropy, so that the naive estimate that the entropy $S(R)$ is given by $S_{\rm eff}(R)$ would lead to an information paradox. This fact is relatively simple to prove in very broad generality; in particular, the argument does not depend on the spacetime dimension, asymptotic boundary conditions, or the details of how the radiation modes $R$ are defined. We shall sketch a version of that argument here; for more details see e.g. \cite{Bousso:2021sji}. 

Let $I'$ be a large subset of the interior of the black hole (containing in particular the interior partners of the Hawking radiation $R$). For an apparent information problem to exist, we need to have
\begin{align}\label{eq:islandnecessary}
S_{\rm eff}(R \cup I') < S_{\rm eff}(R) - \frac{\mathcal{A}_{\rm hor}}{4G}
\end{align}
so that the number of black hole microstates is insufficient to accommodate the entanglement between $R$ and $I'$. If we assume the edge of $I'$ lies (close to) the black hole horizon, we can rewrite \eqref{eq:islandnecessary} as
\begin{align}\label{eq:islandnecessary2}
S_{gen}(R\cup I')< S_{\rm gen}(R) = S_{\rm eff}(R)
\end{align}
If the region $I'$ was extremal, we would now be done: the generalised entropy $S_{gen}(R\cup I')$ is smaller than $S_{\rm gen}(R)$ and so the minimisation in \eqref{eq:islandrule} would necessarily imply that $I'$ (or some other smaller extremal region) is an island for the radiation $R$. In general, finding the exactly extremal region $I'$ is somewhat messy. However, it is much easier to find a region $I'$ that is quantum normal, meaning that any outwards deformation increases $S_{gen}(R\cup I')$, or quantum antinormal, meaning that any outwards deformation decreases $S_{gen}(R\cup I')$. In either case,  maximin arguments \cite{Wall:2012uf,Akers:2019lzs} using the quantum focusing conjecture \cite{Bousso:2015mna} then imply the existence of a quantum extremal island $I$ satisfying
\begin{align}
    S_{gen}(R\cup I) \leq S_{gen}(R\cup I') < S_{\rm gen}(R).
\end{align}
Moreover, if $I'$ is normal, then we have $I \subseteq I'$, while, if $I'$ is antinormal, $I' \subseteq I$.\footnote{The quantum extremal island $I$ satisfying $I' \subseteq I$ or $I \subseteq I'$ is guaranteed to satisfy $S_{gen}(R\cup I)  < S_{\rm gen}(R)$, but it is not guaranteed to have globally minimal $S_{\rm gen}$ among all quantum extremal islands. However, if an antinormal region $I''$ is contained in a normal region $I'$, then we can always find a single island $I$ satisfying $I'' \subseteq I \subseteq I'$.}

It is easy to see that any region $I'$ that a) extends slightly outside the black hole horizon and b) does so less than a scrambling time in the past of the last quanta of radiation escaping the near-horizon region will be quantum normal, because the classical geometry of the black hole dominates over any quantum effects. In any situation where there is an information problem, it will also satisfy \eqref{eq:islandnecessary2} and so an island exists. Moreover, by a simple and universal scaling argument any region $I'$ whose edge is a)  slightly inside the black hole horizon and b) more than a scrambling time in the past of the last radiation emission will be quantum antinormal. It follows that the edge of the island $I$ lies close to the horizon and approximately one scrambling time in the past, in accordance with the predictions for the dynamics of black holes given in \cite{Hayden:2007cs}. All of these arguments apply in generic gravitational setups and do not rely on the existence of an external, non-gravitational reservoir.

\subsection{Entanglement wedge reconstruction and operator dressing}\label{sec:ewrdressingreview}

The claim of entanglement wedge reconstruction \cite{Czech:2012bh,Headrick:2014cta,Wall:2012uf,Jafferis:2015del,Almheiri:2014lwa,Cotler:2017erl,Dong:2016eik} is that the action of a bulk operator (defined in perturbation theory around a given background) can be reproduced (to all orders in $G$) by a boundary operator localised within a boundary region $B$, if and only if the bulk operator is localised within the entanglement wedge $W(B)$. This claim follows from the QES prescription \eqref{eq:QES} by general information-theoretic arguments based on properties of quantum relative entropies and modular Hamiltonians \cite{Jafferis:2015del,Faulkner:2017vdd,Cotler:2017erl,Chen:2019gbt}. It can also be shown more directly by a replica trick computation of the action of the Petz map \cite{Penington:2019kki}.

All of the above arguments apply equal to the entanglement island $W(R)$ associated to a subset of Hawking radiation. The claim then is that, given any operator $\phi$ in perturbative effective field theory that is localised inside the island $W(R)$, we can find an operator $\tilde\phi$ acting only on $R$ whose action in nonperturbative quantum gravity is the same as $\phi$ to the same order in perturbation theory that the operator $\phi$ was defined.

Let us clarify a few points here. The operator $\phi$ is defined in the effective field theory as a perturbative expansion in $G$ (or more precisely in $1/S_{BH}$). Within this perturbation theory $\phi$ commutes with all operators in $R$ because it is spacelike-separated from them, but this need not be true non-perturbatively. Indeed, in a full non-perturbative theory of quantum gravity, the possibility of entanglement wedge reconstruction means that the island cannot exist as independent degrees of freedom from $R$. This is in close analogy with the usual story of holography, where the radial direction present perturbatively is revealed to be an illusion in the full non-perturbative theory.

A second possible source of confusion comes from the perturbative gauge constraints present order by order in quantum gravity. Because of these constraints, for a local bulk operator to remain gauge invariant, we generally need to add new ``gravitational dressing'' at each order in perturbation theory \cite{Donnelly:2015hta,Donnelly:2016rvo}. The dressing may be localised inside $W(B)$ or may extend outside it. In the latter case, the operator stops being reconstructible from $B$ at the order in perturbation theory at which the dressing extends outside $W(B)$. To be clear, this is not a failure of entanglement wedge reconstruction, but rather an expectation arising from it: if the dressed operator is not localised in $W(B)$, it would be a failure of entanglement wedge reconstruction if it could still be reconstructed from $B$.

\begin{figure}[h]
    \centering
    \includegraphics[width=0.4\textwidth]{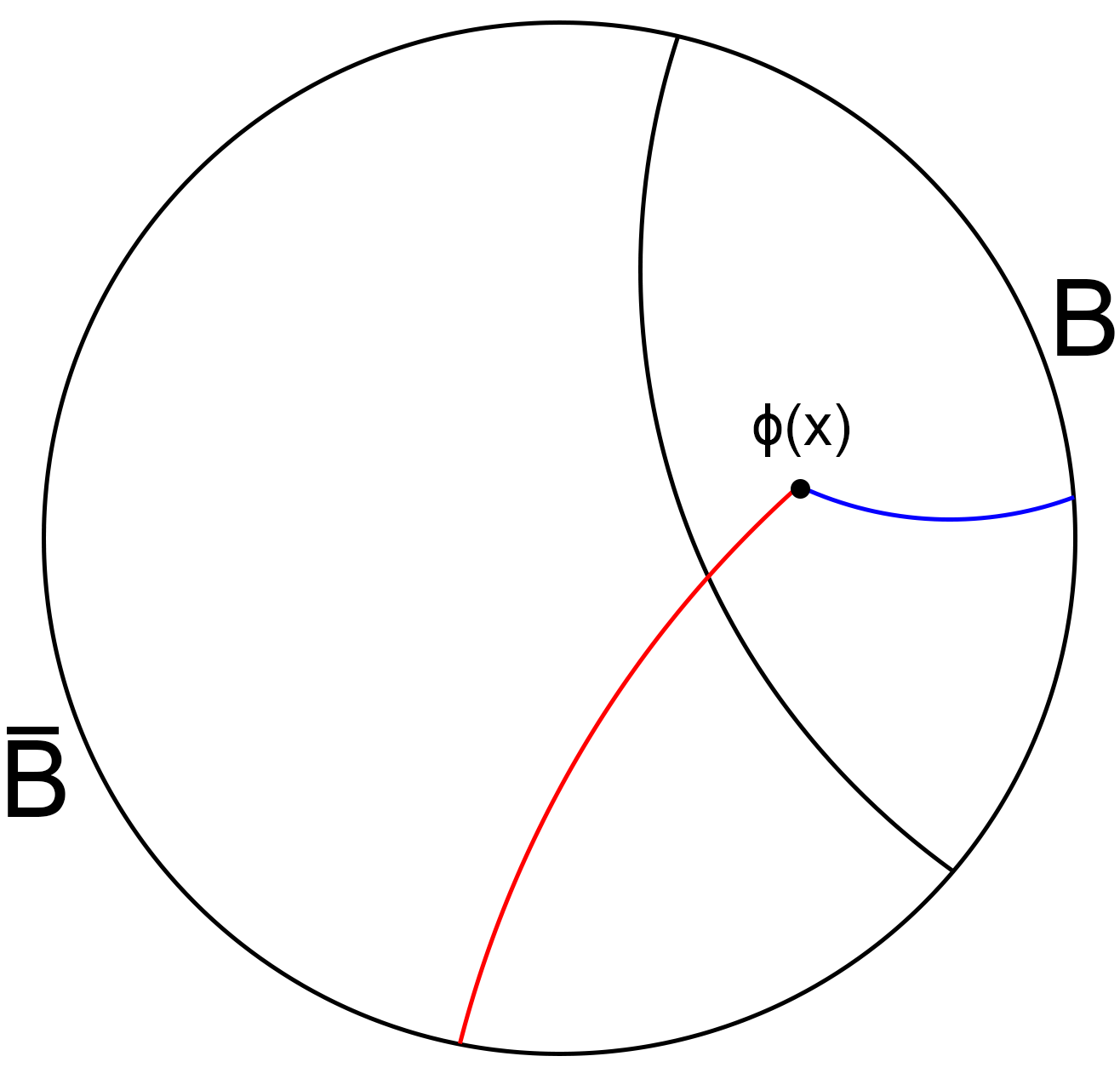}
    \caption{Time slice of pure AdS. A local operator $\phi$ acting at $x\in W(B)$ can be dressed via a gravitational Wilson line either to region $B$ (blue line) or to its complement $\bar{B}$ (red line). In the former case, the dressed operator is reconstructible from $B$ at all orders in perturbation theory; in the latter case it can be reconstructed from $B$ only at leading order, but access to $\bar{B}$ is needed for reconstruction beyond leading order.}
    \label{fig:EWdressing}
\end{figure}

As a simple example let us consider a time slice of pure AdS, a boundary subregion $B$, and a local operator $\phi$ acting at a point $x\in W(B)$, see Figure \ref{fig:EWdressing}. In order to be a well-defined gauge-invariant operator, this needs to be gravitationally dressed, e.g. using a gravitational Wilson line extending to the asymptotic boundary \cite{Donnelly:2015hta,Donnelly:2016rvo}. If the Wilson line is chosen to end at the boundary on region $B$ and to stay entirely within $W(B)$, the dressed operator can be reconstructed from $B$. However, one could choose a Wilson line dressing the operator to the complementary boundary subregion $\bar{B}$. In this case, even though the local operator acts at a point within $W(B)$, the dressed operator cannot be reconstructed from $B$ beyond leading order.

An important question we need to answer when considering entanglement wedges that include an island $I$ is whether, at higher orders in perturbation theory, operators localised entirely within $I$ can exist at all. Indeed, it was argued in \cite{Geng:2021hlu} that the algebra of observables localised entirely within an island is trivial in perturbative gravity, except in special cases, discussed in Section \ref{sec:massivegravitons}, where the AdS boundary conditions make the graviton scaling dimension massive. It is certainly intuitive that it is harder to define gravitationally dressed operators localised inside an island than it is when the entanglement wedge extends to asymptotic infinity: we cannot simply dress the operator to infinity using a gravitational Wilson line as in the above example. However, as we will explain in Sections \ref{sec:linearconstraints} and \ref{sec:BBPSVops}, gravitational dressing within an island is still possible to all orders in perturbation theory because we can always dress operators to features of the gravitational background.

\begin{figure}
    \centering
    \includegraphics[width=0.4\textwidth]{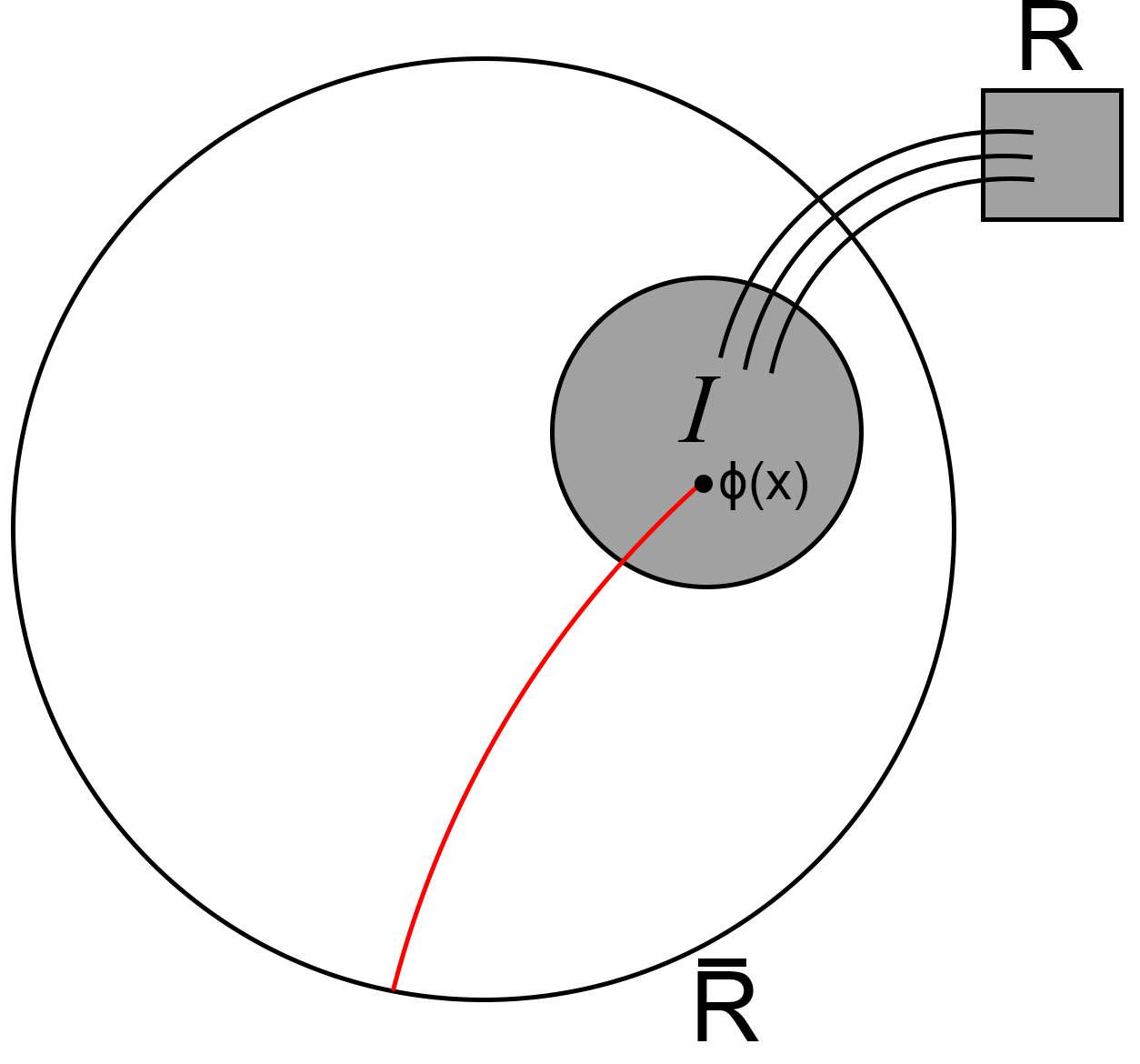}
    \caption{The shaded region $I$ is an island for the reservoir $R$ in an asymptotically AdS setting. A local operator acting in the island and dressed to the asymptotic boundary through a Wilson line cannot be reconstructed from $R$ beyond leading order, because the Wilson line passes through $\overline{W(R)}$ and ends at the asymptotic boundary, where $\bar{R}$ is located.}
    \label{fig:islanddressing}
\end{figure}

\subsection{What is a massive graviton?}
\label{sec:massivegravitons}

The most conventionally understood meaning of massive gravity refers to a classical (or weakly coupled) modification to the field content and action of general relativity, which includes a Fierz-Pauli massive spin 2 field when linearised around Minkowski space (such as the dRGT theory reviewed in \cite{deRham:2014zqa}). But the `massive gravitons' which have been suggested as necessary for islands \cite{Laddha_2021,Geng:2020fxl,Geng:2020qvw,Geng:2021hlu,Raju:2020smc,Geng:2023zhq,Geng:2025rov} are not of this nature. In particular, the theories under consideration are always defined by a conventional diffeomorphism-invariant local action of general relativity coupled to ordinary matter. So to understand these claims we require a more expansive definition of massive gravity. Then once a definition is agreed on, we should be careful to understand precisely how the physics of `massive' gravitons differs  from that of `massless' gravitons. In particular, does this notion of mass make it easier to define compactly supported diffeomorphism invariant operators for entanglement wedge reconstruction on an island?

 The salient examples for which some notion of graviton mass appears in island calculations involve gravity in AdS where the boundary operators are coupled to an external `bath' system, which usually does not have dynamical gravity. The bath provides a convenient way to cleanly separate Hawking radiation (in the bath) from the remaining black hole (understood as the whole asymptotically AdS spacetime, or its dual CFT).  In this setup we do not change the theory governing local dynamics in the bulk of AdS, which is still described by a standard Einstein-Hilbert action coupled to matter; coupling to the bath changes only the boundary conditions.

To understand precisely in what sense changing boundary conditions in AdS can make a field massive, we first recall the meaning of mass in asymptotically flat spacetimes. In interacting quantum field theories, mass terms appearing in the Lagrangian generally need to include a UV-divergent counterterm, while the finite piece is renormalisation scheme dependent. So for the examples of interest where masses are generated by quantum effects, ``Lagrangian mass'' is not a well-defined notion.  However, each on-shell particle has an unambiguous ``pole mass'' $m$ that describes a Casimir of the associated Poincar\'{e} group representation. In particular, this implies a relativistic dispersion relation $m^2 = E^2-p^2$, where massless particles are distinguished by single-particle states with arbitrarily small energy $E$ at low momentum $p$ (and hence long wavelength). The massless representations (for helicity 1 and 2 corresponding to photons and gravitons, for example) are also qualitatively different and are not obtained as the $m\to 0$ limit of a massive representation (because unphysical polarisations appear that are removed by gauge symmetry). A consequence is that $m=0$ is robust under turning on (Poincar\'e-invariant) interactions:  typically interactions change masses even if mass terms in the Lagrangian are held fixed\footnote{Indeed, these corrections will generally be divergent and will need to be renormalised.}, but the massless photon or graviton is protected from such effects.

What about in anti-de Sitter space? In this case, the isometry group of the spacetime background is the conformal group $SO(d,2)$ and the Casimir controlling the minimum energy within a representation is the scaling dimension $\Delta$. In a free theory of a spin $s$ particle with Lagrangian mass $m^2$ (meaning that the bulk linearised equation of motion takes the form $\nabla^2+m^2=0$) and the usual reflecting AdS boundary conditions, we have
\begin{align}\label{eq:deltam}
    \Delta = \frac{d}{2} + \sqrt{\frac{d^2}{4}  + L_\mathrm{AdS}^2 m^2 +s}.
\end{align}
In particular, the minimum energy of a massless scalar particle in AdS is not zero, but instead $\Delta = d$. There is no clear meaning to wavelengths exceeding the curvature scale $L_{AdS}$, and we cannot take arbitrarily small momenta. So in AdS the dispersion relation does not give a clear distinction between massless and massive particles. In fact, even a normal graviton ($s=2,\Delta=d$) is not massless in this sense, but instead has $m^2=-2/L_{AdS}^2$. Nonetheless, there remains a qualitative difference from the  number of physical polarisations. In terms of representation theory, the unphysical polarisations of a massless particle give null descendants (starting at $s=1$, $\Delta=d+1$), leading to a short representation of the conformal group. Under the correspondence between states in AdS and local boundary operators at asymptotic infinity \cite{Witten:1998qj},\footnote{This correspondence is often talked about in the context of AdS/CFT where, on the boundary side of the holographic duality, it becomes the usual state-operator correspondence of conformal field theory. However, it exists more generally for any quantum field theory in AdS.} this leads to the Ward identities for boundary operators $\nabla_a T_{\rm grav}^{ab}$, interpreted in a dual CFT as conservation of energy-momentum. Violation of such a conservation statement is a reasonable definition of a massive graviton and the one used in \cite{Laddha_2021,Geng:2020fxl,Geng:2020qvw,Geng:2021hlu,Raju:2020smc,Geng:2023zhq,Geng:2025rov}, so we will take this to be our precise meaning. Just as for flat space above, this has a certain robustness: generically (i.e., in the absence of other states with precisely tuned dimensions), weak bulk interactions or small changes in boundary conditions which respect conformal invariance cannot lift a massless representation. This protection can be explained by symmetry, namely the \emph{asymptotic symmetries} associated with diffeomorphims which do not decay at infinity. So, our definition parallels the similar definition of a gauge theory in Coulomb phase (a massless photon/gauge boson) via asymptotic symmetries \cite{harlow2019symmetriesquantumfieldtheory}.

One key way in which AdS differs from flat space is that masses (i.e., operator dimensions), and in particular the existence of massless photons/gravitons in the above sense, can depend on boundary conditions. If we change the boundary conditions so that energy can flow in and out of the boundary, either by explicitly breaking time-translation invariance or by coupling to another system, then $\nabla_\mu T^{\mu\nu}_{\rm grav} \neq 0$ and the graviton scaling dimension $\Delta$ will no longer be equal to the dimension $d$. Most famously, this occurs in Karch-Randall braneworlds \cite{Karch:2000ct,Aharony:2003qf,Aharony:2006hz}. A holographic theory CFT${}_d$ is coupled to a higher dimensional holographic theory CFT${}_{d+1}$. The full bulk dual spacetime is a $(d+2)$-dimensional asymptotically AdS spacetime with a $(d+1)$-dimensional end-of-the-world (ETW) brane embedded in it, where the latter can roughly be identified as the bulk dual of the CFT${}_d$. In the case of our interest, the geometry on the brane is also AdS. If the central charges of the two CFTs are comparable in size $c_{d}\approx c_{d+1}$, gravity is necessarily $(d+2)$-dimensional. On the other hand, if $c_{d}\gg c_{d+1}$, gravity approximately localises on the ETW brane, with the $(d+1)$-dimensional graviton acquiring a small mass $m\ll 1/L_{AdS}$ \cite{Karch:2000ct,Cooper:2018cmb,Antonini:2019qkt}. In this case, there exists an effective $(d+1)$-dimensional description of the same physics as a gravitational theory localised on the ETW brane and coupled at the boundary to non-gravitational degrees of freedom associated with the CFT${}_{d+1}$ \cite{Karch:2000ct,Almheiri:2019hni,Antonini:2024bbm}. In this description, the boundary stress-tensor is not conserved and a nonzero `graviton mass' arises because of the coupling to the CFT${}_{d+1}$.

The same phenomenon occurs more generally when a holographic CFT is coupled to a nongravitational bath \cite{Porrati:2001gx,Porrati:2002dt,Porrati:2003sa,Rattazzi:2009ux,Karch:2023wui}. If this coupling is local and respects $SO(d,2)$ boundary conformal symmetry, then operators can still be classified by their scaling dimension. However, translations of the gravitational degrees of freedom that do not also act on the non-gravitational bath do not generate any symmetry. The gravitational stress-energy tensor $T^{\mu\nu}_{\rm grav}$ is therefore not conserved. Perturbatively, we can track the scaling dimension of $T^{\mu\nu}_{\rm grav}$ that results from the coupling and see that it changes. According to the definition \eqref{eq:deltam}, the graviton has become massive. In such cases, we evade the usual robustness of a massless state because the uncoupled theory has extra finely tuned field content (e.g., conformally invariant bulk matter), leading to a spectrum including a double-trace  marginal scalar ($s = 0, \Delta=d$) whose coupling changes the boundary conditions, and a double-trace vector $(s= 1, \Delta=d+1)$ which becomes the flux of stress-energy through the boundary. In a gravitational version of the Higgs mechanism, this latter representation provides the additional `null states' required to lift the $(\Delta=d,J=2)$ representation to become `massive' $\Delta>d$, and violates conservation giving a non-zero right-hand side to $\nabla_\nu T^{\mu\nu}_{\rm grav}$. The fact that the massive representation mixes the stress tensor with a double-trace (interpreted in the bulk as mixing a single graviton with a two-particle matter state) indicates that the bulk explanation for the mass must require quantum effects. And indeed, in \cite{Porrati:2001gx,Porrati:2002dt,Porrati:2003sa,Rattazzi:2009ux,Karch:2023wui} the mass is calculated from a matter loop contribution to the graviton self-energy.

However, all the characterisation of massive gravitons we have discussed refer essentially to single-particle excitations around empty AdS or to asymptotic properties near the boundary. This is for good reason: a graviton mass in this sense does not generally lead to any sharply defined consequences for local physics deep in the bulk, particularly when we consider backgrounds which are far from empty AdS. For example, in the above discussion we can compute the mass induced by non-standard boundary conditions by integrating out the matter at one loop to find an effective action for the graviton to quadratic order in metric fluctuations. But this is not as simple as inducing a Fierz-Pauli mass term and reading off its coefficient: in fact, this effective action is non-local on the AdS scale because we have integrated out a light field. Identifying a local notion of mass from an effective action is possible only if $m$ is large compared to typical curvature scales (so we can consider an action valid on distances much larger than $m^{-1}$ while in an approximately flat region), which is never the case in these examples which have $m L_{\mathrm{AdS}}\ll 1$. The graviton propagator is not substantially changed at distances shorter than $L_\mathrm{AdS}$, and it is only in the far infrared that we can unambiguously read off a mass. This becomes even more stark once we consider a general state (e.g., a black hole background) which breaks the AdS isometries, so there are no symmetries to constrain possible forms of the propagator.

Even in the vacuum state, in any AdS-scale region there is no sense in which the massless and massive vacua are in a different phase. The massless vacuum is mildly excited, but only by the addition of $O(1)$ particles. For example, in the analagous case of a $U(1)$ gauge symmetry broken by boundary conditions, the `mass' induced at one loop does not constitute a Higgs or superconducting phase. Explicitly sourcing a classical field (with a potential, or by Dirichlet boundary conditions with a non-zero boundary value of a charged scalar) can be characterised as a new phase if passing to the new background requires excitation of parametrically many particles. But this is not the case for `double trace' boundary conditions studied in \cite{Karch:2023wui}, for example. Roughly speaking, we can say we are in a different phase if the width of the Higgs wavefunction is much smaller than the expectation value, but in the case of interest they are comparable. Again, this is an example of a sharp distinction between flat space and AdS. In flat space, any translation invariant change in vacuum expectation values (no matter how small) always leads to a distinct phase, because it always requires a parametrically large number of particles to correctly describe the physics of very low-energy, long-wavelength excitations.  In AdS, however, the lowest-energy excitations have wavelength comparable to the AdS-scale and we do not enter a distinct low-energy phase unless there are many particles created in each AdS-sized patch.

Note that one way a double trace definition can lead to a distinct phase is a theory with a parametrically large number of particle species: a `large $c$' bulk matter theory. This species number can then become the large parameter necessary for a phase transition. One example of this is a Karch-Randall braneworld from the perspective of the brane EFT, where gravity is coupled to a semiclassical (i.e. large $c$) holographic CFT. But while such matter theories are sometimes used as a convenience in island calculations, they are certainly not necessary (or realistic).\footnote{Large $c$ matter is convenient because it reduces the spread of the zero modes of the black hole wavefunction (i.e., its mass, location, and velocity), and one does not have to account for a superposition of macroscopically distinct geometries due to uncertainty in those quantities. But this does not change anything substantively, since it is a classical statistical effect with negligible interference between different geometries (these degrees of freedom are thoroughly decohered by the geometry just outside the black hole). In practice, one can treat each different geometry separately and average over them at the end.}

We have only discussed violations of stress-energy conservation coming from a local and conformally invariant coupling to a non-gravitational bath. But of course this is not the only way to violate conservation, and we can couple to another system without retaining all these symmetries. In such cases the meaning of graviton mass becomes less clear. In particular, it is common in AdS/CFT to insert or subtract energy in the boundary theory by turning on sources for given operators. If these sources are localised in the boundary spacetime, they will violate stress-energy conservation, but, since there is no analogue of the global stress-energy tensor of gravity plus bath that is still conserved, it does not make much sense to talk about a perturbed graviton scaling dimension (or mass). Fundamentally, however, the same physics that changes the graviton spectrum coupling a gravitational spacetime to a bath is also present whenever we turn on external sources to excite a CFT state.

As an illustrative example of this, it is helpful to consider a slight variant of the couplings considered in \cite{Porrati:2001gx,Porrati:2002dt,Porrati:2003sa,Rattazzi:2009ux,Karch:2023wui}, where the double trace deformation is chosen to be relevant rather than marginal. This breaks conformal symmetry, and so we cannot define a graviton scaling dimension (except in the boundary UV/bulk IR where we recover the unperturbed theory). Nonetheless, since the gravitational stress-energy tensor is no longer conserved and the minimum energy of a single-graviton state changes, one might still want to say that gravitons have a nonzero mass.\footnote{Because we are breaking conformal symmetry, this can happen without the need for any Higgs analogue to provide the additional states in a $\Delta > d$ representation.} However, because the boundary UV physics (or, equivalently, the bulk IR physics) has not changed, the perturbed vacuum is literally just a mildly excited state of the unperturbed theory. In this case, it isn't just the physics of an AdS-scale patch, but an entire AdS-Cauchy slice (and its domain of dependence/Wheeler-DeWitt patch)) that is identical to the physics of an excited, but finite-energy, state in a theory with unperturbed boundary conditions.

Since \cite{Geng:2021hlu} motivated the necessity of massive gravity by existence of gauge-invariant operators on the island, we will end this section by briefly examining the relationship between these two ideas. Since the mass is defined by asymptotic properties of the theory, and the island is (by definition) a region of finite size in the bulk in a highly excited state, a massive graviton by itself does not tell you anything about the ability to construct dressed diffeomorphism-invariant operators supported on the island. In particular, since we are interested in local diffeomorphism-invariant theories we still have a local Hamiltonian constraint which is a potential obstruction to dressing. Perturbatively around a static background like the vacuum this can be integrated to a `Gauss law' in the sense of \cite{Geng:2021hlu}, which is no different from an ordinary `massless' theory (see around equation \eqref{eq:GaussLaw}). We will revisit this point in Section \ref{ssec:dressingmassive} after discussing the construction of dressed operators.

It is worth noting that there is a possible alternative characterisation of massive gravity in terms of spontaneous symmetry breaking: namely, gravity is massive if and only if the background (which could be a vacuum field configuration or an excited state) breaks all the gauge symmetries. This only makes sense semiclassically (except in the infrared, where it overlaps with the above definition in terms of asymptotic symmetries). Nonetheless, the analogous definition in gauge theory of a massive gauge boson is quite sensible when it applies, and accords with the familiar story of the Higgs mechanism in terms of the background value of a charged scalar, for example. We will also see in Section \ref{sec:linearconstraints} that precisely this notion is directly relevant to the construction of compactly supported gauge-invariant operators. However, once we are discussing gravity in nontrivial backgrounds such as black holes as relevant for discussions of islands, this definition is so expansive as to become essentially meaningless. `Massive gravity' in this sense means working around a background that does not have any isometries. If we were to use this definition, then gravity is massive here and now, since the Earth and other astronomical objects source a classical background without symmetries. This is certainly not what is usually meant by massive gravity (nor does it appear to be the intended meaning in \cite{Laddha_2021,Geng:2020fxl,Geng:2020qvw,Geng:2021hlu,Raju:2020smc,Geng:2023zhq,Geng:2025rov}) and we will not refer to it as such.

\section{Islands exist in massless gravity}\label{sec:masslessislands}

 In this section we provide a number of explicit examples where entanglement islands, and hence a Page curve, arise in setups where gravitons are unquestionably massless. In each case, we focus primarily on the definition of the ``radiation'' degrees of freedom to which the island is associated. Once this is done, the detailed calculations showing the existence of an island and its approximate location are essentially identical in each case and can be found in the original literature \cite{Penington:2019npb,Almheiri:2019psf}. Indeed, as explained in \cite{Bousso:2021sji} and briefly reviewed in Section \ref{sec:review_island}, to show that an island exists, we only need to find a quantum normal region $I'$ (e.g. any region extending outside the horizon less than a scrambling time before the last radiation was emitted) satisfying $S_{\rm gen}(R \cup I') < S_{\rm gen}(R)$.  
 
\subsection{Islands for boundary subregions in AdS/CFT}
\label{sec:boundryislands}

We first consider examples where the entanglement wedge of a boundary subregion $B$ in a holographic CFT includes a bulk region disconnected from the boundary (i.e. an island) even in the absence of an external bath (and hence any possible issues with massive gravitons).\footnote{Because the entanglement wedge $W(B)$ associated to a boundary region $B$ always contains the asymptotic bulk region near $B$ itself, $W(B)$ (unlike the examples in later subsections) can never contain an island by the formal definition given in \cite{Geng:2021hlu}; see Footnote \ref{ft:islanddef}. However, it will contain an island by our definition, which we consider to be more standard and certainly more natural, and that island will lead to a Page curve.} This avoids any subtleties involved in defining a subset of radiation degrees of freedom $R$.  

Suppose we consider a state in a holographic CFT that is dual to an AdS spacetime containing a small black hole. This black hole should be sufficiently small that it is microcanonically unstable, and will evaporate without any need for a nongravitational bath. The goal is to find a boundary subregion whose entanglement wedge will naturally include more than half of the radiation, but not include the black hole. The problem, of course, is that the radiation will naturally surround the black hole in a spherical shell, making this hard to achieve.

\begin{figure}
    \centering
    \includegraphics[width=0.5\linewidth]{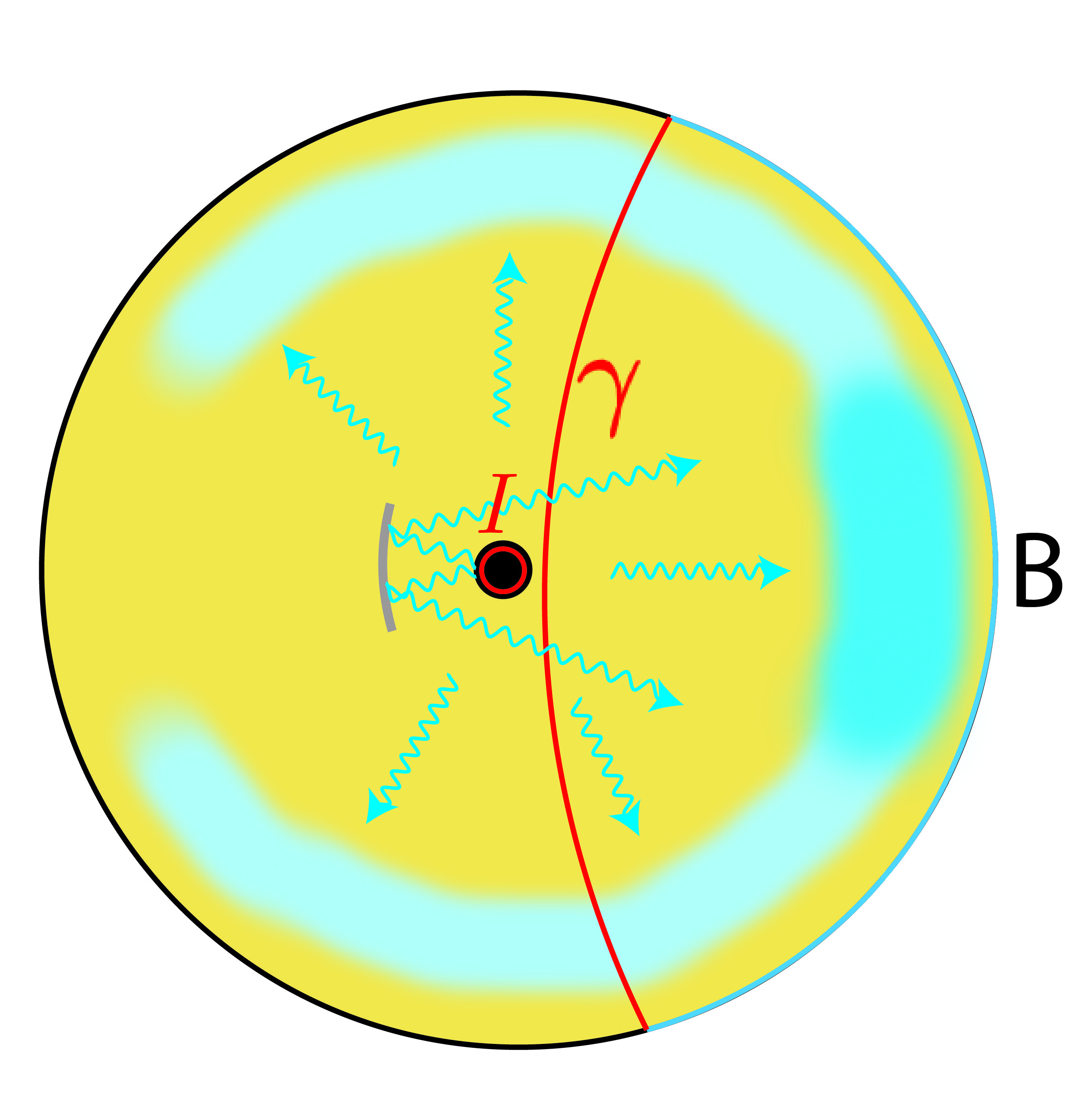}
    \caption{A small black hole in AdS ($r_H\ll L_{AdS}$), is evaporating. A mirror (depicted in grey) is built on one side of the black hole, deflecting Hawking radiation (depicted in light blue) to the other side. Consider a boundary subregion $B$ smaller than half of the CFT and its naive entanglement wedge $W_0(B)$ (without quantum corrections, bounded by the HRT surface $\gamma$ depicted in red). Once the thermal entropy of the radiation contained in $W_0(B)$ is greater than the entropy of the radiation outside $W_0(B)$ plus the horizon area of the black hole, an island $I$ for $B$ will form in the black hole interior (depicted in red) and the entropy $S(B)$ will start decreasing.}
    \label{fig:mirror}
\end{figure}

The solution is to use the semiclassical bulk dynamics to break the spherical symmetry and move the majority of the radiation to one side. Since the radiation consists of low energy modes, this is easy enough to do without running into issues with gravitational backreaction etc. As a simple example, one could construct a gigantic mirror in the bulk on one side of the black hole.\footnote{A very similar version of this argument first appeared in \cite{Bousso:2019ykv}, where the role of the mirror was played by a Dyson sphere surrounding the black hole.}  Even if this mirror were only able to reflect a small fraction of Hawking radiation, it would be enough that the naive entanglement wedge of a boundary region $B$ (bounded by $B$ and a HRT surface $\gamma$ \cite{Hubeny:2007xt}, without taking quantum corrections into account) containing slightly less than half of the boundary contains slightly more than half of the radiation, as shown in Figure \ref{fig:mirror}. By the arguments in Section \ref{sec:review_island}, the true entanglement wedge $W(B)$ as computed by the QES prescription will then include an island in order to reduce $S_{\rm gen}(W(B))$.

\begin{figure}
    \centering
    \includegraphics[width=0.8\linewidth]{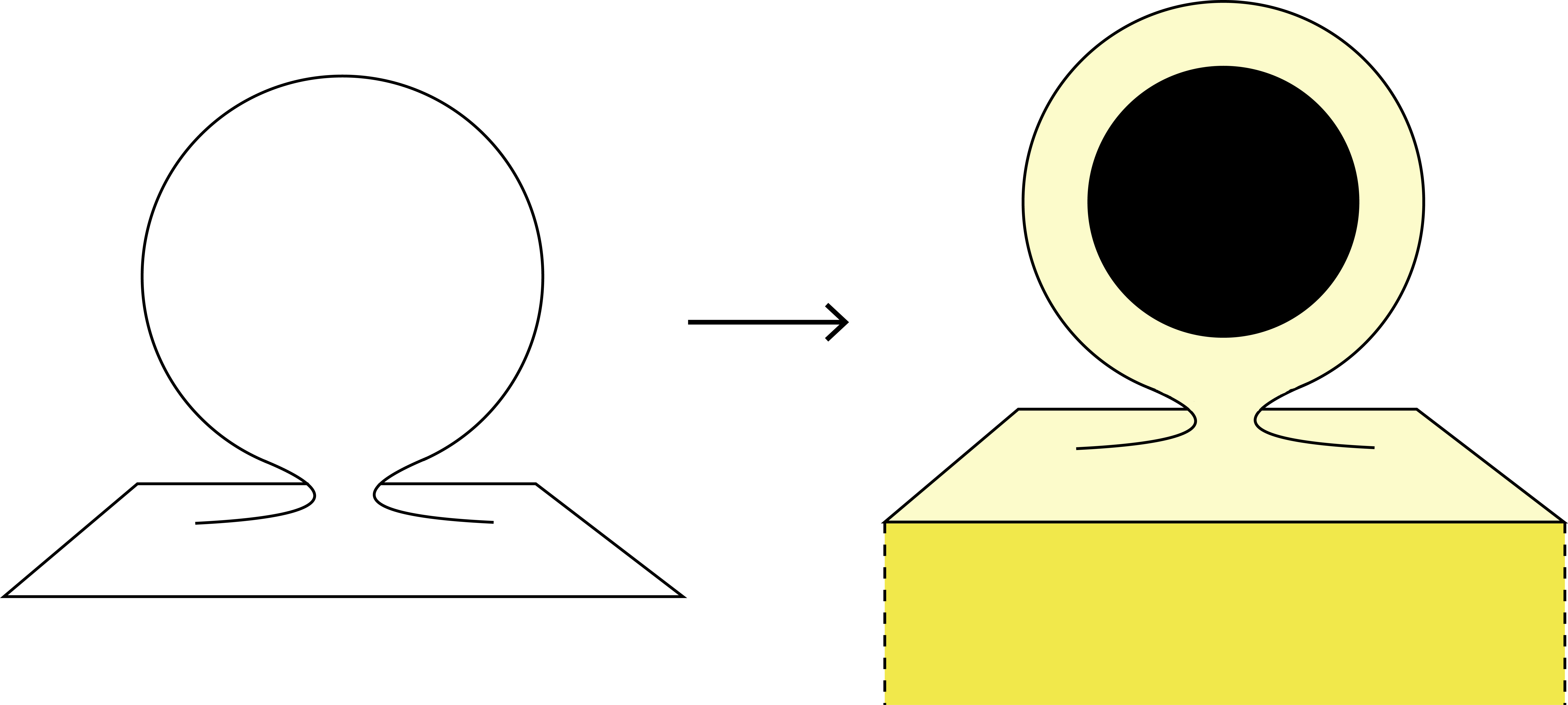}
    \caption{Left: a time slice of the boundary geometry on which the holographic CFT lives consists of a bulb connected to a non-compact, asymptotically flat region by a small neck. Right: bulk dual geometry with a large AdS black hole in the bulb region connected to Poincar\'e-AdS by a narrow throat.}
    \label{fig:narrowthroat}
\end{figure}

While it should certainly be possible in principle to build a mirror that reflects most Hawking modes in any holographic theory rich enough to describe local physics similar to the real world, the details of its construction out of the fields present in ten-dimensional supergravity or string theory would be fairly involved. To describe a setup with our desired properties more explicitly, we will therefore switch tack and instead change the geometry of the boundary spacetime. Consider a holographic CFT living on a spacetime manifold $\partial\mathcal{M}$ for which each spatial slice is non-compact and given by a bulb connected to an asymptotically flat region by a small neck, as shown in Figure \ref{fig:narrowthroat} (left panel).
In the limit where the average curvature radius of the bulb region $L_{bulb}$ is much larger than the radius of the neck $L_{bulb}\gg r_{neck}$, we can zoom in near the neck region, which effectively connects two planes. A simple ansatz for the metric of the boundary manifold in this region is given by
\begin{align}
    ds^2_\text{bdry} = - dt^2+dx^2+\big(r_{neck}^2+x^2)d\Omega_{d-1}^2,\quad -\infty < x<\infty.
    \label{eq:bdymetric2}
\end{align} The neck sits at $x=0,$ and the sign of $x$ indicates either the upper ($x>0$) or lower($x<0$) plane.

If the CFT admits a holographic dual description, the resulting bulk geometry will be an asymptotically AdS spacetime with conformal boundary given by the manifold $\partial\mathcal{M}$. The symmetries of the neck region mean that the bulk metric can be locally written in the form
\begin{equation}\label{eq:neckansatz}
    ds_\text{bulk}^2 = -f(x,\rho) dt^2 + g(x,\rho)dx^2 + h(x,\rho) d\Omega_{d-1}^2 + d\rho^2 + k(x,\rho) d\rho dx,
\end{equation}
for some functions $f(x,\rho)$, $g(x,\rho)$, $h(x,\rho)$ and $k(x,\rho)$ that are also invariant under $x \to - x$ (except for $k \to -k$). The ansatz \eqref{eq:neckansatz} does not completely fix the diffeomorphism gauge symmetry, because we are to introduce a $\rho$-dependent translation of $x$. A convenient gauge choice is to require every $x,t=const$ surface to be extremal.

With this gauge choice, the Einstein equations form a set of coupled partial differential equations for $f$, $g$, $h$, and $k$ that is expected to form a well-posed boundary value problem. For our purposes, the required boundary conditions are that, as $x \to \pm \infty$, the geometry should approach Poincar\'{e}-AdS. In our set of coordinates this corresponds to
\begin{equation}
    f(x,\rho)\overset{x\to\pm\infty}{\longrightarrow} \frac{\cosh^2(\rho)}{x^2},\quad g(x,\rho)\overset{x\to\pm\infty}{\longrightarrow} \frac{\cosh^2(\rho)}{x^2}, \quad h(x,\rho)\overset{x\to\pm\infty}{\longrightarrow} \sinh^2(\rho),\quad k(x,\rho)\overset{x\to\pm\infty}{\longrightarrow} 0.
    \end{equation}
Meanwhile, as $\rho \to \infty$, it should approach AdS asymptotics with boundary metric \eqref{eq:bdymetric2}, i.e.
\begin{equation}
    f(x,\rho)\overset{\rho\to\infty}{\longrightarrow} e^{2\rho},\quad g(x,\rho)\overset{\rho\to\infty}{\longrightarrow} e^{2\rho},\quad h(x,\rho)\overset{\rho\to\infty}{\longrightarrow} e^{2\rho}(r^2_{neck}+x^2),\quad k(x,\rho)\overset{\rho\to\pm\infty}{\longrightarrow} 0.
\end{equation}
We have not attempted to solve the Einstein equations subject to these boundary conditions explicitly, but there is no reason that it could not be done numerically. In the limit where the neck is much smaller than the bulb, we can then glue this neck geometry to a large AdS-Schwarzschild black hole in the bulk and to Poincar\'{e}-AdS in the asymptotically flat region, as shown in Figure \ref{fig:narrowthroat} (right panel).\footnote{One potential obstruction to putting the dual CFT on such a geometry is that the regions of negative curvature could make it unstable (so in the bulk, the energy is lowered by condensation of some matter field in the neck region). For example, a free conformally coupled scalar on the metric \eqref{eq:neckansatz} is unstable. We would find it extremely surprising if such an instability was a universal feature of all holographic CFTs.}

Let us now consider the evaporation process of the black hole in the bulb. Black holes in AdS with a radius comparable to or larger than the AdS radius do not generically evaporate, because reflective boundary conditions mean that they quickly reach thermal equilibrium with their Hawking radiation. However, in our setup radiation $R$ can ``leak'' through the narrow throat and escape into the Poincar\'e-AdS region. In the regime we are considering, evaporation will be very slow, because only a small flux of Hawking radiation crosses the narrow throat. However, it will be much faster than any other relevant decay process (e.g. the entire black hole itself tunneling through the throat).

\begin{figure}
    \centering
    \includegraphics[width=0.75\linewidth]{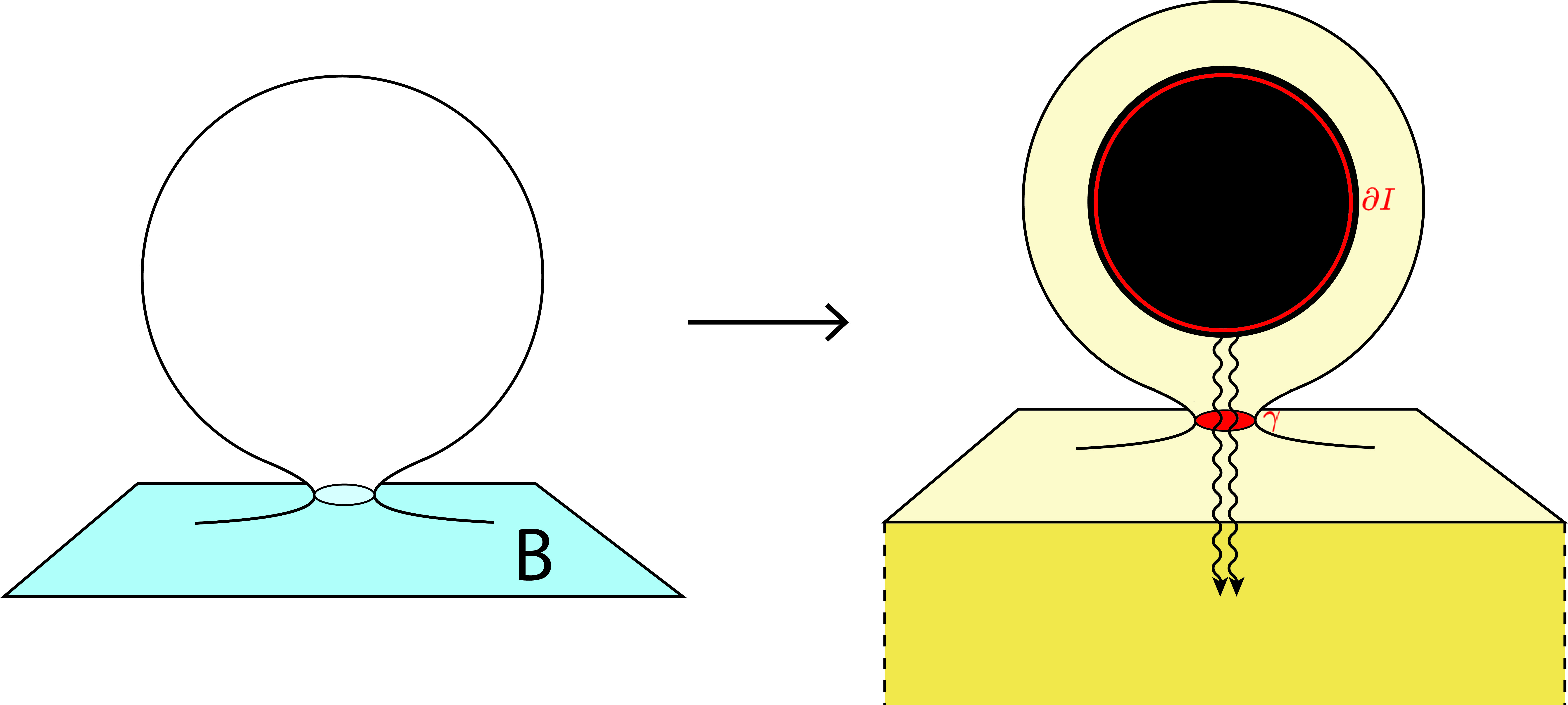}
    \caption{Left: we are interested in computing the entanglement entropy of the boundary subregion $B$ including the boundary manifold up to the narrowest part of the neck $x=0$ (depicted in light blue). Right: when the black hole evaporates and Hawking radiation escapes into the Poincar\'e-AdS region, there are two candidate QES's. The first one is given by the throat surface $\gamma$, dominant when less than half of the radiation has escaped. The second one is given by $\gamma\cup \partial I$, where $\partial I$ is the boundary of an island just inside the black hole horizon. This second QES becomes dominant when more than half of the radiation has escaped, yielding a page curve for $S(B)$.}
    \label{fig:narrowpage}
\end{figure}

We can use the QES prescription to compute the entanglement entropy of the boundary subregion $B$ given by the entire asymptotically flat region up to the narrowest part of the neck at $x=0$. There are two candidate entanglement wedges, see Figure \ref{fig:narrowpage}.
The first one includes the Poincar\'{e}-AdS region and the part of the throat with $x<0$. It is bounded by the classical extremal surface at $x=0$, which will require only perturbatively small corrections to be quantum extremal. This QES dominates when less than half of the Hawking radiation has escaped to the AdS-Poincar\'e region. The second candidate entanglement wedge includes an entanglement island $I$ in the black hole interior, which again exists by the arguments reviewed in Section \ref{sec:review_island}.\footnote{One might worry that a third relevant QES, with a connected entanglement wedge that includes the interior of the black hole, could also exist. But in the limit where the throat is narrow compared to the bulb, the relevant physics is identical to the entanglement wedge transition found for thermal states in global Schwarzschild AdS (since our throat geometry is parametrically well-approximated by Schwarzschild AdS in the relevant region). For comparison with our construction, we should take the boundary subregion $\bar{B}$ to be almost all of the spherical boundary, excluding a small `polar cap' (or equivalently, we take the complementary region $B$ to be the union of a small polar cap on the sphere and the purifying system, for example the other side of the thermofield double). In such a case, the dominant QES is disconnected, with one compact component wrapping the black hole horizon (see figure 2 of \cite{Headrick:2007km}, for example). A connected QES would inevitably have much larger area (and hence generalised entropy) than the island QES. This is because such a surface must wrap around almost the entire black hole horizon, but  it would also need to include a long tube connecting the black hole to the throat.  And in fact, for $d>2$ (as required for our discussion) such a surface does not even exist when the polar cap is smaller than some order one fraction of the boundary \cite{Hubeny:2013gta}.} This island dominates when more than half of the radiation has escaped in the Poincar\'e-AdS region. As a result, the entanglement entropy of $B$ is given by
\begin{equation}
    S(B)=\min\left\{\frac{\mathcal{A}_\gamma}{4G}{+S_{\rm eff}(R)},\frac{\mathcal{A}_\gamma+\mathcal{A}_{\partial I}}{4G}{+S_{\rm eff}(R\cup I)}\right\}
\end{equation}
and follows a Page curve.
Again, there is no auxiliary non-gravitational bath in this setup: the role of the bath is played by the gravitating Poincar\'e-AdS region, in which Hawking radiation can escape through the narrow throat. The CFT stress-energy tensor is (covariantly) conserved, and so the graviton is massless in the sense discussed in Section \ref{sec:massivegravitons}.

\subsection{Radiation at null infinity in asymptotically flat spacetimes}
\label{sec:flat}

In asymptotically flat spacetimes, all black holes are microcanonically unstable, and we do not need to introduce a bath, or neck connecting to a larger AdS region, in order to make them evaporate. Hawking radiation in an asymptotically flat spacetime escapes to null infinity, where it is described by free quantum field theory. A natural object to consider is the algebra of operators acting on matter quantum fields that reach future null infinity prior to some retarded time $u_0$
\begin{align}\label{eq:aqft}
    \mathcal{A}_{QFT,u_0} =\operatorname{span}\{\text{smeared functions of }O_\textit{QFT}(u,\Omega)\}, \quad\quad u\in (-\infty,u_0].
\end{align}
Because gravity is free in the IR, this algebra is well defined (and indeed is isomorphic to the free field algebra) even at finite $G$. 

The entropy of $\mathcal{A}_{QFT,u_0}$ will be divergent due to vacuum entanglement, but after subtracting that divergence one can compute the entanglement entropy as a function of retarded time $u_0$ using a gravitational replica trick, where the relevant matter modes are swapped between replicas in a Schwinger-Keldysh path integral. This leads to the usual QES prescription/island rule. Note that we do not expect the entropy of $\mathcal{A}_{QFT,u_0}$ to return to zero (after vacuum subtraction) in the limit $u_0 \to \infty$. This is because radiation modes in $\mathcal{A}_{QFT,u_0 \to \infty}$ can (and will) be entangled with graviton radiation that is not included in $\mathcal{A}_{QFT,u_0}$. However, because gravitons only contribute a small portion of the entropy of Hawking radiation \cite{Page1}, unitarity does require the entropy of matter fields in $\mathcal{A}_{QFT,u_0}$ to be much smaller than the naive semiclassical answer at sufficiently late times, because the latter becomes much larger than the sum of the entropy of graviton radiation plus the remaining Bekenstein-Hawking entropy of the black hole. As a result, the replica trick calculation leads to a nontrivial entanglement island. The generalised entropy of the island will be equal to the sum of the Bekenstein-Hawking entropy plus the entanglement between gravitons in the island and the radiation, matching the answer expected from unitarity.

In many ways, the algebra $\mathcal{A}_{QFT,u_0}$ includes all the most practically relevant operators at null infinity, since gravitons are considerably harder to measure than photons or even neutrinos. However, we can also construct a larger asymptotic algebra that does include gravitons \cite{Laddha_2021, Prabhu:2022zcr}.
To do so, it is helpful to first briefly review the phase space of null infinity. In an asymptotically flat spacetime near future null infinity, the metric can be written in Bondi-Sachs coordinates as
\begin{align}
    ds^2 = - \left(1 - \frac{2 m_B(u,\Omega)}{r}\right)du^2 - 2 (1 + O(r^{-2})) du dr + r^2\left(\delta_{AB} + \frac{C_{AB}(u, \Omega)}{r}\right) d \Omega^A d \Omega^B.
\end{align}
The Bondi mass $m_B$ is related to the gravitational and matter energy flux through null infinity by the constraint \cite{Prabhu:2022zcr}
\begin{align}\label{eq:gaugeconstr}
     \partial_u m_B =\frac{1}{4} \mathscr{D}^A \mathscr{D}^B N_{AB}-\frac{1}{8}N_{AB}N^{AB}-4\pi G T_{uu}^{(0)},
\end{align}
where $\mathscr{D}^A$ is the covariant derivative on the unit shpere, $N_{AB} = \partial_u C_{AB}$ the news tensor, and $T_{uu}^{(0)}=\lim_{r\rightarrow\infty} \bigg[r^2 T_{uu}^{\rm matt}\bigg]$ is the null energy flux of matter fields at infinity. 

The only dynamical gravitational degrees of freedom are therefore encoded in the $N_{AB}$, which can be canonically quantised using the usual techniques for free field theories. Including gravitons, the full algebra describing Hawking radiation escaping before retarded time $u_0$ is then
\begin{align}\label{eq:agrav}
    \mathcal{A}_{rad,u_0} =\operatorname{span}\{\text{smeared functions of }O_\textit{QFT}(u,\Omega), N_{AB}(u,\Omega)\}, \quad\quad u\in (-\infty,u_0].
\end{align}
For any finite $u_0$, this is a proper subalgebra of the full set of gravitational observables. In particular, its commutant includes all operators at future null infinity at retarded time $u > u_0$. Its entropy can again be computed using a replica trick, and follows a Page curve due to the contributions of entanglement islands. Because gravitons are now included in $\mathcal{A}_{rad,u_0}$, its entropy will return to (approximately) the vacuum value once the evaporation is complete.

A crucial point is that the algebra $\mathcal{A}_{rad,u_0}$ does not include the ADM mass $H_{ADM}$ or, for that matter, the Bondi mass $m_B$. An easy way to see that this is true is that all operators in $\mathcal{A}_{rad,u_0}$ commute with radiation operators at $u > u_0$, while $H_{ADM}$ does not. Of course, we could always enlarge $\mathcal{A}_{rad,u_0}$ to a new algebra $\mathcal{A}_{grav}$ that includes $H_{ADM}$, as was done in \cite{Laddha_2021}. However, once we do so, the algebra no longer depends on $u_0$: by conjugation with $\exp(i H_{ADM} t)$, we can translate operators along null infinity into the future and so all radiation operators, at any retarded time $u$ are included in $\mathcal{A}_{grav}$. Since the algebra $\mathcal{A}_{grav}$ is independent of $u_0$, so is its entropy and it has no associated Page curve.

Which algebra, $\mathcal{A}_{rad,u_0}$ or $\mathcal{A}_{grav}$, is correct? It was argued in \cite{Laddha_2021} (see also \cite{Raju:2021lwh,Raju:2024gvc}) that the answer is $\mathcal{A}_{grav}$: because the corrections to the Minkowski metric from $H_{ADM}$ (or equivalently from $m_B$) and for the news $N_{AB}$ both scale as $1/r$, they argued that both are comparably easy to measure at infinity, and so there is no reason to include one in the algebra of observables associated to asymptotic infinity but not the other.\footnote{Note, however, that, if we keep the semiclassical black hole radius fixed while taking $G \to 0$, then, in order to measure $H_{ADM}$ to $O(T_{BH})$ precision, we need to measure the metric to $O(G)$ precision. On the other hand, measuring gravitational Hawking radiation only requires us to measure the metric to $O(\sqrt{G})$ precision.} They therefore argued that the $u_0$-independence of the entropy of $\mathcal{A}_{grav}$ means that the Page curve is wrong (or at least misleading) even though they expected that $\mathcal{A}_{rad,u_0}$ would follow a conventional Page curve.

From our perspective, and we believe the perspective of most of the field, the Page curve was always supposed to describe the entropy of (a subset of) Hawking radiation -- and, in particular, \emph{only} of radiation. The news operators $N_{AB}$ describe gravitational radiation, while $H_{ADM}$ describes the non-propagating gravitational Coulomb field. The latter is related via constraints to the total energy of the entire spacetime, but does not directly describe or only depend on radiative modes. So $\mathcal{A}_{rad,u_0}$, and not $\mathcal{A}_{grav}$, is the algebra that is ``supposed'' to follow a Page curve.\footnote{Of course, if we are only interested in operators that can be easily measured at infinity, the most relevant algebra is probably $\mathcal{A}_{QFT,u_0}$, which also follows (a slight variant of) a Page curve.} Using entanglement islands, we can see that indeed it does.

\subsection{Radiation inside a gravitating spacetime}\label{sec:bulkradiation}

In the real world, experiments are never carried out at asymptotic infinity, of either the flat space or AdS varieties: infinity is merely a convenient theoretical idealisation. To be physically relevant, asymptotic observables must provide an approximate description of experiments at large but finite distance. In the LHC, for example, ``infinity'' is the detector -- a handful of metres from the particle collision. In this section, we explain how entanglement islands arise even for bulk radiation at finite distance from the black hole, explaining how the entropy of bulk radiation is well defined and follows a Page curve for sufficiently small black holes in AdS/CFT. 

Let us start with AdS/CFT. A sufficiently small black hole, with entropy $S_{BH}$, is microcanonically unstable and will evaporate into a cloud of Hawking radiation heading out towards the boundary. Since no radiation is escaping the spacetime, the black hole evaporation process conserves boundary stress-energy and gravitons are massless. Since all the Hawking radiation is in the causal wedge of the boundary CFT, we can use the HKLL dictionary \cite{Hamilton:2006az} to reconstruct it on the boundary order-by-order in perturbation theory. Non-perturbatively, uncontrolled quantum gravity effects will of course eventually come into play, but for microcanonically unstable black holes such corrections will be much smaller than $\exp(-S_{BH})$. In particular, given two copies of a small evaporating black hole, we can define a bulk operator in effective field theory that swaps some subset of radiation (with an appropriate UV cutoff, etc.) between the two copies. We can then reconstruct that operator on the boundary to all orders in perturbation theory. Any particular boundary reconstruction will be a well defined nonperturbative operator, even though its action is only constrained up to nonperturbative corrections by the requirement that it is  a ``good'' reconstruction. By analogy with the swap test, it is natural to define the ``second R\'{e}nyi entropy of the bulk radiation'' as the expectation value of this boundary operator on two copies of the CFT state. See \cite{Bousso:2023kdj} for a more detailed discussion. There is a straightforward generalisation of this to integer-$n$ R\'{e}nyi entropies for $n>2$ where the swap operator is replaced by a cyclic permutation of $n$ copies of the radiation. These bulk R\'{e}nyi entropies will follow a Page curve because, for the usual reasons, the dominant saddle will include a replica wormhole whenever the majority of the radiation is included in the swapped subset. Note that the reconstructed swap operator, like essentially any matter boundary operator, breaks boundary stress-energy conservation. However, it does so only in a region that is spacelike separated from the entanglement island; the presence of the swap operator cannot causally influence the local physics in the island.

An interesting question is to what extent this swap-operator ``entropy'' is genuinely the R\'{e}nyi entropy of some nonperturbatively well defined degrees of freedom. An initial simple observation is the following. Suppose we use HKLL-style reconstruction to find a boundary isometry $V_R: \mathcal{H}_{\rm CFT} \to \mathcal{H}_{\rm CFT} \otimes \mathcal{H}_R$ that takes the radiation we are interested in and moves it into a reference system $R$, leaving the original radiative modes in their vacuum state. We can use $V_R$ to swap radiation between two black holes as above: to do so, we simply apply the operator ${V_R^\dagger}^{\otimes 2} {\rm SWAP}_R V_R^{\otimes 2}$ to first move all the radiation (from both copies) into reference systems, swap the reference systems, and finally return the radiation to the original spacetimes. It follows immediately from this that the R\'{e}nyi entropies of $R$, in the state created after applying $V_R$ to extract the radiation, are exactly the swap-operator entropies defined above, and hence that the von Neumann entropy of $R$ is given by the island rule. 

However, the isometry $V_R$ again introduces an auxiliary nongravitational bath, which is exactly what we are trying to avoid in order to demonstrate the irrelevance of massive gravitons. What we would really like to do is to define a nonperturbative subalgebra acting on the original Hilbert space $\mathcal{H}_{\rm CFT}$, whose R\'{e}nyi entropies can be computed using the bulk swap operators. One way to do this is the following. We first take the algebra of observables $\mathcal{B}(\mathcal{H}_R)$ and conjugate it with $V_R$ to get a subspace of CFT observables $V_R^\dagger \mathcal{B}(\mathcal{H}_R) V_R \subset \mathcal{B}(\mathcal{H}_{\rm CFT})$. This subspace will always contain the identity $\mathds{1}_{\rm CFT} = V_R^\dagger V_R$ and is preserved when taking adjoints, but it will not form an exact subalgebra because $V_R^\dagger a V_R V_R^\dagger b V_R$ will in general be very close but not exactly equal to $V_R^\dagger a  b V_R$. Fortunately, the map from $\mathcal{B}(\mathcal{H}_R)$ to $\mathcal{B}(\mathcal{H}_{\rm CFT})$ defined by conjugation by $V_R^\dagger$ does form a $\delta$-homomorphism between $C^*$ algebras in the sense of Kitaev \cite{Kitaev:2024qak}. Since we can always take $\mathcal{H}_R$ to be finite dimensional, we can then apply Lemma 8.2 of \cite{Kitaev:2024qak} (in the special case $\varepsilon = 0$) to argue that we can always perturb the $\delta$-homomorphism induced by $V_R$ by an $O(\delta)$ amount and thereby obtain an exact homomorphism between the two algebras.\footnote{We would like to thank Daniel Ranard for pointing out this argument to us.} The image of $\mathcal{B}(\mathcal{H}_R)$ under this exact homomorphism will be a nonperturbative finite-dimensional subalgebra of $\mathcal{B}(\mathcal{H}_{\rm CFT})$ and will satisfy the property that, for any $a \in \mathcal{B}(\mathcal{H}_R)$, the image of $a$ will be $O(\delta)$-close in operator norm to $V_R^\dagger a V_R$. If we take $N$ very large while keeping $S_{BH}$ fixed, the nonzero $\delta$ in the original approximate homomorphism induced by $V_R$ can be arbitrarily small compared to $\exp(-S_{BH})$. It then follows that the von Neumann entropy of the exact radiative subalgebra in $\mathcal{B}(\mathcal{H}_{\rm CFT})$ will be approximately the same as that of $R$ above and, consequently, will be given by the generalised entropy of the entanglement island.

\subsection{Measuring the Page curve}
\label{sec:experiment}

In the previous subsection, we considered radiation inside a gravitating spacetime, but worked within the framework of AdS/CFT and implicitly assumed the existence of a boundary superobserver able to make arbitrary measurements without restriction from locality or causality. What about when the observer, like the radiation and like us, is inside the gravitating spacetime? Can they then see a Page curve or black hole unitarity at all? Clearly it would be fairly hard to do, but we will argue that it should not be impossible.

The first step is to build a Dyson sphere around the black hole. This is no harder than building a Dyson sphere at the same radius around a star of the same mass and is certainly not disallowed by the laws of physics. The second step is to coherently collect photons in the S-wave (and other low angular momentum modes) and channel them into e.g. a fibre optic cable (or the analogous one-dimensional waveguide for the relevant frequency).\footnote{A slight difficulty here is that at scales where neutrinos but no other Standard Model (SM) matter particles are massless, Hawking radiation is dominated by neutrinos, which are considerably harder to collect than photons. This feels like a detail of the SM and not a fundamental obstruction so we won't focus on it more, except to say that it is not true for either very large black holes (where neutrinos are frozen out) or very small black holes (where most of the radiation is made of other SM particles).} This is difficult, but it should still be achievable with an error rate per photon that is much smaller than one -- perhaps $O(1\%)$.\footnote{One way to see this is that there is clearly no upper bound on the precision with which the radiation can be collected within the framework of quantum field theory in curved spacetime. As a result, for large black holes in quantum gravity, any lower bound on the minimum error rate per photon must go to zero as $G \to 0$ (or equivalently as $S_{BH} \to \infty$).} Once the photons are safely in the fibre optic cable, we can use quantum error correction techniques to encode them into the logical state of a fault tolerant quantum computer. In an infinite universe, if the quantum computer in which they are stored is large enough, the threshold theorem \cite{Shor:1996qc,Aharonov:1999ei,Knill:1997mr,Kitaev:1997wr} allows the logical error rate (once successfully encoded) to be arbitrarily small, and, in particular, much smaller than $\exp(-S_{BH})$.\footnote{In our universe cosmological constraints will eventually come into play, but these can be avoided by making the black hole relatively small. The ratio of the Planck and Hubble scales is very large, and error rates in a fault tolerant quantum computer decay exponentially with the number of physical qubits used, so there is a lot of room to play with here.} 

We can now safely carry out e.g. quantum state tomography in order to determine the exact quantum state of the radiation in the computer. Doing so will require an exponential number of experiments, but if an experiment can be done once, it can also be done many times in precisely the same way. We can also collect multiple identical copies of the radiation at the same time, and then swap/permute them with exponential precision. The expectation value of these measurements (which can be determined by repeating it exponentially many times) gives the R\'{e}nyi entropies of the radiation in the computer and can be computed using saddles that will contain replica wormholes when more than half of the total radiation is in the computer. The von Neumann entropy of the radiation in the computer can therefore be computed using an entanglement island and follows a Page curve.

A crucial point here is that we only required the ability to carry out logical operations with exponential precision once the radiation had been successfully collected and encoded in a fault tolerant quantum computer. While collecting the radiation and uploading it into the quantum computer, we only required the error rate per photon to be controlled by some small parameter $\varepsilon \ll 1$. Because there are $O(S_{BH})$ photons in the radiation at the Page time, the fidelity between the ``true'' radiation state and the actual state in the quantum computer will be $\exp(-O(\varepsilon S_{BH}))$. If we wanted this fidelity to be close to one, we would need $\varepsilon \ll 1/S_{BH}$. We would then have to worry about quantum gravity effects making our protocol fail.
Fortunately, for essentially the same reason that Hawking's original information problem was robust to ordinary perturbative corrections, an uploading error rate $\varepsilon$ that is small but $O(1)$ is completely sufficient if we just want to observe the effects of entanglement islands in creating a Page curve. Essentially, this is because we can lose a small but $O(1)$ fraction of the radiation and still see a nontrivial Page curve. Losses, or other errors, during the collection process are therefore not a problem so long as a) the lost radiation has much less entropy than the radiation that is successfully collected and b) we eventually end up with the state being encoded in a fault tolerant quantum computer where operations can be carried out with arbitrary precision.

A more precise technical argument is the following. The noisy collection process can always be described by some quantum channel $\mathcal{N}$. We can describe this channel in a Stinespring picture by purifying the output using an environment subsystem $E$. Over long timescales, the collection process will approximately factorise as $\mathcal{N} \approx \prod_i \mathcal{N}_i$ where each channel $\mathcal{N}_i$ acts on radiation collected over a time period $i$, which we assume to be much longer than the lightcrossing time but much shorter than the Page time. So long as the error rate per photon $\varepsilon\ll 1$ is small, we have
\begin{align}
    S(E_i) \ll S(R_i)
\end{align}
where $S(R_i)$ is the entropy of the radiation $R_i$ collected during time period $i$ (i.e. the output of the channel $\mathcal{N}_i$) and $S(E_i)$ is the entropy of the corresponding purifying subsystem. Semiclassically, the radiation $R = R_1 \dots R_n$ is thermal, so that
\begin{align}\label{eq:sumsri}
S(R) \approx \sum_i S(R_i)
\end{align}
But, by unitarity, we have
\begin{align}\label{eq:sumsei}
    S(R) = S(B E_1 \dots E_n) \leq S(B) + \sum_i S(E_i),
\end{align}
where the entropy $S(B)$ of the black hole is bounded by the Bekenstein-Hawking entropy $S_{BH}$ after the radiation is emitted. In the inequality, we used the subadditivity of von Neumann entropy. Sufficiently long after the Page time, \eqref{eq:sumsri} is much bigger than $S_{BH}$. Hence, \eqref{eq:sumsri} is inconsistent with \eqref{eq:sumsei} and we have an information problem that is resolved, as usual, by the entanglement island. The presence of the environment entropy on the right hand side of \eqref{eq:sumsei} somewhat delays the precise Page time at which the entanglement island becomes dominant and the Page curve begins to drop, but this delay is small compared to the full evaporation time of the black hole. The basic shape of the Page curve is unchanged by the presence of mild noise during the collection process.

\section{Constructing compactly supported operators in gravity}
\label{sec:linearconstraints}

In this section we give a very general argument which explicitly shows that any compactly supported operator built from matter fields can be gravitationally dressed to become a gauge-invariant compactly supported operator, to all orders in perturbation theory in $G$ around a generic background spacetime.

\subsection{Why should localised gauge-invariant operators exist?}
\label{sec:opsexist}

Before delving into the details, we discuss the intuition for why gauge-invariant operators entirely localised in a compact bulk subregion should exist at all orders in perturbation theory.
The operators we are looking for are localised in a bulk subregion, and must thus commute at all orders in perturbation theory with the asymptotic charges as well as with the algebra of operators in the complementary bulk subregion. The crux of the arguments in \cite{Geng:2021hlu} for the inconsistency of islands in theories of massless gravity is that such operators do not exist.

Such a result can be true for backgrounds with a lot of symmetry, such as vacuum AdS \cite{Chowdhury:2021nxw} or Minkowski space. In these cases any deformation whatsoever must raise the energy, which is an asymptotic charge which can be measured at infinity. This immediately tells us that there is no way to dress a matter operator with compact support: the dressing must include `Coulomb tails' extending to infinity. This can be understood as a consequence of time translation symmetry of the background. Intuitively, a gauge-invariant operator requires a diffeomorphism-invariant specification of the time at which it acts, which must be measured relative to some feature of the background. But a stationary background lacks any such features, so a gauge-invariant operator must either be time-independent (commuting with the Hamiltonian, so it does not change the total energy) or dressed to infinity (where the time translation becomes a physical asymptotic symmetry rather than a gauge transformation). Similar comments apply to other symmetries: for example, on a spherically symmetric background compactly supported gauge-invariant operators must also be spherically symmetric, having zero angular momentum.

Nonetheless, such isometries are the only obstruction to dressing operators in perturbation theory. If we consider generic backgrounds, namely those breaking all symmetries, localised gauge invariant operators can be defined. The intuitive reason, which we will make quantitatively precise in the following subsections and in Section \ref{sec:BBPSVops}, is that the spacetime features breaking the symmetries (for example the backreaction caused by an asymmetric lump of matter) can be used to uniquely specify the bulk location where operators act. These ``relational observables'' were discussed extensively in past literature, see e.g. \cite{Donnelly:2015hta,Donnelly:2016rvo,Isham:1992ms,Giddings:2005id,Marolf:1994wh,Geng:2024dbl}.

In the context of an evaporating black hole, which is the most relevant one for a discussion of entanglement islands, the background does indeed break all symmetries in generic settings. For simplicity, let us focus our attention on time translation symmetry and the associated asymptotic charge, the ADM Hamiltonian. The same reasoning can be applied to other symmetries and charges. A black hole formed from collapse clearly breaks time translation symmetry. A scalar feature of the collapsing star, for example its volume, can be used as a clock with respect to which we measure time for other observables. We can use this feature to dress bulk operators to the collapsing star rather than to the asymptotic region. The resulting dressed operator has support only within the island, so in particular will commute with the ADM Hamiltonian. In terms of charges, an operator dressed to the star might create an excitation increasing the energy at one point, but will ``borrow'' energy from the star  so that the ADM energy of the whole spacetime is unchanged.

Nonetheless, this important example is more subtle than a generic background because  we are often interested in parametrically long times (of order $G^{-1}$) after the black hole forms, where there is locally an approximate time translation symmetry (broken only at order $G$). We will comment on this after describing the main construction.

It is worth noting briefly that there is another very simple and intuitive argument why bulk operators localised in the interior of black holes should exist to all orders in perturbation theory. The exterior of an equilibrated black hole is uniquely characterised by its mass, charge and angular momentum. However, it is easy to construct exponentially many states for the interior of such a black hole. For example, we can allow the black hole to evaporate for a while and then measure the Hawking radiation, projecting the Hawking radiation (and hence also its interior partners) onto some typical product state. The number of such states is exponential in the entropy of the Hawking radiation (and hence exponential in $S_{BH}$ close to the Page time) and they are all approximately orthogonal in the black hole interior. But to all orders in perturbation theory there are only polynomially many values the charge, mass and angular momentum can take, meaning that some of the interior states must be indistinguishable in the exterior. Consequently, there must therefore exist gauge-invariant unitary operators, localised in the interior, that simply exchange indistinguishable microstates. We remark that the above argument is only valid in perturbation theory: the holographic principle tells us that all microstate data is encoded in the exterior, but this is understood to be true only non-perturbatively.

\subsection{Formulating the problem}

With this intuition in mind, we turn to the analysis of compactly supported diffeomorphism-invariant perturbations. We will begin with a classical discussion, later explaining how to adapt the argument to the quantum theory. The main ideas are closely related to the `linearisation stability' problem, studied in classical general relativity around 50 years ago \cite{fischer1973linearization,moncrief1975spacetime}. This problem is to determine whether a solution to the linearised Einstein equations around a given background solution approximates a family of exact nonlinear solutions. We refer to these papers for more mathematical details and explicit calculations for Einstein gravity (in particular, \cite{moncrief1975spacetime} contains essentially all the necessary ingredients for us).\footnote{We would like to thank Don Marolf for pointing out the relevance of this literature to our problem.}

We consider a theory where the Lagrangian $\mathcal{L}$ splits as a gravitational (or gauge) term $\mathcal{L}_\mathrm{g}$ (perhaps the Einstein-Hilbert Lagrangian) and a matter term $\mathcal{L}_\mathrm{m}$ with
\begin{equation}\label{eq:laggm}
    \mathcal{L} = \frac{1}{G} \mathcal{L}_\mathrm{g}+\mathcal{L}_\mathrm{m}.
\end{equation}
We will be working perturbatively in the gravitational coupling $G$, with other parameters (particularly the background field configuration) held fixed. This is `semiclassical' in the sense that $G\to 0$ plays the role of a classical $\hbar\to 0$ limit for the gravitational sector but not necessarily for the matter. We take $\mathcal{L}_\mathrm{g}$ to depend on gravitational fields only (usually just the metric, though we could include other fields which have a classical background value such as a dilaton in a two dimensional theory), while $\mathcal{L}_\mathrm{m}$ also depends on additional matter fields.

We will work in the Hamiltonian formulation, which makes the passage from the classical to the quantum theory most transparent. Denote the phase space variables of the gravitational sector by fields $z^I$, where $I$ is a collective index which includes the spatial coordinates $x$ (for example, different values of $I$ correspond to components of the spatial metric $g_{ij}(x)$ and its conjugate momentum $\Pi^{ij}(x)$ at all possible positions). For these phase space coordinates we choose to define momenta which are independent of $G$, which means that a factor of the coupling appears in the Poisson brackets (and similarly for the canonical commutation relations in the quantum theory), which take the form
\begin{equation}
    \{z^I,z^J\} = G \, \omega^{IJ}.
\end{equation}
Matter sector phase space coordinates $y^M$ are defined similarly, but we do not scale their Poisson brackets with any factor of $G$. For example, in Einstein gravity we define momenta $\Pi^{ij}(x)$ in terms of extrinsic curvature $K^{ij}(x)$ as $\Pi^{ij} = -\sqrt{\det g}(K^{ij}-g^{ij}\tr K)$, with no factors of $G$. These factors instead go into the Poisson bracket:
\begin{equation}
    z^I \longrightarrow (g_{ij}(x), \Pi^{ij}(x)),  \qquad \{g_{ij}(x),\Pi^{kl}(y)\} = 8\pi G (\delta^i_k\delta^j_l + \delta^i_l\delta^j_k)\delta^{(3)}(x-y).
\end{equation}

Now, the only possible obstruction to constructing compactly supported operators comes from the constraints, specifically the Hamiltonian $\mathcal{H}(x)$ and momentum $\mathcal{P}_i(x)$ constraints associated with diffeormorphisms of the temporal and spatial coordinates respectively. We collectively denote these by $\Phi_A$ (where again $A$ includes the spatial coordinate $x$), and given \eqref{eq:laggm} we can split these into gravitational and matter pieces as\footnote{There could be higher order terms in $G$, for example if the matter Lagrangian contains derivative couplings of the gravitational field or self-coupling of order $G$, but these do not materially effect what follows.}
\begin{equation}
    \Phi_A(y,z) = \Phi_A^\mathrm{g}(z)+ G \Phi_A^\mathrm{m}(y,z),
\end{equation}
where $\Phi_A^\mathrm{g}$ depends only on gravitational fields $z^I$, while the matter term $\Phi_A^\mathrm{m}$ (which is nothing but the matter energy-momentum density) can depend on all fields. Classically, states are required to satisfy $ \Phi_A=0$. Additionally, the constraints generate gauge transformations with gauge parameters $\lambda^A$ (for gravity, a vector field generating a diffeomorphism), which act on any function of phase space $F$ as the Poisson bracket
\begin{equation}
    \delta_\lambda F = \left\{  \lambda^A\Phi_A,F\right\}.
\end{equation}
Concretely, for Einstein gravity we have
\begin{equation}
    \Phi_A^g \longrightarrow \left(\mathcal{H}=(\det g)^{-1/2}\left(\Pi^{ij}\Pi_{ij}-\tfrac{1}{2}(\tr \Pi)^2\right)-(\det g)^{1/2} (R-2\Lambda), \quad \mathcal{P}^i = -2\nabla_j \Pi^{ij}\right),
\end{equation}
where everything on the right hand side is a function of spatial coordinates.

For the gravitational variables, we will always be working perturbatively around some classical background $z=z_0$ which solves the constraints. Note that this background could include some matter sources (e.g., a star that collapses to form a black hole) with energy-momentum of order $G^{-1}$: in such a case, the   $y^M$ variables parameterise configurations of the matter which change $\Phi_A^\mathrm{m}$ by an amount which is parametrically $O(1)$ in $G$, and $\Phi_A^\mathrm{m}$ is defined as the difference from some arbitrary reference state $y_0$. We will be working entirely in a compact region $\island$, so for all fields ($z^I$, $y^M$) we can take the indices to range only over values corresponding to points within $\island$ (with fields outside held fixed to the choice of background). More specifically, requiring compact support means that any allowed ($z^I$, $y^M$) must leave the background unchanged in a neighborhood of the boundary of $\island$. Similarly, $\Phi_A$ parameterises constraints in $\island$, but in this case the gauge parameter $\lambda^A$ (the diff-generating vector field) is not required to vanish at the boundary $\partial\island$. In particular, this means that operators which commute with $\Phi_A$ will also commute with any global diff, including asymptotic charges like the Hamiltonian (where $\lambda^A$ is the restriction of $\partial_t$ to $\island$ for some choice of time coordinate $t$).

Now, the ultimate aim  is to show that any compactly supported (quantum) operator acting on matter variables can be dressed order-by-order in perturbation theory to become a physical compactly supported gauge-invariant operator which commutes with the constraints. But it will be helpful to warm up with slightly simpler classical problems. First, we show that an arbitrary finite change of the matter state (moving to a different value of $y$) can be accompanied by a perturbative and compactly supported change of gravitational variables $z^I$ to solve the constraints, order-by-order in perturbation theory. Then, we show the corresponding statement for operators (classically meaning a function of phase space): an arbitrary compactly supported  matter operator $O^{(0)}$ (a function of $y$ only) can be dressed perturbatively to a compactly supported physical operator $O$ which has vanishing Poisson brackets with the constraints. With these classical problems understood, it will then be straightforward to implement the same idea in the quantum theory.

\subsection{Changes to the state}

Consider first changing the classical state of matter fields (i.e., choosing some arbitrary $y$ in the matter sector phase space describing the region $\island$), and looking for an accompanying change in the gravitational variables $z$ so that the constraints are solved to leading order in $G$. This means that we would like to find $\delta z^I$ (supported in $\island$) to satisfy
\begin{equation}\label{eq:linearisedconstr}
    \frac{\partial \Phi^g_A}{\partial z^I }(z_0) \delta z^I  = -G  \Phi_A^m(y,z_0),
\end{equation}
for an arbitrary source $\Phi_A^m$ appearing on the right hand side. For this to be possible, we need the derivative of the gravitational constraints on the background $z_0$ (which we denote by $D\Phi^g_0$) to be surjective.  $D\Phi^g_0$ is the linear operator from the tangent space of the gravitational phase space at $z=z_0$  to the space of constraints (indexed by $A$), with matrix elements given by $\frac{\partial \Phi^g_A}{\partial z^I }$ evaluated at $z=z_0$.

Concretely, for Einstein gravity this maps a variation $(\delta g_{ij}(x),\delta \Pi^{ij}(x))$ of the metric and extrinsic curvature  to a scalar density and a vector density defined on the spatial manifold, corresponding to the variation of the Hamiltonian and momentum constraints respectively. Explicitly, these are given by the following \cite{moncrief1975spacetime}:
\begin{align}\label{eq:DPhiex}
    D\mathcal{H}_{g,\Pi}(\delta g,\delta \Pi) = &  (\det g)^{-1/2}\left[-\tfrac{1}{2}(\Pi^{ij}\Pi_{ij}-\tfrac{1}{2}( \Pi^i_i)^2)\delta g^k_k +(2\Pi^{ij}\delta \Pi_{ij}-\Pi^i_i \delta\Pi^j_j) + (2\Pi^{ik}\Pi_k^j-\Pi^k_k\Pi^{ij})\delta g_{ij} \right]\nonumber \\
    &\qquad - (\det g)^{1/2}\left[\nabla^i\nabla^j \delta g_{ij}-\nabla^2 \delta g^i_i - (R^{ij}-\tfrac{1}{2}g^{ij}R+\Lambda g^{ij})\delta g_{ij}\right] \\
    D\mathcal{P}^i_{g,\Pi}(\delta g,\delta \Pi) =& -2\nabla_j\delta \Pi^{ij} - \Pi^{jk}(\nabla_k \delta g^i_j+\nabla_j \delta g^i_k-\nabla^i \delta g_{jk}). \nonumber
\end{align}

The conclusion from the above is that this notion of perturbative dressing is possible if and only if $D\Phi^g_0$ is  surjective for the background  $z_0$ in question. This is a constraint on the background: for example, it fails if $z_0$ corresponds to initial data on a partial Cauchy surface in empty Minkowski or AdS spacetime. Fortunately, there is a simple characterisation of such spacetimes: the derivative $D\Phi^g_0$ is surjective when the background spontaneously breaks the gauge symmetry, meaning that there is no choice of gauge parameters $\lambda^A$ which leaves the background invariant \cite{moncrief1975spacetime}. For our gravitational application, this means that $D\Phi^g_0$ fails to be surjective if and only if the background in the region $\island$ in question is invariant under some diffeomorphism (which is \emph{not} required to vanish at the boundary of $\island$), i.e. if the background has an isometry.

To see this, suppose that there is some set of sources for which no solution $\delta z^I$ exists. Then (under some technical conditions) there's a nonzero $\lambda^A$ which is orthogonal to the subspace of obtainable sources, satisfying $\lambda^A T_A =0$ for all $T_A$ in the image of $D\Phi^g_0$ (which we can think of as `dressable' stress-energy sources).\footnote{This pairing means that $\lambda^A$ should be thought of as belonging to the dual space of $T_A$. In particular, since $T_A$ is always compactly supported there is no such condition on $\lambda^A$, so the diff need not vanish on the boundary of $\island$.} So, $\lambda^A \frac{\partial \Phi^g_A}{\partial z^I }$ vanishes at $z=z_0$. This implies that  $\lambda^A \Phi_A$ generates a gauge symmetry (with $\omega^{IJ}\lambda^A \frac{\partial \Phi^g_A}{\partial z^I}\frac{\partial }{\partial z^J } $ tangent to the gauge orbits) which leaves the background $z=z_0$ invariant. More abstractly, this argument says that an operator $D\Phi^g_0$ is surjective iff its adjoint (with respect to some non-degenerate bilinear form) is injective. The bilinear form in question for the linearised perturbations $\delta z^I$ is the symplectic form for the gravitational sector, and the adjoint $(D\Phi^g_0)^\dag$ maps gauge parameters $\lambda^A$ to the infinitesimal gauge transformation that they generate. See \cite{moncrief1975spacetime} for more technical details.

In practice, to compute the components of the adjoint $(D\Phi^g_0)^\dag$ we first integrate the linearised constraints \eqref{eq:DPhiex} against a gauge transformation ($\eta,\xi_i$) which represent the temporal and spatial components of a diffeomorphism to give $\int (\eta \,D\mathcal{H}+\xi_i\,D\mathcal{P}^i)$. Then, integrate by parts so that the deformations $(\delta g_{ij},\delta \Pi_{ij})$ appear without derivatives, and read off their coefficients. For a simple illustrative example, consider the case where the background has $\Pi_{ij}=0$ (e.g., the initial data surface is a moment of time-reflection symmetry) and look for the component of $\delta \Pi^{ij}$ corresponding to gauge transformations of the spatial metric $g$ (the symplectic conjugate of $\delta \Pi$). This arises purely from the variation of the momentum constraint $D\mathcal{P}^i = -2\nabla_j\delta\Pi^{ij}$, which after taking the adjoint becomes  $\nabla_i\xi_j+\nabla_j\xi_i$. This vanishes when $\xi$ generates an isometry of the background spatial metric. So in this case, if there are no such isometries in $\island$ then for any matter energy-momentum source $\mathcal{P}^i_m$, the linearised momentum constraint $2\nabla_j\delta\Pi^{ij} = G \mathcal{P}^i_m$ has a solution for $\delta\Pi^{ij}$ with support in $\island$. 

Similar considerations relate the dressing of energy (solution of the Hamiltonian constraint) to the existence of timelike isometries. In fact, the `Gauss law' introduced in \cite{Geng:2021hlu} is precisely of this form, where the background corresponds to a constant $t$ slice of a static background (which has $\Pi_{ij}=0$ by time), and we choose the gauge generator $\lambda^A$ to be the lapse $N(x)$. Following their definitions, we have a Gauss law if $N D\mathcal{H}$ can be written as a total derivative for arbitrary variations $\delta g_{ij}$ (there is no dependence on $\delta \Pi_{ij}$ for a background with $\Pi_{ij}=0$). In this case $\int_\island N D\mathcal{H}$ always vanishes, meaning that we cannot dress a source of energy density $\rho$ within the region $I$ if the total energy $\int_\island N\rho$ is nonzero. To determine the condition for $\int_\island  N D\mathcal{H}$ to vanish for all perturbations, we integrate by parts (using \eqref{eq:DPhiex}) and strip off $\delta g_{ij}$ to arrive at the vanishing of the adjoint operator acting on $N$:
\begin{equation}\label{eq:GaussLaw}
    D\mathcal{H}^\dag N = -(\det g)^{1/2}(\nabla^i\nabla^j  - g^{ij}\nabla^2 -(R^{ij}-\tfrac{1}{2}g^{ij}R+\Lambda g^{ij}))N = 0.
\end{equation}
But this is precisely the $ij$ component of the Einstein equations for the background under the assumption that the metric is time-independent, as noted in \cite{Geng:2021hlu}. The upshot is that  we have a Gauss law constraint of this type if and only if the background is stationary. Since our argument is very general, the conclusion applies to any diffeomorphism-invariant local theory.

Note that absence of an isometry of the background in $\island$ implies existence of the perturbation $\delta z^I$, but certainly not uniqueness. Indeed, in general $D\Phi^g_0$ will have a large kernel corresponding to compactly supported purely gravitational deformations (which we are not considering here), some from compactly supported gauge transformations ($\delta z^I = \omega^{IJ}\lambda^A \frac{\partial \Phi^g_A}{\partial z^J}$) and some physical (e.g., gravitational waves). This simply means there are many ways to dress any given matter operator. As a consequence, our general argument will only show existence and not explicitly construct any particular choice of dressing.

It is straightforward to continue this construction of a state satisfying the constraints to all orders in $G$ perturbation theory. For this, write
\begin{equation}
    z=z_0 + G z_1 + G^2 z_2 + \cdots
\end{equation}
and expand the constraints $\Phi_A(z,y)=0$ in powers of $G$ using Taylor series for the $\Phi^g$ and $\Phi^m$ at $z=z_0$. To satisfy the constraints at $n$th order in $G$, we find an equation giving $D\Phi^g_0$  acting on $z_n$ in terms of lower-order terms. Given the condition that $D\Phi^g_0$ is surjective, there is no further obstruction to finding a classical state satisfying the constraints to all orders in $G$.

So far, we have shown that the classical space of physical gauge-invariant states contains a copy of the phase space of matter in the region $\island$, to all orders in perturbation theory, working around a background $z_0$ without symmetries in $\island$. Perturbative quantisation of the fluctuations $\delta z^I$ (and the matter) yields an analogous quantum space of states. We expect this perturbative Hilbert space to be embedded as a `code subspace' of the full Hilbert space (the precise definition depends on the size of allowed fluctuations). In this subspace, operators supported outside $\island$ (defined independently of $G$) -- such as asymptotic charges -- will have correlation functions which are independent of the state to all orders in $G$.  Given such a subspace, there are of course many operators which act on it (defined to be zero on the orthogonal complement, for example), and these will commute with operators outside $\island$ to all orders in perturbation theory. However, this argument is not very concrete so it is worthwhile talking directly about operators.

Before moving onto operators, we first compare with a similar classical problem from  \cite{Folkestad:2023cze}. That work studied the question of whether linearised constraints can be solved independently in two regions, so that an arbitrary perturbation in one region can be dressed without disturbing the other. A particular instance of such independence --- between an arbitrary subregion of $\island$ and the complement of $\island$ --- would imply our criterion, but is a stronger requirement since it can fail in circumstances where our result holds. The main difference is that we separate out an independent matter sector and allow arbitrary gravitational perturbations to dress matter sources, while \cite{Folkestad:2023cze} considers fixing all fields in a region (including gravitational), leaving freedom to choose a dressing only outside the regions in question. So while there are commonalities in motivation and ideas, the precise questions we are studying are somewhat different and complementary.

\subsection{Gauge-invariant operators}

Similarly to the discussion of the state above, we will discuss operators by starting with an arbitrary operator $O^{(0)}$ acting only on the matter sector (classically a function of $y$), and construct a corrected operator $O$ order-by-order in $G$. Classically,  gauge-invariance means that $O$ should be a function on phase space whose Poisson bracket with the constraints vanishes. We define this perturbatively around the background $z=z_0$, so we have an expansion not only in $G$ but also in powers of $\delta z^I = z^I-z^I_0$, which we formally regard as being of order $G$ (since its Poisson brackets are of that order). Explicitly, we can write
\begin{gather}
     O = O^{(0)}+ O^{(1)}+O^{(2)}+\cdots , \\
      O^{(n)}(y,\delta z) = G^n  O^{(n,0)}(y) + G^{n-1}  O^{(n,1)}_I(y) \delta z^I + \tfrac{1}{2} G^{n-2}  O^{(n,2)}_{I,J}(y) \delta z^I  \delta z^J +  \cdots,
\end{gather}
where each $O^{(n,k)}_{I_1,\ldots,I_k}(y)$ (for $k=0,1,\ldots,n$) is a function only of the matter variables. We then demand that the Poisson brackets $\{\Phi_A,O\}$ vanish to all orders in this expansion.

At leading order we have 
\begin{equation}
   \{\Phi_A,O\} =  \{\Phi_A^g,O^{(1)}\} + \{\Phi_A^m,O^{(0)}\}+\cdots = G \left(\omega^{IJ}\left.\frac{\partial \Phi^g_A}{\partial z^I}\right|_{z=z_0} \!\!\!\!\!O^{(1,1)}_J +\{\Phi_A^m,O^{(0)}\} \right) + O(G^2).
\end{equation}
We would like to be able to set this to zero for any matter operator $O^{(0)}$, which sources arbitrary energy-momentum $\{\Phi_A^m,O^{(0)}\}$. As above, this is possible if there are no isometries on the background, so that the linearised gravitational constraint operator $D\Phi^g_0$ is surjective. Notice that the $z$-independent piece $O^{(1,0)}$ of $O^{(1)}$ is completely arbitrary, being part of the ambiguity in choice of dressing.

Once again, the same condition allows us to continue to higher orders. At $n$th order, we get an equation relating the linear operator $D\Phi^g_0$ acting on $\omega^{IJ}\frac{\partial}{\partial z_J} O^{(n)}$ to terms of lower orders. At higher orders, the `source' terms on the right hand side can come either from the matter term $\Phi^m_A$ in the constraints (e.g., $\{\Phi^m_A(z_0,y),O^{(n-1)}\}$)  or from expanding the gravitational constraint $\Phi^g(z_0+\delta z)$ to higher than linear order in $\delta z$ (which can be interpreted as purely gravitational sources of energy-momentum). In either case, the source terms are determined by previous orders in the expansion. Surjectivity of $D\Phi^g_0$ means that there's some choice of $\frac{\partial}{\partial z_J} O^{(n)}$ to make the Poisson bracket at this order, and we can integrate up to get an $n$th order dressing $O^{(n)}$ which is supported in $\island$.

Fortunately, the quantum version of this argument requires very little alteration: we need only replace Poisson brackets by commutators. The constraints and operators are subject to operator-ordering ambiguities, but since the commutators of $\delta z^I$ are suppressed by $G$ these simply show up as changes in the sources at higher orders. Similarly, if there are higher-derivative corrections (such as those generated by renormalisation) they will be suppressed by powers of $G$ (or another small parameter such as the string scale), and only show up as additional sources.

The upshot is that any matter operator $O^{(0)}$ in the region $\island$ can be `dressed' to become a gauge-invariant operator $O$ supported in $\island$ to all orders in perturbation theory, under the condition that the background is not invariant under any diffs in $\island$.

The condition for \emph{any} operator to have a perturbative dressing is that the linearised constraint operator $D\Phi^g_0$ is surjective, or equivalently that its adjoint $(D\Phi^g_0)^\dag$ is injective. This fails if the background fields within the island are left invariant by some residual gauge transformation, so the background has some symmetries (which may be broken outside $\island$). In such a case, the results above tell us that we can still dress operators which are invariant under the symmetry in question, meaning that they commute with the charge which generates the isometry. For example, in a spherically symmetric background we can dress operators with vanishing angular momentum. We comment further on examples with symmetries in Section \ref{ssec:subtle}.

Another way to express this condition for dressing is that the background spontaneously breaks all of the gauge symmetries. One might be tempted to say that this is also a condition for gravity to be massive, but as already commented at the end of Section \ref{sec:massivegravitons} this definition of massive gravity would be wildly expansive, satisfied even for Einstein gravity in generic situations, and certainly not what is usually meant (nor what was meant in \cite{Geng:2020fxl,Geng:2020qvw,Geng:2021hlu,Raju:2020smc,Geng:2023zhq,Laddha_2021}) by this term.

\subsection{An explicit example}\label{ssec:dressingex}

In this section we give an illustration of the ideas above by considering dressing just the momentum  constraints with one spatial dimension. This example is very simple and concrete, but indicative of what happens in more general cases.  Take the gravitational fields to be a spatial metric $ds^2 = a(x)^2 dx^2$ and a scalar $\varphi(x)$ (for example, a dilaton). The following analysis would also apply unchanged in higher dimensions with spherical symmetry for example, where we take $x$  to be a radial coordinate and $\varphi$ the radius or area of the transverse sphere, and we are dressing the radial component of momentum. The momentum constraint is
\begin{equation}
    \Phi^g(x) = p_\varphi(x) \varphi'(x) - p_a'(x) a(x).
\end{equation}
This generates diffeomorphisms of $x$, for example $[\int dy \lambda(y) \Phi^g(y),\varphi(x)]=-i\lambda(x) \varphi'(x)$.

For definiteness and simplicity choose a background with vanishing momenta $p_\varphi =p_a=0$, and with $a(x)=1$ (choosing $x$ to be proper distance on the background), but arbitrary background $\varphi_0(x)$. The linearised constraints depend only on variations of momenta, giving a linear functional mapping two functions $\delta p_\varphi,\delta p_a$ to a single function:
\begin{equation}
    D\Phi^g_0[\delta p_\varphi,\delta p_a ](x) = \varphi_0'(x) \delta p_\varphi(x) - \delta p_a'(x).
\end{equation}
Locally we can straightforwardly solve $D\Phi^g_0 = \mathcal{P}$ for an arbitrary source function $\mathcal{P}$, and in fact we have complete freedom in choosing $\delta p_\varphi(x)$:
\begin{equation}
    \delta p_a(x) = \int_{x_0}^x (\varphi_0'(y) \delta p_\varphi(y)-\mathcal{P}(y))dy.
\end{equation}
However, this solution is not necessarily allowed since we need $\delta p_a$ to be compactly supported on the interval of interest $\island$. We can choose $x_0$ to be the left end of the interval so that $\delta p_a$ vanishes for $x<x_0$ (for any choice of $\delta p_\varphi$ compactly supported on $\island$), but for the solution to vanish on the right we need an integral to vanish: $\int_\island (\varphi_0' \delta p_\varphi-\mathcal{P})=0$. Now, if $\varphi_0'$ is nonzero anywhere in $\island$ we can choose $ \delta p_\varphi$ to satisfy this: we can dress to the background value of the scalar if it is not constant. But if $\varphi_0$ is constant, then we can only dress sources that satisfy the additional constraint of vanishing total momentum, $\int_\island \mathcal{P}=0$.

As an alternative to this explicit solution (which would typically be much more complicated), we can instead follow the above argument and look at the adjoint $(D\Phi^g_0)^\dag$. This is a map from a single function $\lambda(x)$ to a pair of functions,
\begin{equation}
    (D\Phi^g_0)^{\dag}[\lambda](x) = (\varphi_0'(x)\lambda(x),\lambda'(x)).
\end{equation}
To obtain this we integrate the linearised constraints against $\lambda(x)$, integrate by parts to move the derivatives off $\delta p_\varphi$ and $\delta p_a$ (with no boundary terms since these functions are taken to be compactly supported), and read off the coefficients of each of these. Now, for $\lambda$ to be in the kernel of this map we must have $\varphi_0'(x)\lambda(x)=\lambda'(x)=0$. These are of course the conditions for the infinitesimal diffeomorphism generated by the vector field $\lambda(x)\partial_x$ to leave $\varphi$ and $a$ invariant respectively. This has a solution with nonzero $\lambda$ if and only if $\varphi_0'=0$ everywhere in $\island$ (in which case, constant $\lambda$ is a solution). So, if $\varphi_0$ is not a constant then any momentum source can be perturbatively dressed. If $\varphi_0$ is a constant then there is a single obstruction  coming from invariance of the background under a constant $\lambda$ gauge transformation, which requires the source to satisfy $\int_\island \mathcal{P}=0$ exactly as we found above.

In this example we can also give an explicit prescription for dressing operators. If $\varphi_0'$ is nonzero everywhere (take $\varphi_0$ to be an increasing function of $x$ in $\island$, for example), then we can write an exact all orders dressing by `using $\varphi$ as a coordinate'. This means that we can write a dressing of a smeared local matter operator $O^{(0)} = \int f(x)\chi(x)dx$ (where $\chi$ is a matter scalar and $f$ a smooth compactly supported function on $\island$) as
\begin{equation}\label{eq:exDressed}
    O = \int f(x) \chi(\varphi^{-1}(\varphi_0(x)))dx.
\end{equation}
Alternatively, we can write this as
\begin{equation}
    O = \int \chi(x) g(\varphi(x)) \varphi'(x) dx,\qquad \text{where} \quad f(x) = g(\varphi_0(x))\varphi_0'(x),
\end{equation}
and this expression is somewhat nicer because it does not require $\varphi$ to be invertible. In perturbation theory (i.e.,  to all orders when we expand $\varphi$ around the background  $\varphi_0$), these are compactly supported operators in $\island$. Explicitly, to all orders this is precisely the solution we get from the above construction by choosing a solution to the linearised constraints in which only $\delta p_\varphi$ is nonzero (and not adding any additional pure matter operator at each step). For example, to linear order we have
\begin{equation}
    O^{(1)} = -\int f(x) \partial\chi(x) \frac{\delta\varphi(x)}{\varphi_0'(x)} dx.
\end{equation}
This has $[O^{(1)},\Phi^g(x)] = -i f(x) \partial\chi(x)\frac{\varphi(x)}{\varphi_0(x)}$, which at leading order cancels the commutator of $O^{(0)}$ with matter momentum density, $i f(x) \partial\chi(x)$.

The dressed operator in \eqref{eq:exDressed} is a simple example of a more general construction known for more than 60 years  \cite{DeWitt:1962cg,Giddings:2005id}, where we write a diffeomorphism invariant operator as an integral over spacetime:
\begin{equation}\label{eq:integralop}
    O = \int d^{d+1}x \, \chi(x) g(X^\mu(x)) \det (\partial_\nu X^\mu).
\end{equation}
Here, $X^\mu$ are $d+1$ scalars built from background fields (e.g., the dilaton $\varphi$ above, or scalar curvature invariants) and $g$ is a smearing function such that $g(X^\mu)$ is non-zero in a compactly supported region (contained within $\island$) for the background in question. For example, if we take $g$ to be a $\delta$ function this operator instructs us to insert $\chi$ at the point where the field $X^\mu$ all vanish. To connect to the above discussion, use the equations of motion (from the Hamiltonian constraint in perturbation theory) to write operators at time $t$ in terms of operators acting on a specified $t=0$ Cauchy surface. This operator is clearly not compactly supported in general, but it \emph{is} compactly supported to all orders in perturbation theory around a given background (with no symmetries) for judicious choices of $X^\mu$ and $g$.

\subsection{Subtleties}\label{ssec:subtle}

There are several subtleties which prevent the analysis so far from applying to reconstruction of some operators in the important example of an evaporating black hole. Our analysis shows that symmetries of the background are obstructions to dressing operators. A typical black hole formed from collapse has no isometries in the island $\island$, perhaps excepting rotational symmetries for a perfectly spherically symmetric collapse (in which case we can restrict attention to operators with vanishing total angular momentum). So we might conclude that all matter operators can be dressed within $\island$ as described above. However, this does not immediately follow, because there is an \emph{approximate} time-translation symmetry in a large region (generated by Schwarzschild time translation $\partial_t$) once the collapsing star has settled down. This symmetry is broken in two ways, both of which allow us to evade the obstruction to dressing, but both with their own subtleties.

\subsubsection*{Dressing to the star}

First, the time-translation symmetry is broken by the collapse which formed the black hole initially. This allows us to `dress to the star', defining the $t$-coordinate of the insertion with relation to the collapsing matter which breaks the symmetry. This possibility is visible in the simple example of Section \ref{ssec:dressingex}, where we can think of $x$ as something like the (spacelike)  Schwarzschild $t$ coordinate in the black hole interior. If the background $\varphi_0$ is constant in an interval containing a point $x$ we cannot use the local dressing like \eqref{eq:exDressed} for a matter operator $\chi(x)$. Nonetheless, as long as $\varphi_0'(y)\neq 0$ for some other $y$ we can use a non-local dressing to $y$: we  simply need to include a `gravitational Wilson line' between $x$ and $y$. Explicitly, we can choose the linearised dressing operator to be $ O^{(1)} = \left(\int_y^x  \delta a -\frac{\delta \varphi(y)}{\varphi_0'(y)}\right) i\partial \chi(x)$, with the `Wilson line' term $\int_y^x  \delta a$ accounting for fluctuations in proper distance between the operator insertion point and the dressing.

However, the subtlety for evaporating black holes is that our construction above only directly applies for operators defined in a region which is held fixed, independent of $G$. But the Page time is parametrically large, of order $G^{-1}$, and  the size of the island region $\island$ (taken by the volume of a maximal-volume Cauchy surface, for example) will also scale as $G^{-1}$. This means that operators at sufficiently late times and the star which breaks the symmetry do not both lie in any fixed region as $G\to 0$, so showing validity of dressing such late-time operators to the star requires  extra work. This reflects a genuine potential obstruction: the dressing operator (e.g., $\int_y^x \delta a$ in the above paragraph with $x-y\sim G^{-1}$) could have large quantum fluctuations which smear the physical insertion point. In some cases, these fluctuations could potentially push the support of the dressed operator outside $\island$. In the context of an evaporating black hole, the relevant fluctuations are due to the statistical uncertainty in the evaporation rate, which leads to uncertainties in the geometry at time $t$ of order $\sqrt{t}$. For example, this means that an operator at time $t$ dressed to the star using the radial  proper distance along a maximum volume Cauchy surface will have fluctuations in its location of order $G t^\frac{3}{2}$, which is large (of order $G^{-1/2}$) when $t$ is comparable to the Page time. This means that a dressed operator defined in this way may fail to be supported only within $\island$ for operators this close to the edge of the island.

\subsubsection*{Dressing to slowly-varying backgrounds}

Even away from the star, the time-translation symmetry of an evaporating black hole is broken by the evaporation process itself causing the black hole to shrink. However, this breaking is parametrically small, of order $G$, which means we cannot immediately apply the ideas of the above section.

Nonetheless, since we have an approximate symmetry, we can still dress operators which are approximately invariant. This means operators which change the energy by a sufficiently small amount, or vary only over a sufficiently long time. We can immediately apply our above results for operators which change the energy by an amount of order $G$, or break the symmetry on timescales of order $G^{-1}$, since this pushes the relevant dressing terms to a higher order in the expansion. But we can do much better than this using the timescale $\Delta t$ of the operator as a perturbative parameter, getting a sensible perturbative expansion as long as $\frac{1}{\Delta t}$ is small compared to the Hawking temperature. Given this regime of validity of perturbation theory, we should expect that truly local operators cannot be dressed adequately, but we \emph{can} dress operators smeared over a sufficiently large region with $\Delta t$ of order $\beta=T^{-1}$.

Again, this limitation reflects a real physical effect. For some intuition, the sort of operator we would like to define in the black hole interior might change the polarisation of some Hawking partner (i.e., the mode entangled with some quantum of Hawking radiation). But these modes have wavelengths of order $\beta$, so the corresponding operator will be delocalised over precisely the scale identified by the breakdown of perturbation theory.

To make these ideas more explicit, we can consider the example of integral operators like \eqref{eq:integralop}, where one of the coordinate functions $X^\mu$ is slowly varying at a rate of order $G$ (for example, in an evaporating black hole we can choose a curvature scalar which gives the mass on a Schwarzschild background). If the smearing function $g$ has a fixed width of order unity, then the operator is defined independent of $G$ so our usual perturbation theory goes through unchanged, and we get an operator smeared over a distance of order $G^{-1}$. We can do better if we allow the width of $g$  to scale parametrically in the $G\to 0$ limit (or equivalently we scale $X^\mu$). This leads to enhanced terms in the perturbative expansion from derivatives of $g$ (or factors of $X^\mu$), but perturbation theory remains under control as long as the with of $g$ is much larger than $G$. This leads precisely to the expected minimum smearing width identified above.

\subsubsection*{Dressing to the quantum state?}

While the above constructions allow us to define many diffeomorphism invariant observables in the island of an evaporating black hole, there are examples where the background geometry in the island has an exact symmetry so none of these approaches can work. One such example is an eternal black hole in equilibrium with radiation, as studied in \cite{Almheiri:2019yqk}, which has an exact time-translation symmetry.

To define gauge-invariant operators on this background to which bulk reconstruction applies, we must somehow find a way to break the symmetry so that we can specify a time at which the operator acts. To do this, note that while the geometry has an exact symmetry, this is broken by the quantum state of the island and radiation together. Specifically, the pattern of entanglement between modes in the black hole interior and the radiation is not invariant under time translation acting in the interior alone. This raises the possibility of `dressing to the radiation'. Heuristically, we choose a quantum of Hawking radiation, identify the interior partner mode which purifies it (which is localised on the thermal scale $\beta=T^{-1}$), and insert the desired matter operator at the location of the partner.

Note that the resulting dressed operator is not localised only in the island, but acts on both the island and the radiation; however, since the entanglement wedge of the radiation always contains the radiation itself, this is not a problem for our purposes. The relevant gauge symmetry is time translation acting only in the interior (not on the radiation), since the radiation is either in a non-gravitating auxiliary system, or, if the radiation is in a region with gravity, it must be accompanied by some `clock' which breaks the time-translation symmetry.\footnote{The existence of a clock is not an additional assumption, since it is required to even define the radiation subsystem in the first place.}

Since the breaking of the symmetry is provided not by classical background fields but by a quantum state, this dressing goes beyond the discussion above. We will not attempt to make it more precise here, leaving the details of this construction as an interesting problem for the future.

\subsubsection*{Perturbing the background}

In any case with symmetries (exact or approximate), there is something much more straightforward we can do to construct operators which are supported  on the island, and act nontrivially in a code subspace of states close to the original background. The idea is that while the original background may lack any features to dress to, we can add such a feature which breaks the symmetry by hand. This means perturbing the state by an amount which is sufficiently weak to give a small change in the background (staying within the space of states defined perturbatively around the original spacetime), but nonetheless sufficiently strong to break the symmetry enough so that gauge-invariance operators can be defined in relation to the deformation.

For example, we can consider throwing a `lab' into the black hole from the outside (so that it passes well inside the island), and then define an operator which corresponds to some measurement made in the lab. If the mass of the lab is small compared to the mass of the black hole $ m_\mathrm{lab} \ll m_\mathrm{BH}$, then the backreaction is small so this can still be regarded as occurring in the original code subspace of states. In particular, unless the original background was right at a phase transition (i.e., another quantum extremal surface has equal generalised entropy at order $G^{-1}$) or the lab is close to the edge of the island so that its backreaction can create new relevant extremal surfaces, the existence and location of the island is stable to such perturbations. And if  the typical energy of the measurement operator $O$ is small compared to the mass of the lab $E_O\ll m_\mathrm{lab}$, then we can use the location of the lab to dress the operator $O$.

For a simple specific example of this, we can take the `lab' to be a particle which is charged under some approximate global symmetry (perhaps broken by perturbative effects of order $G$, or even non-perturbative effects). Then we take the operator $O$ to act as an internal symmetry generator on the particle.

The existence of the compactly supported operators described in this section demonstrates that the statements of entanglement wedge reconstruction always have real content, and are not vacuous due to a dearth of operators supported in the island.

\subsection{Dressing in massive gravity}\label{ssec:dressingmassive}

We conclude by commenting on gravitational dressing in theories of massive gravity like those discussed in Section \ref{sec:massivegravitons}. In particular, if we are working around a background with an approximate or exact symmetry, does the `graviton mass' in these examples help to make the construction of dressed operators any easier?

First, we can consider a linearised Fierz-Pauli massive graviton on a static background such as the AdS vacuum. From one perspective, there is no problem with dressing because we no longer gauge diffeomorphism invariance: there is still a constraint from the $h_{00}$ equation of motion, but it is second class so does not arise from a gauge symmetry. Alternatively, we can restore gauge-invariance at the cost of introducing St\"uckelberg fields (after which the constraints become first class once again). In that case we must choose a classical background for the St\"uckelberg, and there is no choice which leaves the background invariant under any gauge symmetry, hence we never have an obstruction to dressing. We can always solve the constraints by turning on the momentum conjugate to the St\"uckelberg. More directly in terms of operators, we can always make an operator gauge-invariant by adding compensating functions of the St\"uckelberg. To illustrate with the massive vector analogue, the St\"uckelberg field $\theta$ transforms to $\theta + \lambda$ under a  $U(1)$ gauge transformation $\lambda$, so a charge $q$ operator $O$ can be dressed to become the gauge-invariant $e^{-iq\theta}O$.

However, this is not the situation of interest, where the mass is induced for a  conventional diff-invariant theory by choice of boundary conditions. In such a case, there is some dynamical field (perhaps composite) which plays the role of the St\"uckelberg above: for example, in the abelian gauge theory analogue $\theta$ could be the phase of a complex scalar Higgs field. For gravity, the St\"uckelberg vector field might be constructed from the gradients of some scalar fields with varying background values. But these are not always well-defined --- for example, the phase of the Higgs does not make sense as the symmetry-invariant point where the field vanishes --- so dressed operators like $e^{-iq\theta}O$ are not well-defined. Nonetheless, if we were working around a classical background which explicitly broke all symmetries (e.g., a non-zero background value for the Higgs in the $U(1)$ example), then these fields (and hence the dressing) are good to all orders in perturbation theory.

In contrast, simply coupling the boundary of AdS to an auxiliary system leading to a graviton/photon mass induced by one-loop effects, does not change the classical background so it is not sufficient to guarantee the existence of gauge-invariant, compactly supported operators by the arguments we have given. The comparison to the Pauli-Fierz theory is not justified, and there is no obvious symmetry breaking of the background which allows us to solve the constraints perturbatively in some parameter.  In other words, it is far from clear that adding this sort of graviton mass even helps with dressing as argued in \cite{Geng:2021hlu}.

Having said this, dressing could be easier in examples where we couple many fields to an auxiliary system, and we dress perturbatively in the inverse species number. This is strongly suggested by the `doubly holographic' perspective in Karch-Randall models, where we can dress through the higher-dimensional bulk (though it should be noted that the resulting dressed operator is supported not just in the island, but also in the auxiliary system). In fact, this picture can be interpreted as a geometrisation of the `dressing to the quantum state' discussed in the previous subsection, where it is the pattern of entanglement between radiation and interior partners that breaks the isometry of the background. But we emphasise that it is the symmetry-breaking pattern of the quantum state rather than the graviton mass \emph{per se} which is required for such dressing to be possible.

\section{A microscopic construction of compactly supported operators}
\label{sec:BBPSVops}

The results of Section \ref{sec:linearconstraints} show that it is possible to build gauge-invariant operators entirely localised in the island to all orders in perturbation theory, so long as the background spacetime has no symmetries. In this section we will explicitly build a class of operators which commute by construction with the ADM Hamiltonian (and can be generalised to commute with other asymptotic charges) at all orders in perturbation theory. Their advantage is in their complete generality and in that they are built to act at leading order like the corresponding local operators.
This is possible because the commutators between operators dressed to asymptotic features and the asymptotic charges are suppressed in $G$ (or $1/N$ in a holographic dual microscopic theory) with respect to the leading order: a modification of the operators at subleading orders is sufficient to make their commutator with asymptotic charges vanish, while leaving their leading order action unchanged and equal to that of the original undressed local operators. These operators are based on a construction first introduced by the authors of \cite{Bahiru:2022oas,Bahiru:2023zlc} (BBPSV) (see also \cite{Papadodimas:2015jra,Papadodimas:2015xma} for an earlier construction based on similar ideas), which builds operators that a) commute with the Hamiltonian at all orders in $G$ and b) have the same one-point functions as the corresponding undressed operators at leading order. 

We will first review the construction of \cite{Bahiru:2022oas,Bahiru:2023zlc} and emphasise a significant limitation of it, namely that it only correctly reproduces one- and not higher-point correlators. We will then refine the BBPSV construction\footnote{See also \cite{Jensen:2024dnl} for related recent work generalising \cite{Bahiru:2022oas,Bahiru:2023zlc}.} to obtain operators that act at leading order like the undressed operators at the level of the state (and therefore reproduce all correlation functions with an $O(G^0)$ number of operator insertions). We will also comment on the generalisation of this construction to operators commuting with all asymptotic charges and other spacelike separated, boundary-dressed operators.

\subsection{The BBPSV construction}
\label{sec:bbpsvreview}

For simplicity, let us focus our attention on a generic AdS/CFT setting. Let us consider a holographic state $\ket{\psi_0}$ corresponding to a spacetime geometry which breaks all symmetries at the classical level (e.g. the geometry evolves in time, breaking time-translation symmetry). A state with geometry different from the AdS vacuum has energy of order $N^2$, with $N^2\sim 1/G$ the number of boundary degrees of freedom:
\begin{equation}
    \bra{\psi_0}H\ket{\psi_0}\equiv E_0=O(N^2),
    \label{eq:avgen}
\end{equation}
where $H$ is the Hamiltonian of the boundary CFT (or, from a bulk perspective, the ADM Hamiltonian). The time dependence of the geometry also guarantees that the variance of the energy scales with $N^2$ \cite{Bahiru:2022oas,Bahiru:2023zlc}
\begin{equation}
    \bra{\psi_0}(H-\langle H\rangle)^2\ket{\psi_0}\equiv \sigma^2=O(N^2).
    \label{eq:variance}
\end{equation}
For this class of states, the return amplitude (i.e., expectation value of the time evolution operator) is expected to be decaying exponentially in $N^2$ at early times $T=O(N^0)$:
\begin{equation}
    \bra{\psi_0}e^{iHT}\ket{\psi_0}=e^{-N^2f(T)+iTE_0}
    \label{eq:return}
\end{equation}
where $f(T) = O(T^2)$ for $T \lesssim 1$.\footnote{At later times the behavior is more complicated and it is generally not exponentially decaying \cite{Bahiru:2022oas,Bahiru:2023zlc,Saad:2019lba,Stanford:2022fdt,Iliesiu:2024cnh}. At very late (doubly exponential) times, quantum recurrence is expected to occur, with the return probability going to one. We will only be interested in the behaviour for times of order $N^0$.} For instance, the properties $\eqref{eq:avgen}-\eqref{eq:return}$ are satisfied for the thermofield double state at high temperature when $H=H_L$ or $H=H_R$.\footnote{Notice that the difference $H_R-H_L$ does not satisfy these properties because of the symmetries of the thermofield double state. This also implies that operators cannot be made commute with both $H_R$ and $H_L$ at the same time.} This is a mathematical formulation of the intuitive statement that quantum gravity states associated with two different classical geometries (in this case, the spatial geometries of the $t=0$ and $t=T$ time slices, which are different by assumption) are orthogonal to each other up to non-perturbative corrections. 

Expanding $\ket{\psi_0}$ in the energy eigenbasis $\ket{\psi_0}=\frac{1}{\sqrt{\zeta}}\sum_nc_n\ket{n}$, we can write
\begin{equation}
    \bra{\psi_0}e^{\pm iHT}\ket{\psi_0}=\frac{1}{\zeta}\sum_n|c_n|^2e^{\pm iTE_n}\to \frac{1}{\zeta}\int dE p(E)e^{\pm iTE}\approx e^{-\alpha N^2 T^2\pm iTE_0},
    \label{eq:return2}
\end{equation}
where $p(E) = e^{S(E)} |c_n|^2$ is the probability that the state $\ket{\psi_0}$ has energy $E$. The distribution $p(E)$ is therefore determined by taking an inverse Laplace transform of \eqref{eq:return}. In the large $N$ limit, $p(E)$ can also be determined by a saddle point approximation. In particular, near the peak of the distribution
\begin{equation}
    p(E)\approx \frac{1}{2 \sqrt{\alpha\pi} N} e^{-\frac{(E-E_0)^2}{4\alpha N^2}}
    \label{eq:saddlepoint}
\end{equation}
is approximately Gaussian. This approximation is valid for $E - E_0 \sim N$. 

Following \cite{Bahiru:2022oas,Bahiru:2023zlc}, let us consider the code subspace $\mathcal{H}_0$ at boundary time $t=0$ generated by acting on $\ket{\psi_0}$ with a small number $k=O(N^0)$ of CFT single trace operators of small dimension:
\begin{equation}
    \mathcal{H}_0=\textrm{span}\{\ket{\psi_0},\mathcal{O}_1(t_1,x_1)\ket{\psi_0},...,\mathcal{O}_1(t_1,x_1)...\mathcal{O}_k(t_k,x_k)\ket{\psi_0}\},
    \label{eq:codesub}
\end{equation}
where we labeled by $x$ the CFT spatial coordinates. Note that the code subspace contains states obtained by acting with operators on $\ket{\psi_0}$ at different times. In the Schr\"odinger picture, this corresponds to evolving $\ket{\psi_0}$ to a different time slice, acting with the operator, and then evolving it back to the original time slice. Since the code subspace on a given time slice should be connected by time evolution to the code subspace on a different time slice, this choice is completely natural. This property will prove important later on to show that commutators between locally gauge invariant and boundary-dressed operators vanish.

These operators do not backreact at leading order on the geometry, and the different states in the code subspace should be thought of as different states for the quantum fields on top of the same background geometry. Quantitatively, the energy difference between any two states in the code subspace is $E_i-E_j=O(N^0)$. The energy variance \eqref{eq:variance} for different states in the code subspace is the same up to subleading corrections of order $N^0$. The return probability for a state $\ket{\psi_i}$ is therefore given by equation \eqref{eq:return} with $E_0\to E_i$. Let us call $P_0$ the projector on this code subspace. The BBPSV operators acting on the code subspace $\mathcal{H}_0$ at time $t=0$ are then defined as
\begin{equation}
    \hat{\phi}=c\int_{-t_*}^{t_*}dTe^{-iHT}P_0\phi P_0e^{iHT}
    \label{eq:BBPSVop}
\end{equation}
with
\begin{equation}
    c^{-1}=\int_{-t_*}^{t_*}dT\bra{\psi_0}e^{-iHT}P_0e^{iHT}\ket{\psi_0},
     \label{eq:BBPSVnorm}
\end{equation}
where $t_*=O(N^0)$ ($t_*$ must be within the range of times for which equation \eqref{eq:return} holds) and $\phi$ is a bulk operator dressed to the asymptotic boundary, which does not commute with $H$.

Let us gain some intuition about the operator \eqref{eq:BBPSVop}. The projector $P_0$ plays a crucial role in specifying the point (in this case the time, $t=0$) where to act with the operator $\phi$, i.e. in dressing the operator. To see this, let us act with the operator $\hat{\phi}$ on a time-evolved state $e^{-iH\tau}\ket{\psi_j}$, with $\ket{\psi_j}\in \mathcal{H}_0$ and $\tau<t_*$. Using equation \eqref{eq:return}, $P_0e^{iH(T-\tau)}\ket{\psi_j}\approx e^{-\alpha N^2 (T-\tau)^2+i(T-\tau)E_i}\ket{\psi_j}$\footnote{In this and all the following calculations in this section, we use $P_0e^{\pm iH(T-\tau)}\ket{\psi_j}\approx e^{-\alpha N^2 (T-\tau)^2\pm i(T-\tau)E_j}\ket{\psi_j}$ for generic $\tau$, which is valid only for $T-\tau=O(1/N)$. For larger $T-\tau$, states $\ket{\psi_i}$ with $i\neq j$ can contribute to the projector $P_0$ at leading order \cite{Bahiru:2023zlc}. When dealing with operators of the form \eqref{eq:BBPSVop}, this approximation is well-justified at leading order because the integral over $T$ is dominated by values of $T$ such that $T-\tau=O(1/N)$ due to the behavior of the return amplitude \eqref{eq:return}.} and the integral is dominated by values of $T$ such that $T=\tau+O(1/N)$. In other words, the operator \eqref{eq:BBPSVop} roughly acts as $\hat{\phi}\sim e^{-iH\tau}\phi e^{iH\tau}$: it evolves the state back to the time slice $t=0$ where we want to act with $\phi$, it acts with $\phi$, then it evolves the state forward again. In doing so, it does not refer to the asymptotic boundary, but rather it exploits the features of the state itself, namely its breaking of time translation symmetry captured by the return probability \eqref{eq:return}.

The operator \eqref{eq:BBPSVop} is gauge-invariant by construction ($\phi$ is, and the definition of $\hat{\phi}$ only involves $\phi$ and other gauge-invariant quantities). Moreover, it is straightforward to check that its commutator with the Hamiltonian vanishes up to non-perturbative corrections on any state in the code subspace or obtained by evolving a state in $\mathcal{H}_0$ for a time $\tau$ such that $|\tau|<t_*$ with $|\tau|-t_*=O(1)$ \cite{Bahiru:2022oas,Bahiru:2023zlc}:
\begin{equation}
\begin{aligned}
    &[\hat{\phi},H]e^{-iH\tau}\ket{\psi_i}=\left.\left(i\frac{d}{ds}e^{isH}\hat{\phi}e^{-isH}\right)\right|_{s=0}e^{-iH\tau}\ket{\psi_i}\\&=-ic\left[P_{t_*}\phi(t_*) P_{t_*}-P_{-t_*}\phi(-t_*) P_{-t_*}\right]e^{-iH\tau}\ket{\psi_i}
    =e^{-O(N^2)} \quad \quad \forall |\tau|<t_*\wedge |\tau|-t_*=O(1),
    \end{aligned}
\end{equation}
where we defined $\phi(t)=e^{-iHt}\phi e^{iHt}$ and the projector $P_{t}=e^{-iHt}P_0e^{iHt}$ on the code subspace at time $t$, and we used the definition \eqref{eq:BBPSVop} and the assumption \eqref{eq:return}.

Let us now turn to the issue of whether $\hat{\phi}$ acts at leading order like $\phi$---and therefore like the undressed local operator, since the dressing in the operator $\phi$ modifies its action only at subleading order. BBPSV showed that this is the case at the level of the one-point function \cite{Bahiru:2022oas,Bahiru:2023zlc},
\begin{equation}
    \bra{\psi_i}\hat{\phi}\ket{\psi_j}=\bra{\psi_i}\phi\ket{\psi_j}+O\left(\frac{1}{N}\right)
    \label{eq:1pf}
\end{equation}
for any $\ket{\psi_i},\ket{\psi_j}\in \mathcal{H}_0$. However, this property is not enough to guarantee that $\hat{\phi}$ acts like $\phi$ at leading order. 
In order to show this, let us consider $\hat{I}$ obtained by applying the construction \eqref{eq:BBPSVop} to the identity operator $I$ and let us work again in the saddle point approximation \eqref{eq:saddlepoint}. The state obtained by acting with $\hat{I}$ on a state in the code subspace is
\begin{equation}
    \hat{I}\ket{\psi_j}\approx \frac{\sqrt{2}}{\sqrt{\zeta}}\sum_n c_n^{j}e^{-\frac{(E_n-E_j)^2}{4\alpha N^2}}\ket{n}=\sqrt{2}e^{-\frac{(H-E_j)^2}{4\alpha N^2}}\ket{\psi_j},
    \label{eq:Ihataction}
\end{equation}
where $c=\sqrt{2\alpha N^2/\pi}$ can be computed from equation \eqref{eq:BBPSVnorm}, and we extended the integral \eqref{eq:BBPSVop} from $-\infty$ to $\infty$ because the contributions away from the $[-t_*,t_*]$ window are non-perturbatively suppressed. In this and the following expressions the $\approx$ means that we are only keeping track of the leading order term in $1/N$.
Note that $\hat{I}$ does not leave the state unchanged as one would expect from the identity, but rather it changes the variance of the energy distribution at leading order (i.e. at order $N^2$). Since, by construction, all states in the code subspace have the same energy variance up to subleading corrections, the resulting state is not in the code subspace. Nonetheless, by virtue of the choice of normalisation constant $c$, the property \eqref{eq:1pf} is still satisfied and we have $\bra{\psi_i}\hat{I}\ket{\psi_j}=\bra{\psi_i}I\ket{\psi_j}+O(1/N)$ \cite{Bahiru:2022oas,Bahiru:2023zlc}, which can also be checked explicitly in the saddle point approximation. 

However, the state $\hat{I}\ket{\psi_i}$ being outside the code subspace implies that higher point correlators of $\hat{I}$ differ from those of $I$. For instance, using the saddle point approximation \eqref{eq:saddlepoint} it is easy to show that
\begin{equation}
    \bra{\psi_j}\hat{I}^2\ket{\psi_j}\approx\frac{2}{\sqrt{4\pi\alpha N^2}}\int dE e^{-\frac{3(E-E_j)^2}{4\alpha N^2}}=\frac{2}{\sqrt{3}},
    \label{eq:idsquare}
\end{equation}
which clearly differs from $\bra{\psi_j}I^2\ket{\psi_j}=1$. The same is true for any higher point correlation function of $\hat{I}$. It is easy to show that $[H,\hat{I}^2]=[H^2,\hat{I}^2]=0$ still holds on the code subspace, but we violated the requirement that the action of the BBPSV operators is equal to the action of the boundary-dressed operators. In particular, the identity should be unitary, while $\hat{I}$ clearly is not. Its action can be detected by an asymptotic observer because $\hat{I}$ modifies the variance of the energy distribution and so \begin{align}
\bra{\psi_j}\hat{I}H^2\hat{I}\ket{\psi_j}=\bra{\psi_j}\hat{I}^2H^2\ket{\psi_j}\neq \bra{\psi_j}H^2\ket{\psi_j}\,.
\end{align}

It is easy to show that the same problem is inherited by any $\hat{\phi}$, whose action is given in general by
\begin{equation}
    \hat{\phi}\ket{\psi_j}\approx \sqrt{2}e^{-\frac{(H-E_j)^2}{4\alpha N^2}}\phi \ket{\psi_j},
    \label{eq:phihataction}
\end{equation}
again implying $\bra{\psi_i}\hat{\phi}...\hat{\phi}\ket{\psi_j}\neq \bra{\psi_i}\phi...\phi\ket{\psi_j}$. Notice that this also implies that a unitary $\hat{U}_{\hat{\phi}}$ built out of the $\hat{\phi}$ BBPSV operator acts different from the corresponding boundary-dressed unitary $U_\phi$ at leading order, even at the level of the one-point function. This feature is clearly undesirable, because we are ultimately interested in building localised gauge invariant unitary bulk operators.

\subsection{Refined BBPSV operators}
\label{sec:bbpsvrefinement}

A simple modification of the BBPSV operators \eqref{eq:BBPSVop} yields operators that commute with the asymptotic charges at all orders in perturbation theory \textit{and} act on the code subspace at leading order like their boundary-dressed counterparts (and therefore the local, undressed operators). Restricting again our analysis of boundary charges to the Hamiltonian, this is given by
\begin{equation}
    \tilde{\phi}=\tilde{c}\int_{-t_*}^{t_*}dTf(H)e^{-iHT}P_0\phi P_0e^{iHT}f(H)
    \label{eq:ourop}
\end{equation}
with
\begin{equation}
    \tilde{c}^{-1}=\int_{-t_*}^{t_*}dT\bra{\psi_0}f(H)e^{-iHT}P_0e^{iHT}f(H)\ket{\psi_0}
     \label{eq:ournorm}
\end{equation}
and $f(H)$ a function of the Hamiltonian. The simplest choice we can make to obtain operators $\tilde{\phi}$ that act at leading order like $\phi$ is
\begin{equation}
    f(H)=e^{\frac{(H-\bar{E})^2}{8\alpha N^2}}
    \label{eq:fofh}
\end{equation}
where $\bar{E}=\frac{1}{d}\sum_i \bra{\psi_i}H\ket{\psi_i}$ (with $i$ running over a basis of the code subspace and $d$ the dimension of the code subspace) is the average energy of the code subspace.\footnote{Notice that we could replace $\bar{E}$ with the energy $E_i$ of any state $\ket{\psi_i}$ in the code subspace (or in general with some $E$ such that $E-E_i=O(1)$, i.e. with a value of $E$ within the energy window of the code subspace) without changing our result.}

With the choice \eqref{eq:fofh}, a calculation similar to the one carried out in the previous subsection shows that
\begin{equation}
    \tilde{\phi}\ket{\psi_j}=\phi\ket{\psi_j}+O\left(\frac{1}{N}\right),
    \label{eq:actionourop}
\end{equation}
which is the property we were looking for. Notice that the introduction of $f(H)$ in the definition \eqref{eq:ourop} clearly leaves the commutator with the Hamiltonian $H$ unchanged with respect to the BBPSV operators \eqref{eq:BBPSVop}, and therefore $[\tilde{\phi},H]=0$ on the code subspace up to non-perturbative corrections.
We have therefore obtained operators that commute at all orders with the boundary Hamiltonian and act at leading order like the boundary-dressed operators, and therefore like the corresponding local, undressed operators. As a consequence, higher-point correlators of $\tilde{\phi}$ and the action of a unitary $\tilde{U}_{\tilde{\phi}}$ are also equal at leading order to higher-point correlators of $\phi$ and the action of a unitary $U_\phi$, respectively.  

A cleaner way to obtain the same result is to replace $f(H)$ with $\hat{I}^{-\frac{1}{2}}$, where $\hat{I}^{-1}$ is the generalised inverse of the operator obtained by applying the original BBPSV construction \eqref{eq:BBPSVop} to the identity.\footnote{Note that $f(H)$ defined in equation \eqref{eq:fofh} is roughly, but not exactly, $\hat{I}^{-\frac{1}{2}}$ at leading order. The difference is in the explicit presence of $\bar{E}$, which for $\hat{I}^{-\frac{1}{2}}$ would be replaced by the energy of the specific state $\hat{I}^{-\frac{1}{2}}$ is acting on, as it is clear from equation \eqref{eq:Ihataction}.} Since, as we have seen in equations \eqref{eq:Ihataction} and \eqref{eq:phihataction}, the deviation of the action of the BBPSV operators from that of the boundary-dressed operators is independent of the choice of the specific operator $\phi$, the operator 
\begin{equation}
    \bar{\phi}=\hat{I}^{-\frac{1}{2}}\hat{\phi}\hat{I}^{-\frac{1}{2}}
    \label{eq:I-12ops}
\end{equation}
acts at leading order like $\phi$ on the code subspace, similar to the operator \eqref{eq:ourop}. Naturally, because both $\hat{I}$ and $\hat{\phi}$ commute at all orders in perturbation theory with the boundary Hamiltonian, $\bar{\phi}$ does as well. The advantages of the definition \eqref{eq:I-12ops} are that it does not explicitly involve the energy of the code subspace, and it commutes with spacelike-separated operators dressed to the asymptotic boundary, as we will see shortly. The drawback is that $\hat{I}^{-\frac{1}{2}}$ is defined implicitly as a generalised inverse, but its explicit form in generic cases is hard to obtain.

Provided that the background spacetime breaks all symmetries and not only time translation invariance, so that equation \eqref{eq:return} has an equivalent for any unitary $U(g)$ implementing symmetry transformations
\begin{equation}
    |\bra{\psi_0}U(g)\ket{\psi_0}|=e^{-N^2f(g)},
\end{equation}
our result can be generalised to all asymptotic charges \cite{Bahiru:2023zlc}. This can be achieved by replacing the integral in $T$ with an integral over the asymptotic symmetry group ($SO(2,d)$ in our AdS case) with an appropriate measure, and $e^{\pm iTH}$ with $U(g)$. To build the corresponding refined BBPSV operators, we can then either replace $f(H)$ with an appropriate function of the symmetry generators, or simply use the implicit definition \eqref{eq:I-12ops}, with $\hat{\phi}$ now commuting with all asymptotic charges. We leave the details of this generalisation to future work.

\subsubsection*{Commutator with other operators outside the island}

The operators \eqref{eq:ourop} commute with the asymptotic charges because they are not dressed to the asymptotic boundary, but to features of the state (in the case of the Hamiltonian, the time-dependence of the state). As a practical example, let us consider the evaporation of a single-sided black hole formed from collapse. In this setup, time translation symmetry of the background is classically broken due to the backreaction of the collapsing shell, allowing our operators $\tilde{\phi}$ (or $\bar{\phi}$) to be defined.
In particular, consider an operator $\tilde{\phi}(x)$ (or $\bar{\phi}(x)$) acting in the interior at late times, such that $x$ is inside the entanglement island. The action of this operator, including its dressing, can be entirely specified without referring to the asymptotic boundary, as we have seen. But does this operator commute with boundary-dressed operators outside the island?

The intuition that the operator $\tilde\phi$ is dressed to the background spacetime suggests this should be the case. To make this argument more precise, let us consider two boundary-dressed operators $\phi_{\rm in}$ and $\phi_{\rm out}$ acting inside and outside the island, respectively. Clearly, neither of these operators commute with the Hamiltonian $H$. Let us assume, however, that these two operators commute with each other (on the code subspace) at all orders in perturbation theory, and moreover that $[\phi_{\rm in},\phi_{\rm out}(t)]=0$ for $t\in [-t_*,t_*]$, where $\phi_{out}(t)=e^{-iHt}\phi_{out}e^{iHt}$. This is reasonable if $\phi_{out}(t)$ acts only in the exterior (and hence is spacelike separated from $\phi_{\rm in}$) and the dressing from $\phi_{\rm in}$ to the boundary was chosen to commute with everything except asymptotic charges such as $H$.\footnote{We expect that this can be achieved order-by-order in perturbation theory (by modifying the operator at subleading orders if necessary).}

Now consider the BBPSV operator $\hat{\phi}_{in}$ built according to equation \eqref{eq:BBPSVop}. One could worry that the modification of $\phi_{in}$ at subleading orders, which allows $\hat{\phi}_{in}$ to commute with $H$, might introduce a non-vanishing commutator between $\hat{\phi}_{in}$ and $\phi_{out}$. This turns out to not be the case. In fact, from the definition \eqref{eq:BBPSVop} we obtain
\begin{equation}
\begin{aligned}
    \hat{\phi}_{in}\phi_{out}&=c\int_{-t_*}^{t_*}dTe^{-iHT}P_0\phi_{in} P_0e^{iHT}\phi_{out}=c\int_{-t_*}^{t_*}dTe^{-iHT}P_0\phi_{in} P_0\phi_{out}(T)e^{iHT}\\
    &=c\int_{-t_*}^{t_*}dTe^{-iHT}\phi_{out}(T)P_0\phi_{in} P_0e^{iHT}=\phi_{out}\hat{\phi}_{in},
    \end{aligned}
\end{equation}
where we are implicitly assuming the operators are acting on the code subspace. In the third equality we used $[P_0,\phi_{out}(T)]=0$ on the code subspace---which follows from the definition of the code subspace \eqref{eq:codesub}, in particular the fact that the action of $\phi_{out}(T)$ maps states in the code subspace to states in the code subspace---and the assumption $[\phi_{in},\phi_{out}(t)]=0$.

Does this property survive after our refinement of the BBPSV operators? If we choose the simplest refinement \eqref{eq:ourop}, the answer is no, because $[f(H),\phi_{out}]\neq 0$. However, $\phi_{out}$ does commute with the more general refinement \eqref{eq:I-12ops}, because, as we have just shown, it commutes with $\hat{I}$ and therefore with its generalised inverse.

To summarise, we obtained operators $\bar{\phi}$, given by \eqref{eq:I-12ops}, which can be localised in the island, act at leading order like the corresponding local operators, and commute at all orders in perturbation theory with the asymptotic charges and with other boundary-dressed operators in the complementary entanglement wedge. The action of a unitary $U_{\bar{\phi}}$ built out of $\bar{\phi}$ is then undetectable by a boundary observer which cannot detect non-perturbative effects. Thus, these operators entirely localised in the island (at the perturbative level) are reconstructible from the radiation encoding it.

\section*{Acknowledgements}

 We would like to thank the authors of \cite{Geng:2021hlu} (Hao Geng, Andreas Karch,  Carlos Perez-Pardavila, Suvrat Raju, Lisa Randall, Marcos Riojas, Sanjit Shashi) for many constructive discussions and comments on this draft, as well as Alex Belin, Daniel Ranard, Don Marolf and Kyriakos Papadodimas for valuable discussions. SA is supported by the U.S.~
Department of Energy through DE-FOA-0002563. C-HC is supported in part by the Department of Energy through DE-FOA-
0002563 and by AFOSR award FA9550-22-1-0098. HM is supported by DOE grant DE-SC0021085 and a Bloch fellowship from Q-FARM. GP is supported by the Department of Energy through QuantISED Award DE-SC0019380 and an Early Career Award DE-FOA-0002563, by AFOSR award FA9550-22-1-0098 and by a Sloan Fellowship. This research was supported in part by grant NSF PHY-2309135 to the Kavli Institute for Theoretical Physics (KITP).

\appendix

\section{Operators localised in the island for Kourkoulou-Maldacena states}
\label{app:KM}

\begin{figure}
\centering
    \begin{subfigure}{0.49\textwidth}
    \centering
        \includegraphics[height=0.7\textwidth]{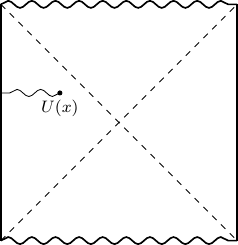}
        \caption{}
    \end{subfigure}
    \begin{subfigure}{0.49\textwidth}
    \centering
        \includegraphics[height=0.7\textwidth]{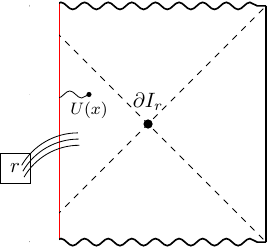}
        \caption{}
    \end{subfigure}
        \caption{(a) Unitary operator $U(x)$ acting in the left wedge and dressed to the left asymptotic boundary. The operator can be reconstructed from the left CFT by a unitary $U_L$. (b) After the isometry \eqref{eq:isometry} from the left CFT to an external reservoir $r$ is implemented, the unitary operator $U(x)$ is dressed to the end-of-the-world brane (in red) cutting off the left asymptotic boundary. What remains of the left wedge is now encoded in the reservoir $r$, and $U(x)$ can be reconstructed by a unitary $U_r$ acting on $r$. The black dot labeled by $\partial I_r$ represents the quantum extremal surface for the entire right boundary (or, equivalently, the reservoir $r$), i.e. the boundary of the island $I_r$.}
    \label{fig:KMop}
\end{figure}

Let us discuss a simple, intuitive example in which operators entirely localised in an entanglement island should exist and we are able to reconstruct them from an external reservoir encoding the island. Consider a double-sided AdS black hole spacetime and a unitary $U(x)$ acting in the left asymptotic region and dressed to the left asymptotic boundary by a gravitational Wilson line, see Figure \ref{fig:KMop} (a). 
This unitary can be reconstructed by a unitary acting on the left CFT:
\begin{equation}
    VU(x)V^{-1}\ket{TFD}_{LR}=U_L\ket{TFD}_{LR},
    \label{eq:leftrec}
\end{equation}
where $V$ is the bulk-to-boundary map and $\ket{TFD}_{LR}$ is the thermofield double (TFD) state of the two boundary CFTs at inverse temperature $\beta$.

Now consider an isometry $\mathcal{O}$ mapping the subspace of the Hilbert space of the left CFT associated with the bulk Hilbert space of a one-sided black hole at inverse temperature $\beta$, to an external reservoir $r$. We can write this isometry as
\begin{equation}
    \mathcal{O}=\frac{1}{d}\sum_i\ket{i}_r\bra{\psi_i}_L,
    \label{eq:isometry}
\end{equation}
where $d=e^{S_{BH}}$ is the dimension of the black hole Hilbert space and $\ket{\psi_i}_L$ is a complete orthonormal basis of black hole microstates. Let us take this basis to be formed by Kourkoulou-Maldacena geometric microstates \cite{Kourkoulou:2017zaj}, where the second asymptotic region is cut off by an end-of-the-world brane carrying a flavor $i$, see Figure \ref{fig:KMop} (b). Applying the isometry \eqref{eq:isometry} to the TFD state we obtain the state
\begin{equation}
    \mathcal{O}\ket{TFD}_{LR}=\sum_ic_i\ket{i}_r\ket{\psi_i}_R
    \label{eq:westcoast}
\end{equation}
for some appropriate coefficients $c_i$.\footnote{The state \eqref{eq:westcoast} is analogous to the state considered in the West Coast model for black hole evaporation \cite{Penington:2019kki}.} Clearly the isometry \eqref{eq:isometry} leaves the thermal reduced density matrix of the right CFT unchanged,\footnote{We are now representing it as a mixture over a complete Kourkoulou-Maldacena basis.} and therefore the entanglement wedge of the right CFT remains restricted to the right wedge. The right CFT is now purified by the reservoir $r$, which has an island $I_r$ given by the left asymptotic region cut off by the end-of-the-world brane, see Figure \ref{fig:KMop} (b). The entanglement allowing the island to exist is the one between degrees of freedom of the brane (labeled by the flavor $i$) and degrees of freedom in the reservoir $r$.

Let us now focus on the unitary operator $U(x)$. Before implementing the isometry, this operator was entirely localised in the left wedge and reconstructible exclusively from the left CFT at all orders in perturbation theory, as we pointed out in equation \eqref{eq:leftrec}. Implementing the isometry \eqref{eq:isometry} we then obtain
\begin{equation}
    \mathcal{O}VU(x)V^{-1}\ket{TFD}_{LR}=\frac{1}{d}\sum_i\leftindex_L{\bra{\psi_i}}U_L\ket{TFD}_{LR}\otimes  \ket{i}_r=\frac{1}{d}\sum_i\leftindex_L{\bra{\psi_i}}\ket{TFD}_{LR}\otimes U_r\ket{i}_r,
\end{equation}
where the existence of $U_r$ (which reconstructs $U(x)$ from the reservoir) is guaranteed by the fact that $\mathcal{O}$ is an isometry.
The bulk interpretation of this result is that the operator $U(x)$, which was initially dressed to the left boundary, is now dressed to the end-of-the-world brane and therefore still entirely localised inside the left wedge, which is now an island for the reservoir, see Figure \ref{fig:KMop} (b). This construction provides a simple example in which operators entirely localised inside an entanglement island exist in a massless gravity setup.

\bibliographystyle{jhep}
	\bibliography{references.bib}

\providecommand{\href}[2]{#2}\begingroup\raggedright\begin{thebibliography}{10}

\bibitem{Hawking:1975vcx}
S.~W. Hawking, {\it {Particle Creation by Black Holes}},  {\em Commun. Math. Phys.} {\bf 43} (1975) 199--220. [Erratum: Commun.Math.Phys. 46, 206 (1976)].

\bibitem{Almheiri:2020cfm}
A.~Almheiri, T.~Hartman, J.~Maldacena, E.~Shaghoulian, and A.~Tajdini, {\it {The entropy of Hawking radiation}},  {\em Rev. Mod. Phys.} {\bf 93} (2021), no.~3 035002, [\href{http://arxiv.org/abs/2006.06872}{{\tt arXiv:2006.06872}}].

\bibitem{Bekenstein:1972tm}
J.~D. Bekenstein, {\it {Black holes and the second law}},  {\em Lett. Nuovo Cim.} {\bf 4} (1972) 737--740.

\bibitem{Bekenstein:1973ur}
J.~D. Bekenstein, {\it {Black holes and entropy}},  {\em Phys. Rev. D} {\bf 7} (1973) 2333--2346.

\bibitem{Hawking:1974rv}
S.~W. Hawking, {\it {Black hole explosions}},  {\em Nature} {\bf 248} (1974) 30--31.

\bibitem{Penington:2019npb}
G.~Penington, {\it {Entanglement Wedge Reconstruction and the Information Paradox}},  {\em JHEP} {\bf 09} (2020) 002, [\href{http://arxiv.org/abs/1905.08255}{{\tt arXiv:1905.08255}}].

\bibitem{Almheiri:2019psf}
A.~Almheiri, N.~Engelhardt, D.~Marolf, and H.~Maxfield, {\it {The entropy of bulk quantum fields and the entanglement wedge of an evaporating black hole}},  {\em JHEP} {\bf 12} (2019) 063, [\href{http://arxiv.org/abs/1905.08762}{{\tt arXiv:1905.08762}}].

\bibitem{Penington:2019kki}
G.~Penington, S.~H. Shenker, D.~Stanford, and Z.~Yang, {\it {Replica wormholes and the black hole interior}},  {\em JHEP} {\bf 03} (2022) 205, [\href{http://arxiv.org/abs/1911.11977}{{\tt arXiv:1911.11977}}].

\bibitem{Almheiri:2019qdq}
A.~Almheiri, T.~Hartman, J.~Maldacena, E.~Shaghoulian, and A.~Tajdini, {\it {Replica Wormholes and the Entropy of Hawking Radiation}},  {\em JHEP} {\bf 05} (2020) 013, [\href{http://arxiv.org/abs/1911.12333}{{\tt arXiv:1911.12333}}].

\bibitem{tHooft:1993dmi}
G.~'t~Hooft, {\it {Dimensional reduction in quantum gravity}},  {\em Conf. Proc. C} {\bf 930308} (1993) 284--296, [\href{http://arxiv.org/abs/gr-qc/9310026}{{\tt gr-qc/9310026}}].

\bibitem{Susskind:1994vu}
L.~Susskind, {\it {The World as a hologram}},  {\em J. Math. Phys.} {\bf 36} (1995) 6377--6396, [\href{http://arxiv.org/abs/hep-th/9409089}{{\tt hep-th/9409089}}].

\bibitem{Maldacena:1997re}
J.~M. Maldacena, {\it {The Large N limit of superconformal field theories and supergravity}},  {\em Adv. Theor. Math. Phys.} {\bf 2} (1998) 231--252, [\href{http://arxiv.org/abs/hep-th/9711200}{{\tt hep-th/9711200}}].

\bibitem{Witten:1998qj}
E.~Witten, {\it {Anti-de Sitter space and holography}},  {\em Adv. Theor. Math. Phys.} {\bf 2} (1998) 253--291, [\href{http://arxiv.org/abs/hep-th/9802150}{{\tt hep-th/9802150}}].

\bibitem{Aharony:1999ti}
O.~Aharony, S.~S. Gubser, J.~M. Maldacena, H.~Ooguri, and Y.~Oz, {\it {Large N field theories, string theory and gravity}},  {\em Phys. Rept.} {\bf 323} (2000) 183--386, [\href{http://arxiv.org/abs/hep-th/9905111}{{\tt hep-th/9905111}}].

\bibitem{Harlow:2013tf}
D.~Harlow and P.~Hayden, {\it {Quantum Computation vs. Firewalls}},  {\em JHEP} {\bf 06} (2013) 085, [\href{http://arxiv.org/abs/1301.4504}{{\tt arXiv:1301.4504}}].

\bibitem{Brown:2019rox}
A.~R. Brown, H.~Gharibyan, G.~Penington, and L.~Susskind, {\it {The Python\textquoteright{}s Lunch: geometric obstructions to decoding Hawking radiation}},  {\em JHEP} {\bf 08} (2020) 121, [\href{http://arxiv.org/abs/1912.00228}{{\tt arXiv:1912.00228}}].

\bibitem{Hayden:2018khn}
P.~Hayden and G.~Penington, {\it {Learning the Alpha-bits of Black Holes}},  {\em JHEP} {\bf 12} (2019) 007, [\href{http://arxiv.org/abs/1807.06041}{{\tt arXiv:1807.06041}}].

\bibitem{Akers:2019wxj}
C.~Akers, S.~Leichenauer, and A.~Levine, {\it {Large Breakdowns of Entanglement Wedge Reconstruction}},  {\em Phys. Rev. D} {\bf 100} (2019), no.~12 126006, [\href{http://arxiv.org/abs/1908.03975}{{\tt arXiv:1908.03975}}].

\bibitem{Akers:2021fut}
C.~Akers and G.~Penington, {\it {Quantum minimal surfaces from quantum error correction}},  {\em SciPost Phys.} {\bf 12} (2022), no.~5 157, [\href{http://arxiv.org/abs/2109.14618}{{\tt arXiv:2109.14618}}].

\bibitem{Engelhardt:2021mue}
N.~Engelhardt, G.~Penington, and A.~Shahbazi-Moghaddam, {\it {A world without pythons would be so simple}},  {\em Class. Quant. Grav.} {\bf 38} (2021), no.~23 234001, [\href{http://arxiv.org/abs/2102.07774}{{\tt arXiv:2102.07774}}].

\bibitem{Akers:2022qdl}
C.~Akers, N.~Engelhardt, D.~Harlow, G.~Penington, and S.~Vardhan, {\it {The black hole interior from non-isometric codes and complexity}},  {\em JHEP} {\bf 06} (2024) 155, [\href{http://arxiv.org/abs/2207.06536}{{\tt arXiv:2207.06536}}].

\bibitem{Geng:2020fxl}
H.~Geng, A.~Karch, C.~Perez-Pardavila, S.~Raju, L.~Randall, M.~Riojas, and S.~Shashi, {\it {Information Transfer with a Gravitating Bath}},  {\em SciPost Phys.} {\bf 10} (2021), no.~5 103, [\href{http://arxiv.org/abs/2012.04671}{{\tt arXiv:2012.04671}}].

\bibitem{Geng:2020qvw}
H.~Geng and A.~Karch, {\it {Massive islands}},  {\em JHEP} {\bf 09} (2020) 121, [\href{http://arxiv.org/abs/2006.02438}{{\tt arXiv:2006.02438}}].

\bibitem{Geng:2021hlu}
H.~Geng, A.~Karch, C.~Perez-Pardavila, S.~Raju, L.~Randall, M.~Riojas, and S.~Shashi, {\it {Inconsistency of islands in theories with long-range gravity}},  {\em JHEP} {\bf 01} (2022) 182, [\href{http://arxiv.org/abs/2107.03390}{{\tt arXiv:2107.03390}}].

\bibitem{Raju:2020smc}
S.~Raju, {\it {Lessons from the information paradox}},  {\em Phys. Rept.} {\bf 943} (2022) 1--80, [\href{http://arxiv.org/abs/2012.05770}{{\tt arXiv:2012.05770}}].

\bibitem{Geng:2023zhq}
H.~Geng, {\it {Graviton Mass and Entanglement Islands in Low Spacetime Dimensions}},  \href{http://arxiv.org/abs/2312.13336}{{\tt arXiv:2312.13336}}.

\bibitem{Geng:2025rov}
H.~Geng, {\it {The Mechanism behind the Information Encoding for Islands}},  \href{http://arxiv.org/abs/2502.08703}{{\tt arXiv:2502.08703}}.

\bibitem{Laddha_2021}
A.~Laddha, S.~Prabhu, S.~Raju, and P.~Shrivastava, {\it The holographic nature of null infinity},  {\em SciPost Physics} {\bf 10} (Feb., 2021).

\bibitem{Donnelly:2015hta}
W.~Donnelly and S.~B. Giddings, {\it {Diffeomorphism-invariant observables and their nonlocal algebra}},  {\em Phys. Rev. D} {\bf 93} (2016), no.~2 024030, [\href{http://arxiv.org/abs/1507.07921}{{\tt arXiv:1507.07921}}]. [Erratum: Phys.Rev.D 94, 029903 (2016)].

\bibitem{Donnelly:2016rvo}
W.~Donnelly and S.~B. Giddings, {\it {Observables, gravitational dressing, and obstructions to locality and subsystems}},  {\em Phys. Rev. D} {\bf 94} (2016), no.~10 104038, [\href{http://arxiv.org/abs/1607.01025}{{\tt arXiv:1607.01025}}].

\bibitem{Giddings:2005id}
S.~B. Giddings, D.~Marolf, and J.~B. Hartle, {\it {Observables in effective gravity}},  {\em Phys. Rev. D} {\bf 74} (2006) 064018, [\href{http://arxiv.org/abs/hep-th/0512200}{{\tt hep-th/0512200}}].

\bibitem{Bahiru:2022oas}
E.~Bahiru, A.~Belin, K.~Papadodimas, G.~Sarosi, and N.~Vardian, {\it {State-dressed local operators in the AdS/CFT correspondence}},  {\em Phys. Rev. D} {\bf 108} (2023), no.~8 086035, [\href{http://arxiv.org/abs/2209.06845}{{\tt arXiv:2209.06845}}].

\bibitem{Bahiru:2023zlc}
E.~Bahiru, A.~Belin, K.~Papadodimas, G.~Sarosi, and N.~Vardian, {\it {Holography and Localization of Information in Quantum Gravity}},  \href{http://arxiv.org/abs/2301.08753}{{\tt arXiv:2301.08753}}.

\bibitem{Kourkoulou:2017zaj}
I.~Kourkoulou and J.~Maldacena, {\it {Pure states in the SYK model and nearly-$AdS_2$ gravity}},  \href{http://arxiv.org/abs/1707.02325}{{\tt arXiv:1707.02325}}.

\bibitem{Bousso:2021sji}
R.~Bousso and A.~Shahbazi-Moghaddam, {\it {Island Finder and Entropy Bound}},  {\em Phys. Rev. D} {\bf 103} (2021), no.~10 106005, [\href{http://arxiv.org/abs/2101.11648}{{\tt arXiv:2101.11648}}].

\bibitem{Ryu2006a}
S.~Ryu and T.~Takayanagi, {\it {Aspects of Holographic Entanglement Entropy}},  {\em JHEP} {\bf 08} (2006) 045, [\href{http://arxiv.org/abs/hep-th/0605073}{{\tt hep-th/0605073}}].

\bibitem{Ryu2006b}
S.~Ryu and T.~Takayanagi, {\it Holographic derivation of entanglement entropy from the anti--de sitter space/conformal field theory correspondence},  {\em Physical review letters} {\bf 96} (2006), no.~18 181602.

\bibitem{Hubeny:2007xt}
V.~E. Hubeny, M.~Rangamani, and T.~Takayanagi, {\it {A Covariant holographic entanglement entropy proposal}},  {\em JHEP} {\bf 07} (2007) 062, [\href{http://arxiv.org/abs/0705.0016}{{\tt arXiv:0705.0016}}].

\bibitem{Engelhardt:2014gca}
N.~Engelhardt and A.~C. Wall, {\it {Quantum Extremal Surfaces: Holographic Entanglement Entropy beyond the Classical Regime}},  {\em JHEP} {\bf 01} (2015) 073, [\href{http://arxiv.org/abs/1408.3203}{{\tt arXiv:1408.3203}}].

\bibitem{Colin-Ellerin:2025dgq}
S.~Colin-Ellerin, G.~Lin, and G.~Penington, {\it {Generalized entropy of gravitational fluctuations}},  \href{http://arxiv.org/abs/2501.08308}{{\tt arXiv:2501.08308}}.

\bibitem{Lewkowycz:2013nqa}
A.~Lewkowycz and J.~Maldacena, {\it {Generalized gravitational entropy}},  {\em JHEP} {\bf 08} (2013) 090, [\href{http://arxiv.org/abs/1304.4926}{{\tt arXiv:1304.4926}}].

\bibitem{Faulkner:2013ana}
T.~Faulkner, A.~Lewkowycz, and J.~Maldacena, {\it {Quantum corrections to holographic entanglement entropy}},  {\em JHEP} {\bf 11} (2013) 074, [\href{http://arxiv.org/abs/1307.2892}{{\tt arXiv:1307.2892}}].

\bibitem{Wall:2012uf}
A.~C. Wall, {\it {Maximin Surfaces, and the Strong Subadditivity of the Covariant Holographic Entanglement Entropy}},  {\em Class. Quant. Grav.} {\bf 31} (2014), no.~22 225007, [\href{http://arxiv.org/abs/1211.3494}{{\tt arXiv:1211.3494}}].

\bibitem{Akers:2019lzs}
C.~Akers, N.~Engelhardt, G.~Penington, and M.~Usatyuk, {\it {Quantum Maximin Surfaces}},  {\em JHEP} {\bf 08} (2020) 140, [\href{http://arxiv.org/abs/1912.02799}{{\tt arXiv:1912.02799}}].

\bibitem{Bousso:2015mna}
R.~Bousso, Z.~Fisher, S.~Leichenauer, and A.~C. Wall, {\it {Quantum focusing conjecture}},  {\em Phys. Rev. D} {\bf 93} (2016), no.~6 064044, [\href{http://arxiv.org/abs/1506.02669}{{\tt arXiv:1506.02669}}].

\bibitem{Hayden:2007cs}
P.~Hayden and J.~Preskill, {\it {Black holes as mirrors: Quantum information in random subsystems}},  {\em JHEP} {\bf 09} (2007) 120, [\href{http://arxiv.org/abs/0708.4025}{{\tt arXiv:0708.4025}}].

\bibitem{Czech:2012bh}
B.~Czech, J.~L. Karczmarek, F.~Nogueira, and M.~Van~Raamsdonk, {\it {The Gravity Dual of a Density Matrix}},  {\em Class. Quant. Grav.} {\bf 29} (2012) 155009, [\href{http://arxiv.org/abs/1204.1330}{{\tt arXiv:1204.1330}}].

\bibitem{Headrick:2014cta}
M.~Headrick, V.~E. Hubeny, A.~Lawrence, and M.~Rangamani, {\it {Causality \& holographic entanglement entropy}},  {\em JHEP} {\bf 12} (2014) 162, [\href{http://arxiv.org/abs/1408.6300}{{\tt arXiv:1408.6300}}].

\bibitem{Jafferis:2015del}
D.~L. Jafferis, A.~Lewkowycz, J.~Maldacena, and S.~J. Suh, {\it {Relative entropy equals bulk relative entropy}},  {\em JHEP} {\bf 06} (2016) 004, [\href{http://arxiv.org/abs/1512.06431}{{\tt arXiv:1512.06431}}].

\bibitem{Almheiri:2014lwa}
A.~Almheiri, X.~Dong, and D.~Harlow, {\it {Bulk Locality and Quantum Error Correction in AdS/CFT}},  {\em JHEP} {\bf 04} (2015) 163, [\href{http://arxiv.org/abs/1411.7041}{{\tt arXiv:1411.7041}}].

\bibitem{Cotler:2017erl}
J.~Cotler, P.~Hayden, G.~Penington, G.~Salton, B.~Swingle, and M.~Walter, {\it {Entanglement Wedge Reconstruction via Universal Recovery Channels}},  {\em Phys. Rev. X} {\bf 9} (2019), no.~3 031011, [\href{http://arxiv.org/abs/1704.05839}{{\tt arXiv:1704.05839}}].

\bibitem{Dong:2016eik}
X.~Dong, D.~Harlow, and A.~C. Wall, {\it {Reconstruction of Bulk Operators within the Entanglement Wedge in Gauge-Gravity Duality}},  {\em Phys. Rev. Lett.} {\bf 117} (2016), no.~2 021601, [\href{http://arxiv.org/abs/1601.05416}{{\tt arXiv:1601.05416}}].

\bibitem{Faulkner:2017vdd}
T.~Faulkner and A.~Lewkowycz, {\it {Bulk locality from modular flow}},  {\em JHEP} {\bf 07} (2017) 151, [\href{http://arxiv.org/abs/1704.05464}{{\tt arXiv:1704.05464}}].

\bibitem{Chen:2019gbt}
C.-F. Chen, G.~Penington, and G.~Salton, {\it {Entanglement Wedge Reconstruction using the Petz Map}},  {\em JHEP} {\bf 01} (2020) 168, [\href{http://arxiv.org/abs/1902.02844}{{\tt arXiv:1902.02844}}].

\bibitem{deRham:2014zqa}
C.~de~Rham, {\it {Massive Gravity}},  {\em Living Rev. Rel.} {\bf 17} (2014) 7, [\href{http://arxiv.org/abs/1401.4173}{{\tt arXiv:1401.4173}}].

\bibitem{harlow2019symmetriesquantumfieldtheory}
D.~Harlow and H.~Ooguri, {\it Symmetries in quantum field theory and quantum gravity},  2019.

\bibitem{Karch:2000ct}
A.~Karch and L.~Randall, {\it {Locally localized gravity}},  {\em JHEP} {\bf 05} (2001) 008, [\href{http://arxiv.org/abs/hep-th/0011156}{{\tt hep-th/0011156}}].

\bibitem{Aharony:2003qf}
O.~Aharony, O.~DeWolfe, D.~Z. Freedman, and A.~Karch, {\it {Defect conformal field theory and locally localized gravity}},  {\em JHEP} {\bf 07} (2003) 030, [\href{http://arxiv.org/abs/hep-th/0303249}{{\tt hep-th/0303249}}].

\bibitem{Aharony:2006hz}
O.~Aharony, A.~B. Clark, and A.~Karch, {\it {The CFT/AdS correspondence, massive gravitons and a connectivity index conjecture}},  {\em Phys. Rev. D} {\bf 74} (2006) 086006, [\href{http://arxiv.org/abs/hep-th/0608089}{{\tt hep-th/0608089}}].

\bibitem{Cooper:2018cmb}
S.~Cooper, M.~Rozali, B.~Swingle, M.~Van~Raamsdonk, C.~Waddell, and D.~Wakeham, {\it {Black hole microstate cosmology}},  {\em JHEP} {\bf 07} (2019) 065, [\href{http://arxiv.org/abs/1810.10601}{{\tt arXiv:1810.10601}}].

\bibitem{Antonini:2019qkt}
S.~Antonini and B.~Swingle, {\it {Cosmology at the end of the world}},  {\em Nature Phys.} {\bf 16} (2020), no.~8 881--886, [\href{http://arxiv.org/abs/1907.06667}{{\tt arXiv:1907.06667}}].

\bibitem{Almheiri:2019hni}
A.~Almheiri, R.~Mahajan, J.~Maldacena, and Y.~Zhao, {\it {The Page curve of Hawking radiation from semiclassical geometry}},  {\em JHEP} {\bf 03} (2020) 149, [\href{http://arxiv.org/abs/1908.10996}{{\tt arXiv:1908.10996}}].

\bibitem{Antonini:2024bbm}
S.~Antonini and L.~G.~C. Bariuan, {\it {Magnetic braneworlds: cosmology and wormholes}},  {\em JHEP} {\bf 09} (2024) 070, [\href{http://arxiv.org/abs/2405.18465}{{\tt arXiv:2405.18465}}].

\bibitem{Porrati:2001gx}
M.~Porrati, {\it {Mass and gauge invariance 4. Holography for the Karch-Randall model}},  {\em Phys. Rev. D} {\bf 65} (2002) 044015, [\href{http://arxiv.org/abs/hep-th/0109017}{{\tt hep-th/0109017}}].

\bibitem{Porrati:2002dt}
M.~Porrati and A.~Starinets, {\it {On the graviton selfenergy in AdS(4)}},  {\em Phys. Lett. B} {\bf 532} (2002) 48--54, [\href{http://arxiv.org/abs/hep-th/0201261}{{\tt hep-th/0201261}}].

\bibitem{Porrati:2003sa}
M.~Porrati, {\it {Higgs phenomenon for the graviton in ADS space}},  {\em Mod. Phys. Lett. A} {\bf 18} (2003) 1793--1802, [\href{http://arxiv.org/abs/hep-th/0306253}{{\tt hep-th/0306253}}].

\bibitem{Rattazzi:2009ux}
R.~Rattazzi and M.~Redi, {\it {Gauge Boson Mass Generation in AdS4}},  {\em JHEP} {\bf 12} (2009) 025, [\href{http://arxiv.org/abs/0908.4150}{{\tt arXiv:0908.4150}}].

\bibitem{Karch:2023wui}
A.~Karch, M.~Wang, and M.~Youssef, {\it {AdS Higgs mechanism from double trace deformed CFT}},  {\em JHEP} {\bf 02} (2024) 044, [\href{http://arxiv.org/abs/2311.10135}{{\tt arXiv:2311.10135}}].

\bibitem{Bousso:2019ykv}
R.~Bousso and M.~Toma\v{s}evi\'c, {\it {Unitarity From a Smooth Horizon?}},  {\em Phys. Rev. D} {\bf 102} (2020), no.~10 106019, [\href{http://arxiv.org/abs/1911.06305}{{\tt arXiv:1911.06305}}].

\bibitem{Headrick:2007km}
M.~Headrick and T.~Takayanagi, {\it {A Holographic proof of the strong subadditivity of entanglement entropy}},  {\em Phys. Rev. D} {\bf 76} (2007) 106013, [\href{http://arxiv.org/abs/0704.3719}{{\tt arXiv:0704.3719}}].

\bibitem{Hubeny:2013gta}
V.~E. Hubeny, H.~Maxfield, M.~Rangamani, and E.~Tonni, {\it {Holographic entanglement plateaux}},  {\em JHEP} {\bf 08} (2013) 092, [\href{http://arxiv.org/abs/1306.4004}{{\tt arXiv:1306.4004}}].

\bibitem{Page1}
D.~N. Page, {\it Particle emission rates from a black hole: Massless particles from an uncharged, nonrotating hole},  {\em Phys. Rev. D} {\bf 13} (Jan, 1976) 198--206.

\bibitem{Prabhu:2022zcr}
K.~Prabhu, G.~Satishchandran, and R.~M. Wald, {\it {Infrared finite scattering theory in quantum field theory and quantum gravity}},  {\em Phys. Rev. D} {\bf 106} (2022), no.~6 066005, [\href{http://arxiv.org/abs/2203.14334}{{\tt arXiv:2203.14334}}].

\bibitem{Raju:2021lwh}
S.~Raju, {\it {Failure of the split property in gravity and the information paradox}},  {\em Class. Quant. Grav.} {\bf 39} (2022), no.~6 064002, [\href{http://arxiv.org/abs/2110.05470}{{\tt arXiv:2110.05470}}].

\bibitem{Raju:2024gvc}
S.~Raju, {\it {How does information emerge from a black hole?}},  \href{http://arxiv.org/abs/2404.00374}{{\tt arXiv:2404.00374}}.

\bibitem{Hamilton:2006az}
A.~Hamilton, D.~N. Kabat, G.~Lifschytz, and D.~A. Lowe, {\it {Holographic representation of local bulk operators}},  {\em Phys. Rev. D} {\bf 74} (2006) 066009, [\href{http://arxiv.org/abs/hep-th/0606141}{{\tt hep-th/0606141}}].

\bibitem{Bousso:2023kdj}
R.~Bousso and G.~Penington, {\it {Islands far outside the horizon}},  {\em JHEP} {\bf 11} (2024) 164, [\href{http://arxiv.org/abs/2312.03078}{{\tt arXiv:2312.03078}}].

\bibitem{Kitaev:2024qak}
A.~Kitaev, {\it {Almost-idempotent quantum channels and approximate $C^*$-algebras}},  \href{http://arxiv.org/abs/2405.02434}{{\tt arXiv:2405.02434}}.

\bibitem{Shor:1996qc}
P.~Shor, {\it Fault-tolerant quantum computation},  in {\em Proceedings of 37th Conference on Foundations of Computer Science}, pp.~56--65, 1996.
\newblock \href{http://arxiv.org/abs/quant-ph/9605011}{{\tt quant-ph/9605011}}.

\bibitem{Aharonov:1999ei}
D.~Aharonov and M.~Ben-Or, {\it {Fault-Tolerant Quantum Computation with Constant Error Rate}},  {\em SIAM J. Comput.} {\bf 38} (2008), no.~4 1207--1282, [\href{http://arxiv.org/abs/quant-ph/9906129}{{\tt quant-ph/9906129}}].

\bibitem{Knill:1997mr}
E.~Knill, R.~Laflamme, and W.~H. Zurek, {\it {Resilient quantum computation: Error models and threshold}},  {\em Proc. Roy. Soc. Lond. A} {\bf 454} (1998) 365--384, [\href{http://arxiv.org/abs/quant-ph/9702058}{{\tt quant-ph/9702058}}].

\bibitem{Kitaev:1997wr}
A.~Y. Kitaev, {\it {Fault tolerant quantum computation by anyons}},  {\em Annals Phys.} {\bf 303} (2003) 2--30, [\href{http://arxiv.org/abs/quant-ph/9707021}{{\tt quant-ph/9707021}}].

\bibitem{Chowdhury:2021nxw}
C.~Chowdhury, V.~Godet, O.~Papadoulaki, and S.~Raju, {\it {Holography from the Wheeler-DeWitt equation}},  {\em JHEP} {\bf 03} (2022) 019, [\href{http://arxiv.org/abs/2107.14802}{{\tt arXiv:2107.14802}}].

\bibitem{Isham:1992ms}
C.~J. Isham, {\it {Canonical quantum gravity and the problem of time}},  {\em NATO Sci. Ser. C} {\bf 409} (1993) 157--287, [\href{http://arxiv.org/abs/gr-qc/9210011}{{\tt gr-qc/9210011}}].

\bibitem{Marolf:1994wh}
D.~Marolf, {\it {Quantum observables and recollapsing dynamics}},  {\em Class. Quant. Grav.} {\bf 12} (1995) 1199--1220, [\href{http://arxiv.org/abs/gr-qc/9404053}{{\tt gr-qc/9404053}}].

\bibitem{Geng:2024dbl}
H.~Geng, {\it {Quantum Rods and Clock in a Gravitational Universe}},  \href{http://arxiv.org/abs/2412.03636}{{\tt arXiv:2412.03636}}.

\bibitem{fischer1973linearization}
A.~E. Fischer and J.~E. Marsden, {\it Linearization stability of the einstein equations}, .

\bibitem{moncrief1975spacetime}
V.~Moncrief, {\it Spacetime symmetries and linearization stability of the einstein equations. i},  {\em Journal of Mathematical Physics} {\bf 16} (1975), no.~3 493--498.

\bibitem{Folkestad:2023cze}
r.~Folkestad, {\it {Subregion Independence in Gravity}},  \href{http://arxiv.org/abs/2311.09403}{{\tt arXiv:2311.09403}}.

\bibitem{DeWitt:1962cg}
B.~S. DeWitt, {\it {The Quantization of geometry}}, .

\bibitem{Almheiri:2019yqk}
A.~Almheiri, R.~Mahajan, and J.~Maldacena, {\it {Islands outside the horizon}},  \href{http://arxiv.org/abs/1910.11077}{{\tt arXiv:1910.11077}}.

\bibitem{Papadodimas:2015jra}
K.~Papadodimas and S.~Raju, {\it {Remarks on the necessity and implications of state-dependence in the black hole interior}},  {\em Phys. Rev. D} {\bf 93} (2016), no.~8 084049, [\href{http://arxiv.org/abs/1503.08825}{{\tt arXiv:1503.08825}}].

\bibitem{Papadodimas:2015xma}
K.~Papadodimas and S.~Raju, {\it {Local Operators in the Eternal Black Hole}},  {\em Phys. Rev. Lett.} {\bf 115} (2015), no.~21 211601, [\href{http://arxiv.org/abs/1502.06692}{{\tt arXiv:1502.06692}}].

\bibitem{Jensen:2024dnl}
K.~Jensen, S.~Raju, and A.~J. Speranza, {\it {Holographic observers for time-band algebras}},  \href{http://arxiv.org/abs/2412.21185}{{\tt arXiv:2412.21185}}.

\bibitem{Saad:2019lba}
P.~Saad, S.~H. Shenker, and D.~Stanford, {\it {JT gravity as a matrix integral}},  \href{http://arxiv.org/abs/1903.11115}{{\tt arXiv:1903.11115}}.

\bibitem{Stanford:2022fdt}
D.~Stanford and Z.~Yang, {\it {Firewalls from wormholes}},  \href{http://arxiv.org/abs/2208.01625}{{\tt arXiv:2208.01625}}.

\bibitem{Iliesiu:2024cnh}
L.~V. Iliesiu, A.~Levine, H.~W. Lin, H.~Maxfield, and M.~Mezei, {\it {On the non-perturbative bulk Hilbert space of JT gravity}},  \href{http://arxiv.org/abs/2403.08696}{{\tt arXiv:2403.08696}}.

\end{thebibliography}\endgroup

\end{document}